\documentclass[a4paper,11pt]{article}
\usepackage{jheppub}
\pdfoutput=1
\usepackage[utf8]{inputenc}
\usepackage{epsfig}
\usepackage{amsmath}
\usepackage{slashed}
\usepackage{amsfonts} 
\usepackage{float} 
\usepackage{amssymb}
\usepackage{xcolor}
\usepackage{pbox}
\usepackage{subfigure}
\usepackage{array,multirow,makecell}
\usepackage{hyperref}
\usepackage{tabularx}
\usepackage{titlesec}
\usepackage{natbib}
\usepackage{array,multirow,makecell}
\usepackage{lipsum}
\usepackage{comment}
\usepackage[section]{placeins}
\usepackage{cancel}
\usepackage[normalem]{ulem}
\usepackage[usestackEOL]{stackengine}
\strutlongstacks{T}

\def\b{\beta}
\def\g{\gamma}

\def\k{\kappa}
\def\l{\lambda}
\def\wt{\widetilde}
\def\m{\mu}

\newcommand{\ba}{\begin{array}}
\newcommand{\ea}{\end{array}}

\def\lsim{\raise0.3ex\hbox{$\;<$\kern-0.75em\raise-1.1ex\hbox{$\sim\;$}}}
\def\gsim{\raise0.3ex\hbox{$\;>$\kern-0.75em\raise-1.1ex\hbox{$\sim\;$}}}
\def\be{\begin{equation}}
	\def\ee{\end{equation}}
\def\bea{\begin{eqnarray}}
	\def\eea{\end{eqnarray}}
\def\nn{\nonumber}
\newcommand{\beq}{\begin{equation}}
	\newcommand{\eeq}{\end{equation}}
\newcommand{\beqn}{\begin{eqnarray}}
	\newcommand{\eeqn}{\end{eqnarray}}

\def\lsim{\raise0.3ex\hbox{$\;<$\kern-0.75em\raise-1.1ex\hbox{$\sim\;$}}}
\def\gsim{\raise0.3ex\hbox{$\;>$\kern-0.75em\raise-1.1ex\hbox{$\sim\;$}}}
\def\be{\begin{equation}}
	\def\ee{\end{equation}}
\def\bea{\begin{eqnarray}}
	\def\eea{\end{eqnarray}}
\def\nn{\nonumber}
\title{\boldmath Electroweak Phase Transition in a Right-Handed Neutrino Superfield Extended NMSSM}
\author[a]{Pankaj Borah,}
\author[a]{Pradipta Ghosh,}
\author[b]{Sourov Roy}
\author[b]{and Abhijit Kumar Saha}
\affiliation[a]{Department of Physics, Indian Institute of Technology Delhi, Hauz Khas 110 016, India}
\affiliation[b]{School of Physical Sciences, Indian Association for the Cultivation of Science, 2A $\&$ 2B Raja S.C. Mullick Road, Kolkata 700 032, India}
\emailAdd{Pankaj.Borah@physics.iitd.ac.in}
\emailAdd{tphyspg@physics.iitd.ac.in}
\emailAdd{tpsr@iacs.res.in}
\emailAdd{psaks2484@iacs.res.in}
\abstract{Supersymmetric models with singlet extensions can accommodate single- or multi-step first-order phase transitions (FOPT) along the various constituent field directions. Such a framework can also produce Gravitational Waves, detectable at the upcoming space-based interferometers, e.g., U-DECIGO. We explore the dynamics of electroweak phase transition and the production of Gravitational Waves in an extended set-up of the Next-to-Minimal Supersymmetric Standard Model (NMSSM) with a Standard Model singlet right-handed neutrino superfield. We examine the role of the new parameters compared to NMSSM on the phase transition dynamics and observe that the occurrence of a FOPT, an essential requirement for Electroweak Baryogenesis, typically favours a right-handed sneutrino state below 125 GeV. Our investigation shows how the analysis can offer complementary probes for physics beyond the Standard Model besides the collider searches.}
\begin{document} 
\maketitle
\section{Introduction}\label{sec:Intro}

Baryon asymmetry of the Universe is a precisely measured quantity by Planck experiment \cite{Planck:2018vyg}. Different kinds of proposals pertaining to baryon asymmetry production mechanism in the early Universe are prevalent in literature (for a brief summary see Ref.\,\cite{Cline:2006ts}). In recent times, baryon asymmetry production during the Electroweak Phase Transition (EWPT), known as the Electroweak Baryogenesis (EWBG) \cite{Anderson:1991zb, Morrissey:2012db} has gained particular attention. The EWBG occurs around the TeV scale and has the potential to be probed in collider experiments \cite{Curtin:2014jma,Artymowski:2016tme,Ramsey-Musolf:2019lsf, Modak:2020uyq}. Irrespective of different baryon asymmetry generation mechanisms, the Sakharov conditions \cite{Sakharov:1967dj}, namely, (i) baryon number violation, (ii) charge (C) and charge-parity (CP) violation and (iii) deviation from thermal equilibrium must be satisfied. 
	
It is well known that the Standard Model (SM) of particle physics fails to provide a sufficient departure from thermal equilibrium \cite{Gavela:1994dt,Bernreuther:2002uj}. Moreover, C and CP violations in the SM are not adequate enough to yield the observed baryon asymmetry of the Universe \cite{Gavela:1994dt,Bernreuther:2002uj}. In principle, a strong first-order EWPT (SFOEWPT) in the early Universe can pave the way for the EWBG by allowing sufficient out-of-equilibrium processes \cite{Morrissey:2012db}. The SM of particle physics with the observed Higgs mass $\sim 125$ GeV \cite{ATLAS:2012yve,CMS:2012qbp}, shows a smooth cross-over pattern along the Higgs field direction without any PT \cite{Kajantie:1996mn,Rummukainen:1998as,Csikor:1998eu} and thus, fails to accommodate the EWBG. This issue can be circumvented by introducing new scalar degrees of freedom having sizable coupling with the SM Higgs boson. In general, the strength of the EW phase transition is determined by both the high and low-temperature behaviour of the scalar potential. Computation of critical temperature reveals displacement of the global minimum for a scalar potential when expressed as a function of the temperature (T) of the Universe. However, a correct description of the EWPT requires the study of bubble nucleation dynamics since PT proceeds via the nucleation of bubbles \cite{Mazumdar:2018dfl}. The dynamics of bubble nucleation, during the first-order EWPT, can yield stochastic Gravitational Waves (GWs) in the early Universe \cite{Apreda:2001us,Grojean:2004xa,Weir:2017wfa,Ellis:2018mja,Alanne:2019bsm,Caprini:2019egz} that may appear detectable at different GW experiments. In fact, the search for GWs for probing different kinds of beyond the SM (BSM) frameworks has long been practiced \cite{Costa:2022oaa,Ahriche:2018rao,Shajiee:2018jdq,Alves:2018jsw,Chatterjee:2022pxf,Wang:2022lxn}, often as a complementary probe besides the collider searches \cite{Ramsey-Musolf:2019lsf,Friedrich:2022cak}.  
	
Supersymmetric models, having a rich scalar sector compared to the SM, carry the necessary ingredients for exhibiting an SFOEWPT. The PT properties in the Minimal Supersymmetric Standard Model (MSSM) (see Ref. \cite{Haber:1984rc} for a review) are exercised in Refs. \cite{Espinosa:1993yi,Carena:1996wj,Delepine:1996vn,Laine:1998qk,Cline:1998hy,Balazs:2004ae,Lee:2004we,Carena:2008vj,Menon:2009mz,Carena:2012np}. Following PTs, the EWBG in the MSSM is also extensively studied in the literature \cite{Cirigliano:2006dg,Chung:2008aya,Chung:2009qs,Cirigliano:2009yd,Morrissey:2012db,Kozaczuk:2012xv,Kozaczuk:2013fga,Katz:2015uja}.
It is shown in Ref.\,\cite{Carena:2012np} that a strong EWPT with a 125 GeV Higgs boson favours a hierarchical stop sector in the MSSM, i.e., one of two stops appears to be much heavier than the EW scale while the lighter one remains around $\mathcal{O}(100$ GeV) \cite{Menon:2009mz,Carena:2012np}. The presence of such a light stop enhances the Higgs production rate through gluon-gluon fusion \cite{Curtin:2012aa,Carena:2012np} and confronts constraints from LHC data \cite{ATLAS:2012yve,CMS:2012qbp}. This tension, nevertheless, can be alleviated by considering a light neutralino with a mass lower than about 60 GeV \cite{Carena:2012np}. However, once again it is challenged by the LHC data of Higgs invisible decay width  \cite{ATLAS:2021gcn,CMS:2022qva,ATLAS:2022yvh,CMS:2022dwd} and neutralino searches from the stop decay \cite{Krizka:2012ah,Delgado:2012eu,Grober:2014aha,Aebischer:2014lfa}.
Besides, the MSSM also suffers from a new kind of naturalness problem known as the $\mu$-problem \cite{Kim:1983dt} and, just like the SM, is incapable of accommodating non-zero neutrino masses and mixing \cite{Esteban:2020cvm,Gonzalez-Garcia:2021dve} in its original form\footnote{MSSM extended with new superfields or new symmetries or R-parity violation \cite{Hall:1983id} (see Refs. \cite{Hirsch:2004he,Barbier:2004ez,Mohapatra:2005wg} for further reading) can accommodate neutrino data \cite{deSalas:2020pgw,Esteban:2020cvm,Gonzalez-Garcia:2021dve}. R-parity is defined as $R_P=(-1)^{3B+L+2s}$ where $L(B)$ denotes the lepton (baryon) number and $s$ represents the spin.}.

The Next-to-Minimal Supersymmetric Standard Model (NMSSM) \cite{Ellwanger:2009dp} provides a dynamical solution to the $\mu$-problem, a challenge that has plagued the MSSM. The superfield content of the NMSSM is enhanced compared to the MSSM as it includes a new SM gauge-singlet superfield $\hat S$. The presence of $\hat S$ offers a dynamical solution to the $\mu$-problem and, simultaneously augments the particle spectrum over the MSSM. Studies related to EWPT in the NMSSM can be found in Refs. \cite{Pietroni:1992in,Davies:1996qn,Menon:2004wv,Huber:2006wf,Carena:2011jy,Balazs:2013cia,Huang:2014ifa,Kozaczuk:2014kva,Baum:2020vfl,Chatterjee:2022pxf,Wang:2022lxn}. It has been observed \cite{Pietroni:1992in,Davies:1996qn,Huber:2006wf,Huang:2014ifa,Kozaczuk:2014kva,Baum:2020vfl} that in the NMSSM soft supersymmetry (SUSY) breaking term involving $S$ and Higgs doublets assists to form the potential barrier even at $T=0$ in contrast to the MSSM where $T\neq0$ effects are essential for barrier formation. Thus, the PT dynamics is more involved in the NMSSM where one needs to consider a three-dimensional field space spanned by three\footnote{The PT dynamics in guided by a two-dimensional field space in the MSSM \cite{Carena:2008rt,Carena:2008vj}.} CP-even scalar fields. 
	
The EWPT could occur either in single-step or multi-step. In the NMSSM, both single-step and multi-step phase transitions are possible as discussed in Ref. \cite{Kozaczuk:2014kva,Baum:2020vfl}. These studies \cite{Kozaczuk:2014kva,Baum:2020vfl} rely on an effective field theory set-up after integrating out heavy stops which yield  potentially large contributions to the one-loop effective potential. Such an effective-theory-based approach reduces the degrees of freedom participating in the EWPT dynamics.  Refs.\,\cite{Kozaczuk:2014kva,Baum:2020vfl} also showed that the NMSSM can accommodate EWBG in some region corners of the NMSSM parameter space.

Shifting our attention to non-zero neutrino masses and mixing \cite{deSalas:2020pgw,Esteban:2020cvm,Gonzalez-Garcia:2021dve}, another experimentally established BSM signature, both MSSM and NMSSM, are futile just like the SM. Extensions of these models with additional ingredients, e.g., right-handed (RH) neutrinos, however, offer a simple elegant way to accommodate massive neutrinos using the popular type-I see-saw mechanism \cite{Minkowski:1977sc,Yanagida:1979as,Gell-Mann:1979vob,Mohapatra:1979ia}. Supersymmetric type-I seesaw mechanism, where the MSSM superfield content is extended with RH-neutrino superfield(s) is well studied, see for example, Refs. \cite{Hisano:1995cp,Hisano:1995nq,Ellis:2002fe}. Incorporating RH-neutrino superfield(s) in the NMSSM provides a minimal model \cite{Kitano:1999qb} where, apart from accommodating none-zero neutrino masses and mixing, one also gets a solution for the $\mu$-problem\footnote{An alternative framework, known as the $\mu\nu$SSM \cite{Lopez-Fogliani:2005vcg,Escudero:2008jg,Ghosh:2008yh}, also simultaneously solves the $\mu$-problem and accommodates non-zero neutrino masses and mixing, even at the tree-level \cite{Ghosh:2008yh}. The $\mu\nu$SSM, at the cost of explicit R-parity violation, relies only on the SM gauge-singlet right-handed neutrino superfields $(\hat N)$ to accomplish these two goals. The NMSSM + RHN model \cite{Kitano:1999qb}, however, needs both $\hat S$ and $\hat N$ for the same as here non-zero neutrino masses and mixing emerges through either spontaneous R-parity violation \cite{Kitano:1999qb} or seesaw mechanism with conserved R-parity \cite{Das:2010wp}. The EWPT in the $\mu\nu$SSM, solely based on critical temperature analysis, has been performed in Ref. \cite{Chung:2010cd}.}
In such a framework, non-zero neutrino masses appear through three sources: (i) type-I seesaw mechanism involving RH-neutrino(s), generally known as the ``canonical seesaw", (ii) type-I and type-III seesaw involving gauginos, popularly known as the ``gaugino seesaw" and, (iii) seesaw involving higgsinos, better known as ``higgsino seesaw" \cite{Kitano:1999qb}. The last two pieces arise when left-handed (LH) and RH sneutrinos acquire vacuum expectation values (VEVs), i.e., R-parity
gets spontaneously broken \cite{Aulakh:1982yn,Masiero:1990uj} and effective bilinear R-parity-violating \cite{Hall:1983id} terms are generated.
It is important to emphasize here the scale of these BSM particles, e.g., gaugino. higgsino, right-handed neutrino, that is instrumental for the lightness of active neutrino mass. For this work, we confine ourselves in the context of the TeV scale seesaw, i.e., BSM states around a TeV or so, such that the hope of probing these states at the LHC survives. This, in turn, suggests smaller values of the associated parameters, e.g., neutrino Yukawas, which hardly affects the key objective of this study, i.e., analysis of the PT dynamics. Before moving towards
the discussion of the PT dynamics, however, we would like to discuss the number of RH-neutrino used for the chosen analysis. As it is possible to accommodate the neutrino oscillation data while keeping the lightest one massless, at least two generations of RH-neutrino superfields are needed if one tries to accommodate neutrino data at the tree-level \cite{Kitano:1999qb,Das:2010wp}. One, however, can achieve the same with only one RH-neutrino by incorporating judicious loop contributions \cite{Tang:2014hna}. The analysis of the PT dynamics being our primary objective, we consider only one generation of RH-neutrino. We refrain from playing with certain model parameters, e.g., neutrino Yukawas, off-diagonal
soft masses for sleptons, etc, as advocated in Ref. \cite{Tang:2014hna}, to accommodate the full neutrino oscillation data \cite{deSalas:2020pgw,Esteban:2020cvm,Gonzalez-Garcia:2021dve}
and consider only the reproduction of the correct neutrino mass scale and the atmospheric mass square difference in this study. If we 
overlook non-zero neutrino masses and mixing for a moment, our conclusion about the PT dynamics hardly alters if we move from one generation to three generations of RH neutrino as only one of them can develop a non-zero vacuum expectation value 
\cite{Kitano:1999qb} \footnote{Neglecting quadratic contributions from tiny neutrino Yukawa couplings \cite{Kitano:1999qb}, thanks to the TeV scale seesaw.}
and remains relevant for our analysis. Hence, we consider the one RH-neutrino case only which offers a nice platform to investigate the PT dynamics and subsequently the predictions for GW emission, besides providing the correct scale for the neutrino mass and the atmospheric mass-square difference.


Restoring the discussion of PT dynamics, the electrically neutral uncoloured scalar sector of the NMSSM extended with one RH-neutrino superfield set-up possesses fourteen degrees of freedom, including the neutral Goldstone mode. However, as we will see later in section \ref{sec:Model}, the effective degrees of freedom appear to be eight owing to weak couplings of the LH-sneutrino states with the remaining states. Out of these eight, only four are CP-even in nature and actively participate in the PT dynamics. Hence, the concerned field space is four-dimensional for the chosen framework. This enhanced field space compared to the MSSM (two-dimensional due to two Higgses) and the NMSSM (three-dimensional owing to two Higgs doublets and one singlet), facilitates the study of EWPT, via single steps and multi-steps.

In the numerical frontier, we adopt a benchmark-based analysis and finally select a few benchmark points (BPs) that appear promising from the viewpoint of EWBG and also exhibit distinct (single-step or two-steps) PT properties in the early Universe along the various constituent field directions. In the later part, we exploit some of the BPs in order to further investigate the role of new parameters that appear in the setup due to the presence of RH-neutrino superfield in the PT dynamics. We also consider various relevant experimental constraints, e.g., collider, charged-lepton flavour violation, etc., while choosing our BPs. In fact, null experimental evidence of sparticles to date has put stringent lower bounds on the concerned states, especially the coloured ones \cite{CMS:2022uby,CMS:2022tqr,CMS:2022bdb,ATLAS:2022rcw}. Thus, for the analysis of EWPT, we integrate such heavy states out and work in the context of a simplified effective model rather than considering the full NMSSM + one RH-neutrino framework. We have adopted both the critical and nucleation temperature analyses to describe the PT properties in our model. This is crucial since earlier studies, e.g., Ref. \cite{Baum:2020vfl}, have reported that the analysis of PT, solely based on critical temperature calculation
does not provide a complete picture. In fact, the critical temperature analysis does not confirm whether a PT has indeed taken place or not. A first-order phase transition (FOPT) proceeds via bubble nucleation and hence computation of nucleation probability and subsequently, nucleation temperature are vital to correctly describe the pattern of a FOPT.
Finally, we discuss the detection prospects of all our BPs in the forthcoming GW interferometers and find that the future space-based experiments: namely, U-DECIGO and U-DECIGO-corr \cite{Kudoh:2005as,Yagi:2011wg}, have the required sensitivities to test a few of our BPs. This possibility gives a complementary detection scope for the NMSSM + one RH-neutrino set-up beyond the conventional  experimental searches, e.g., collider, neutrino, flavour, etc.
	
The paper is organised as follows. In section\,\ref{sec:Model} we discuss the model setup. Next in section\,\ref{sec:param_space}, we talk about the relevant model parameters that are important for studying the PT properties and the possible experimental constraints. Subsequently in section\,\ref{sec:EWPTnu}, we present the dynamics of EWPT in detail along with our numerical findings. Besides, we explore semi-analytical analyses and the issues of gauge dependence for an elucidated understanding of the EWPT dynamics. This same section also addresses the production of the GW and the testability of our framework in upcoming space-based interferometers, e.g., U-DECIGO. Finally, we summarize our analysis and conclude in section\,\ref{sec:conclu}. Some useful formulae and relations are relegated to the appendices.

\section{The Model}\label{sec:Model}
The superpotential for the chosen 	framework is given by
\beq\label{eq:sup-nmssmRhn}
W = W_{\mathrm{MSSM}}'+\lambda\;\widehat{S}\;\widehat{H}_u\cdot\widehat{H}_d + \frac{\kappa}{3}\;\widehat{S}^3 + Y^i_N\;\widehat{N}\;\widehat{L}_i\cdot\widehat{H}_u + \frac{\lambda_N}{2}\;\widehat{S}\;\widehat{N}\;\widehat{N},
\eeq
where $i=1,\,2,\,3$ denotes the generation indices. Eq.(\ref{eq:sup-nmssmRhn}) is nothing but the $\mathbb{Z}_3$ symmetric NMSSM superpotential, extended with one Right-Handed Neutrino (RHN) superfield $(\hat N)$, keeping the initial $\mathbb{Z}_3$ symmetry unbroken. Here $W_{\mathrm{MSSM}}'$ denotes the MSSM superpotential (see reviews \cite{Nilles:1983ge,Haber:1984rc,Sohnius:1985qm,Martin:1997ns}) without the bilinear $\mu$-term, $\hat H_u=	\,(\hat H^+_u,\,\hat H^0_u)^{T}, \hat H_d=\, (\hat H^0_d,\,\hat H^-_d)^T, \hat L_i=\, (\hat \nu_i,\,\hat l_i)^T$ are the $SU(2)_L$ doublet up-type Higgs, down-type Higgs, and lepton superfields, respectively and the ``$\cdot$" notation is used to express $SU(2)$ product, e.g., $\hat L_i \cdot \hat H_u 	= \hat \nu_i \hat H^0_u - \hat l_i \hat H^+_u$. The  superpotential in Eq. (\ref{eq:sup-nmssmRhn}) cannot be made invariant under a global $U(1)$ symmetry, e.g., $U(1)$ of the Lepton number. This in turn ensures the disappearance of a Nambu-Goldstone boson which results from the spontaneous breaking of a global symmetry. The $\hat{N}$ is considered to be odd under $R_P$ while the $\hat{S}$ transforms as even.
$R_P$ is violated spontaneously in this model when, along with the other neutral scalars, the RH-sneutrino $(\widetilde {N})$ also acquires a non-zero VEV. These VEVs yield the effective $\mu$-term ($\mu=\lambda \langle S\rangle$), the effective bilinear $R_P$-violating couplings $(\epsilon_i=Y^i_N\langle \widetilde N\rangle)$, and the Majorana mass term for 	the RHN ($\lambda_N \langle S\rangle$). One should note the presence of four extra couplings (three neutrino Yukawa couplings $Y^{1,2,3}_N$ and another trilinear coupling $\lambda_N$) in Eq. (\ref{eq:sup-nmssmRhn}), apart from the known $\mathbb{Z}_3$ invariant NMSSM couplings, $\lambda$ and $\kappa$ (see for example Refs. \cite{Ellwanger:2009dp,Maniatis:2009re}).
	
We would like to re-emphasize here that with only one $\hat N$, of course, one cannot reproduce the observed neutrino mass squared differences and mixing \cite{deSalas:2020pgw,Esteban:2020cvm,Gonzalez-Garcia:2021dve}, even after including loop corrections \cite{Ibarra:2003up}. However, even this simple choice can predict the absolute mass scale and atmospheric mass squared difference for the active neutrinos, besides 
giving interesting information about the EWPT and GW, the primary goals of this article. We plan to explore the possible correlations between neutrino observable with the EWPT and GW sectors in the context of a two or three $\hat N$ scenario \cite{Kitano:1999qb} in future work. 
	
Following Eq. (\ref{eq:sup-nmssmRhn}), in a similar way, we can write down $\mathcal{L}_{\rm soft}$, the piece of Lagrangian density that contains soft-SUSY breaking terms:
\bea
-\mathcal{L}_{\rm soft} &&= -\mathcal{L'}_{\rm soft} +  m^2_{S} ~S^* S + {M^2_N\widetilde{N}^*\widetilde{N}} + \Big(\lambda A_{\lambda}S\; H_u\cdot H_d + h.c. \Big)\nn \\
&&+ \Big(\frac{\kappa A_{\kappa}}{3} S^3 + (A_N Y_N)^i\;\widetilde{L}_i\cdot H_u\;\widetilde{N} + \frac{A_{\lambda_{N}}\lambda_N}{2}\;S \widetilde{N} \widetilde{N} + h.c. \Big),
\label{eq:softSUSY}
\eea
where $\mathcal{L}'_{\rm soft}$ contains the MSSM soft-supersymmetry breaking terms, excluding the $B_\mu$ term \cite{Haber:1984rc,Simonsen:1995cf,Drees:2004jm, Martin:1997ns, Nilles:1983ge, Sohnius:1985qm}. The remaining terms are typical to that of the $\mathbb{Z}_3$ symmetric NMSSM, except the terms involving $\widetilde N$. Soft terms, as depicted in Eq. (\ref{eq:softSUSY}), are written in the framework of supergravity mediated SUSY breaking \cite{Hall:1983iz}. All the trilinear $A$-terms and the soft squared masses are assumed to lie in the TeV regime and consequently, all VEVs are expected to appear also in the same regime. In other words, the scale of RHN mass, which is determined solely by the scale of soft-SUSY breaking terms will also lie in the TeV regime assuming $\lambda_N\sim \mathcal{O}(1)$. This assures neutrino mass generation via the TeV scale seesaw mechanism which is also testable at colliders \cite{Roy:1996bua,Porod:2000hv,Hirsch:2008ur,Bartl:2009an,Fidalgo:2011ky,Ghosh:2012pq}. Further, the TeV scale seesaw immediately suggests $Y^i_N \sim \mathcal{O}$ $(10^{-6}-10^{-7})$ and left-handed sneutrino VEVs, $\langle \widetilde \nu_i \rangle \sim \mathcal{O}$ $(10^{-4}-10^{-5})$ GeV. These values of $Y^i_N, \, \langle \widetilde \nu_i \rangle$ indicate (i) tiny $R_P$ violation ($\sim \mathcal {O}$ $(10^{-3}-10^{-4})$ GeV, typical for the  bilinear $R_P$ violation \cite{Hirsch:2000ef}) and, (ii) weak mixing of the left-handed leptons and sleptons (neutral and charged) with the concerned sectors, e.g., charged and neutral gauginos, higgsinos, Higgses, right-handed neutrino and sneutrino, etc. One can use the advantage of such weak mixing to perform a simplified analysis without the loss of generality, e.g., using a set of four fields ($H_u, H_d, S,\tilde{N}$) instead of seven ($H_u, H_d, S,\widetilde{N},\widetilde{L_i}$) while investigating the PT phenomena.
	
The tree-level neutral scalar potential is the sum of $F$-term $(V_F)$, $D$-term $(V_D)$ and the soft-SUSY breaking terms and is given by
\beq
V_{\rm tree}=V_F+V_D+V_{\rm soft},
\label{eq:TotScalar}
\eeq
where $V_{\rm soft}\equiv$ $-\mathcal{L}_{\rm soft}$ is given by Eq. (\ref{eq:softSUSY}).
$V_F$, following the usual prescription from Eq. (\ref{eq:sup-nmssmRhn}), is written as
\bea\label{eq:vf-1}
V_F&&=\Big|-\l H^0_u H^0_d + \k S^2 + \frac{\l_N}{2} \wt{N}^2\Big|^2
+ \sum^3_{i=1} |Y^i_N|^2\;|H^0_u|^2 |\widetilde{N}|^2 +|\lambda|^2 |S|^2|H^0_u|^2 \nn \\
&& + \Big|\sum^3_{i=1} Y^i_N \wt{\nu}_i H^0_u + \l_N S \wt{N} \Big|^2 
+ \Big|\sum^3_{i=1} Y^i_N \wt{\nu}_i \wt{N} - \l S H^0_d \Big|^2,
\eea
and $V_D$, again using the standard procedure is read as
\beq
V_D=\frac{g_1^2+g_2^2}{8}\left( |H^0_d|^2 + \sum^3_{i=1} |\wt{\nu_i}|^2 -|H^0_u|^2 \right)^2,
\label{eq:Dterm}
\eeq
with $g_1,\,g_2$ as the $U(1)_Y,\,SU(2)_L$ gauge couplings, respectively.

The neutral CP-even scalar components\footnote{Here we adhere to CP-conservation. Further, we do not consider the possibility of charge and colour-breaking minima for this study (see e.g., Ref. \cite{Krauss:2017nlh} in the context of the NMSSM) and hence, assign vanishing VEVs to charged and coloured scalars.}, after the EW-symmetry breaking (EWSB), develop the following zero-temperature VEVs:
\beq
\langle H^0_u\rangle=v_u,\,\,\langle H^0_d\rangle=v_d,\,\,\langle S\rangle=v_S,\,\,\langle \wt{\nu}_i\rangle=v_i, \,\,\langle \wt{N}\rangle=v_N,\quad i=1,\,2,\,3 \quad {\rm or}
 \quad e,\,\mu,\,\tau.
\label{eq:VEVs}
\eeq
The first three VEVs are typical to the NMSSM while the last two VEVs appear for the 
chosen framework as a consequence of the spontaneous $R_P$ violation. One can use these VEVs to trade off the concerned soft squared masses as depicted in Eq. (\ref{eq:softSUSY}). 
The VEVs $v_S,\,v_N$, being governed by the TeV scale soft-terms, also lie in the same regime whereas $v_i$ appears to be much smaller $\sim \mathcal {O} (100$ MeV) for $v_N,\,v_S\sim \mathcal{O}(1$ TeV) \cite{Masiero:1990uj}.
Generation of the neutrino mass via a TeV scale seesaw mechanism, as already advocated, however, offers a more stringent constraint on $v_i$ ($\sim \mathcal{O}$ $(10^{-4}-10^{-5})$ GeV), similar to models studied in Refs. \cite{Georgi:1981pg,Aulakh:1982yn,Santamaria:1987uq,Santamaria:1988zm,Santamaria:1988ic}. One can write down minimization conditions for $v_N,\,v_i$, using Eq. (\ref{eq:TotScalar}), as:
\bea
&& \frac{\partial V_{\rm tree}}{\partial \wt{N}}\Big|_{\rm VEVs~as~Eq.~(\ref{eq:VEVs})}
=\l_N  v_N\Big(\l v_u v_d +\k v^2_S + \frac{\l_N}{2} v^2_N\Big) + |Y^i_N|^2 v^2_u v_N\nn \\
&& + \l_N v_S \Big(\sum^3_{i=1} Y^i_N v_i v_u + \l_N v_S v_N\Big) + \sum^3_{i=1} Y^i_N v_i
\Big(\sum^3_{j=1} Y^j_N v_j v_N - \l v_S v_d\Big) \nn \\ 
&&+ M^2_N v_N + \sum^3_{i=1}(A_N Y_N)^i v_i v_u
+ {A_N \l_N} v_S v_N, \nn\\
&& \frac{\partial V_{\rm tree}}{\partial \wt{\nu_i}}\Big|_{\rm VEVs~as~Eq.~(\ref{eq:VEVs})}
=Y^i_N v_u \Big(\sum^3_{j=1} Y^j_N v_j v_u + \l_N v_S v_N\Big)
+ Y^i_N v_N \Big(\sum^3_{j=1} Y^j_N v_j v_N - \l v_S v_d\Big) \nn \\
&&+ \sum^3_{j=1}  m^2_{\wt{L}_{ij}} v_j 
+ (A_N Y_N)^i v_u v_N + \frac{g_1^2+g_2^2}{4}\left( v^2_d + \sum^3_{j=1} v^2_j -v^2_u \right) v_i,
\label{eq:minimizationNL}
\eea
where $m^2_{\wt{L}_{ij}}$ denotes soft-squared masses for sleptons \cite{Nilles:1983ge,Haber:1984rc,Sohnius:1985qm,Martin:1997ns} and all the concerned parameters are assumed to be real. It is apparent from Eq.(\ref{eq:minimizationNL}) that if one neglects terms like $Y^i_N Y^j_N,\, Y^i_N v_i$ for smallness, then $v_S \rightarrow 0$ suggests $v_N \to 0$ and consequently $v_i \to 0$. Thus, a non-zero $v_S$ is indirectly connected to a non-zero $v_i$. The smallness of $v_i$, compared to $v_u,\,v_d$, also assures that one can still safely use the MSSM relations $v^2 = v^2_u + v^2_d$ and $\tan\beta = v_u/v_d$.

The presence of tiny but non-zero $Y^i_N,\,v_i$, as already stated, generates mixing between left-handed neutrinos and neutral gauginos. These new mixing terms in the EW sector enhance the size of neutral scalar, neutral pseudoscalar, charged scalar, neutral fermion and charged fermion mass matrices. Being explicit, $R_P$-violating mixing of $H^0_u,\, H^0_d, S$ states with $\wt{N}$ and three families of $\wt{\nu_i}$, enlarges the NMSSM CP-even and CP-odd neutral scalar mass matrices from $3\times 3$ to $7 \times 7$. Similar augmentation appears (i) in the charged scalar sector ($2\times 2$ in the NMSSM to $8\times 8$ due to $R_P$-violating mixing of $H^\pm_u,\, H^\mp_d$ states with the three families of left- and right-handed charged sleptons), (ii) in the neutral fermion sector ($5\times 5$ in the NMSSM to $9\times 9$ due to $R_P$-violating mixing among neutral gauginos, $\wt{H^0_u},\,\wt{H^0_d},\, \wt{S}$ states with the right-handed neutrino and the three families of left-handed neutrinos), and (iii) in the charged fermion sector ($2\times 2$ in the NMSSM to $5\times 5$ due to $R_P$-violating mixing among the charged higgsino, gaugino states with the three families of the  left- and right-handed handed leptons).
However, because of tiny values of $Y^i_N,\,v_i$, 
one can easily decompose the aforesaid mass matrices in blocks for approximate analytical studies. For example, for all practical purposes, the neutral scalar mass matrix can be decomposed into two diagonal blocks: (i) a $4\times 4$ one consisting of CP-even $H^0_u,\, H^0_d,\, S,\,\wt{N}$ states, (ii) another $3\times 3$ one consisting of CP-even left-handed sneutrino states, and off-diagonal blocks containing tiny mixing terms between the two aforementioned states. A similar observation holds true for the neutral pseudoscalar, charged scalar, neutralino and chargino mass matrices, which can be effectively considered as having dimensions $3\times 3,\, 2\times 2,\,6 \times 6$ and $2\times 2$, respectively\footnote{One can easily identify the remaining three neutralinos and three chargions, lying at the bottom of the mass spectrum, as three LH-neutrino dominated states and the charged leptons, $e,\,\mu,\,\tau$.}, without any loss of generality, leaving the almost pure left-handed CP-odd sneutrino, charged slepton, left-handed neutrino and charged leptons states aside. For the purpose of analyzing the chosen model numerically, it is convenient to express the aforesaid mass matrices in the extended Higgs basis \cite{Carena:2015moc,Georgi:1978ri,Donoghue:1978cj,Lavoura:1994fv,Botella:1994cs,Gunion:2002zf,Gunion:1989we,Gunion:1992hs} which will be introduced subsequently. Entries of these mass matrices are detailed in appendix\,\ref{app:FM}, along with the full uncoloured scalar potential.

\subsection{A convenient basis choice}{\label{susec:effective pot}}
We have already introduced the tree-level neutral scalar potential in Eq.(\ref{eq:TotScalar}), using Eqs.(\ref{eq:softSUSY}), (\ref{eq:vf-1}) and (\ref{eq:Dterm}). However, to study the phenomena of PT we need to move beyond the tree-level contribution. For this purpose, as we already mentioned, it is useful to work in the extended Higgs basis \cite{Carena:2015moc,Georgi:1978ri,Donoghue:1978cj,Lavoura:1994fv,Botella:1994cs,Gunion:2002zf,Gunion:1989we,Gunion:1992hs}, given as:
\bea
H_d &&=  \begin{pmatrix}
\frac{1}{\sqrt{2}}(c_{\beta} H_{\rm SM}-s_{\beta} H_{\rm NSM}) + \frac{i}{\sqrt{2}}(-c_{\beta} G^0 + s_{\beta} A_{\rm NSM}) \nn \\
-c_{\beta} G^{-} + s_{\beta} H^{-}
\end{pmatrix}, \nn \\
H_u&&= \begin{pmatrix}
s_{\beta} G^{+} + c_{\beta} H^{+}\\
\frac{1}{\sqrt{2}}(s_{\beta} H_{\rm SM}+c_{\beta} H_{\rm NSM}) + \frac{i}{\sqrt{2}}(s_{\beta} G^0 + c_{\beta} A_{\rm NSM})
\end{pmatrix},\nn \\
S &&= \frac{1}{\sqrt{2}}(H_S + i A_S),\nn\\
\widetilde{N} &&=\frac{1}{\sqrt{2}}(N_R+i\,N_I),
\label{eq:Higgsbasis}
\eea
where $c_\beta (s_\beta)=\cos\beta (\sin\beta)$ with $\tan\beta=v_u/v_d$. Note that one trades off the scalar, the pseudoscalar and the charged components of the relevant four fields $\{H_u, H_d, S,\wt{N}\}$ with the four neutral CP-even interaction states ($H_{\rm SM}$, $H_{\rm NSM}$, $H_S$, $N_R$), three CP-odd interaction states ($A_{\rm NSM}$, $A_{\rm S}$, $N_I$), one charged Higgs pairs ($H^\pm$), along with the neutral and charged Goldstone modes ($G^0,\, G^\pm)$ in the extended Higgs basis. This particular basis choice assures the SM-like couplings between $H_{\rm SM}$ with the up-type SM fermions, the down-type SM fermions and the SM vector bosons. In addition, the aforementioned basis choice also predicts vanishing couplings between $H_S,\, N_R$ with the same aforesaid SM states. Furthermore, from Eq. (\ref{eq:Higgsbasis}), in the light of Eq.(\ref{eq:VEVs}) and $v^2=v^2_u+v^2_d$, one can see that $\langle H_{\rm SM} \rangle = \sqrt{2} v,\, \langle H_{\rm NSM} \rangle = 0$, $\langle H_{S} \rangle = \sqrt{2} v_S$ and $\langle N_{R} \rangle = \sqrt{2} v_N$, i.e., non-vanishing VEVs appear only in certain field directions leaving the SM-direction undisturbed. These interaction states later mix to produce the mass eigenstates. However, one of the CP-even states with a mass in the ballpark of $125$ GeV (see Ref. \cite{ParticleDataGroup:2022pth} and references therein) contains the predominant $H_{\rm SM}$ component. This alignment between the 125 GeV SM-like Higgs in the mass basis and $H_{\rm SM}$ of the extended Higgs basis implies negligible admixing among various states in the extended Higgs basis. Mathematically, after the EWSB, in the $H_{\rm SM}$, $H_{\rm NSM}$, $H_S$, $N_R$ basis:
\bea
|\mathcal{M}^2_{S,1i}| \ll |\mathcal{M}^2_{S,ii}-\mathcal{M}^2_{S,11}|,
\label{eq:decoupling}
\eea
where $i=2,\,3,\,4$ and $\mathcal{M}^2_{S,1i}$, the entries of the CP-even scalar squared mass matrix,  are given in appendix \ref{ap:MassmMatricesEWSB}. It is now apparent that 
in order to satisfy Eq.(\ref{eq:decoupling}) one either needs small $\mathcal{M}^2_{S,1i}$ or large $|\mathcal{M}^2_{S,ii}-\mathcal{M}^2_{S,11}|$, i.e., decoupling of $H_{\rm SM}$ from the three remaining  states. The latter, in terms of the mass eigenstates, predicts three significantly heavier states dominated by $H_{\rm NSM}$, $H_S$, $N_R$ compositions, and one $\sim \mathcal{O} (125~\rm{GeV})$ state controlled by $H_{\rm SM}$ composition. In reality, for the SFOEWPT, singlet-like states lighter than 125 GeV are favoured. Besides, heavier singlet-dominated states create a kind of  ``push-down" effect \cite{Escudero:2008jg,Ghosh:2014ida} which makes it difficult to achieve an SM-like Higgs state around 125 GeV. Thus, for our numerical studies, we consider regions of the parameter space that can accommodate one or more singlet-like states lighter than 125 GeV. These light singlet-dominated states are helpful in accommodating a 125 GeV SM-like   Higgs through the ``push-up effect" \cite{Escudero:2008jg,Ghosh:2014ida}. 
  
{\textcolor{black}One can use Eq.(\ref{eq:decoupling}) subsequently to derive a few approximate relations, useful for parameter space scanning. For example, using appendix \ref{ap:MassmMatricesEWSB} and assuming $\mathcal{M}^2_{S,11} = m^2_{h_{125}}$, the condition $\mathcal{M}^2_{S,12}\rightarrow 0$, i.e., vanishing mixing between the 
$H_{\rm SM}$ and $H_{\rm NSM}$ states, implies
\bea
\lambda^2 \simeq \frac{m^2_{h_{125}} - m^2_Z\,{\rm \cos\,2 \beta}}{2 v^2 \, {\rm \sin^2 \beta}}.
\label{eq:M12zero}
\eea
As $m_{h_{125}}, m_Z$ (mass of the SM $Z^0$-boson), $v$ are known, $\lambda$ approximately appears to be a function of $\tan\beta$ only. A similar relation like Eq.(\ref{eq:M12zero}) holds also for the NMSSM \cite{Baum:2020vfl}. Applying the same   procedure to minimize the mixing between $H_{\rm SM}$ and $H_{\rm S}$ states, i.e., $\mathcal{M}^2_{S, 13} \rightarrow 0$, one gets
\bea
M^2_A \simeq \frac{4 \mu^2}{ {\rm \sin^2\,2 \beta}} \left(  1 - \frac{\kappa}{ 2 \l} {\rm \sin\,2 \beta} + \frac{\lambda \l_N v^2_{N}}{4 \mu^2} {\rm \sin 2 \beta}\right),
\label{eq:AnsmSqMass}
\eea
choosing $M^2_A \simeq \frac{2 \mu}{ {\rm \sin \,2 \beta}} \left(  A_{\l} + \frac{\kappa \mu}{\l} + \frac{\l \l_N v^2_{N}}{2 \mu} \right)$ \footnote{At the limit $\lambda_N \to 0$, Eq. (\ref{eq:AnsmSqMass}) reproduces the known NMSSM result \cite{Baum:2020vfl}. If one further considers $\kappa \to 0$, Eq. (\ref{eq:AnsmSqMass}) matches the well-known MSSM relation \cite{Haber:1984rc}.}. 
The last term in the Eq. (\ref{eq:AnsmSqMass}) appears due to mixing with 
the RH-sneutrino. In the limit of $\kappa\ll\lambda$, using Eq. (\ref{eq:AnsmSqMass}), it turns out that $M^2_A \simeq M^2_{H} \simeq M^2_{H^\pm}\simeq {4\mu^2}{\csc^2 2\beta}\left(1+\frac{\lambda\lambda_N v_{N}^2}{4\mu ^2}\sin 2\beta \right)$ where $M_H$ represents mass of a state with dominant $H_{\rm NSM}$ contribution. The presence of $v_N$ shows that these mass eigenstates possess contributions from the RH-sneutrino. These kinds of mixing may appear sizable depending on $\lambda_N$ and $v_S$ values.
  
Adopting a similar analysis for $\mathcal{M}^2_{S, 14} \rightarrow 0$, 
i.e., effacing the mixing between $H_{\rm SM}$ and $N_R$ states, it is hardly possible
to get a simple relation. A light state below 80 GeV with dominant RH-sneutrino
contribution hints for a sizable mixing between the $H_{\rm SM}$ and $N_R$ states. 
This effect, via one-loop, makes it easy to assure a 125 GeV SM-like Higgs, even with stop mass below $\mathcal{O}(1$ TeV) \cite{Huitu:2014lfa}. By choosing the parameters carefully, one can of course consider a heavier stop mass to secure a 125 GeV SM-like Higgs having negligible admixing with a lighter RH-sneutrino-dominated state. This is precisely what we have done while scanning the parameter space since a lighter sneutrino,
as also stated earlier, is advantageous for SFOEWPT. We will discuss this aspect in detail later. We note in passing that so far we have discussed only the tree-level aspects of the scalar potential. In reality, the scalar potential receives considerable contributions from radiative effects involving various SM particles and their SUSY partners \cite{Derendinger:1983bz,King:1995vk,Masip:1998jc,Ellwanger:2009dp}. Some of these higher-order contributions have observable consequences, e.g., effects of the top and stop loops 
to procure a 125 GeV SM-like Higgs.

\subsection{Higher order contributions}{\label{susec:higher-order-pot}}

It is relevant to investigate various sources critically before implementing 
higher-order effects arising from the different SM and BSM states on the tree-level scalar potential. The effect of higher-order contributions, especially via SUSY partners, is crucial for yielding the observed SM mass spectrum, e.g., the Higgs mass. These effects, however, are diluted for the analysis of EWPT. Hence, we concentrate only on the leading one-loop effects which can arise from various SM and BSM sources. Regarding the latter, one needs to consider the following facts: (i) BSM Higgs masses, i.e., states with dominant $H_{\rm NSM}, H_S, N_R$, $A_{\rm NSM}, A_S$ and $N_I$ components, must not remain very far from the EW scale for a successful SFOEWPT and, (ii) hitherto unseen experimental evidence of SUSY searches have set lower limits on sparticle masses. These limits are stringent for the coloured sector, e.g., gluinos and squarks, $\gsim \mathcal{O}$ (1 TeV) (see, for example, the latest CMS \cite{CMS:2021beq,CMS:2022uby,CMS:2022tqr,CMS:2022bdb}
and ATLAS \cite{ATLAS:2021twp,ATLAS:2021nka,ATLAS:2022rcw,ATLAS:2022ihe} limits). On the other hand, for the uncoloured sparticles, e.g., sleptons, LH-sneutrinos, etc, experimental lower bounds are rather flexible \cite{ATLAS:2019gti,CMS:2021ruq,ATLAS:2022hbt}. For convenience, however, we consider heavy sleptons and LH-sneutrinos, $\gsim \mathcal{O}$ (1 TeV), for this study\footnote{Unlike the coloured sector, $\gsim \mathcal{O}$ (1 TeV) sleptons and sneutrinos do not introduce large higher-order corrections to the scalar sector owing to small values of the concerned lepton Yukawa couplings.}. A careful range of relevant parameters was considered so that even with these heavy sleptons one can satisfy the latest result on the anomalous magnetic moment of muon \cite{Muong-2:2021ojo} which typically favours the aforesaid states to be lighter than a TeV.

With the above mentioned facts and assumptions, one ends up with a situation where 
one encounters $\gsim \mathcal{O}$ (1 TeV) sleptons, LH-sneutrinos, squarks \& gluinos
together with other BSM states, e.g., scalar and pseudoscalar Higgses, neutralinos, and charginos, in the ballpark of the EW scale. Clearly, now one can integrate out these $\gsim \mathcal{O}$ (1 TeV) states to yield an effective theory with BSM scalar, pseudoscalar, charged Higgses, neutralinos, charginos and, of course, the SM particles. Here we would like to point out again that the neutralino and the chargino sector for the concerned model are enhanced compared to the NMSSM, owing to the presence of $Y^i_N$ in the superpotential (see Eq. (\ref{eq:sup-nmssmRhn})) and non-zero LH-sneutrino VEVs (see Eq. (\ref{eq:VEVs})). However, these parameters are compelled to remain tiny ($\sim \mathcal{O}$ $(10^{-6}-10^{-7})$ and $\sim \mathcal{O}$ $(10^{-4}-10^{-5})$ GeV), thanks to the constraints arising from the neutrino experiments and the assumption of a TeV scale seesaw. A similar observation, as already stated, also holds true for the BSM Higgs sector. In summary, the effective number of contributing states are four CP-even Higgses ($S^0_i$), three CP-odd Higgses ($P^0_i$), two charged Higgses ($H^\pm$), six neutralinos ($\widetilde{\chi}^0_i$), two charginos ($\widetilde{\chi}^\pm_i$), charged and the neutral Goldstone bosons ($G^\pm,\, G^0$), and, the relevant SM particles ($t,\, W^\pm, \, Z^0$)\footnote{Contributions from the remaining SM fermions are sub-leading due to the sizes of concerned Yukawa couplings.}. This set of nineteen particles including the two Goldstone bosons, together with the $t,\, W^\pm, \, Z^0$, will be considered as the dynamical degrees of freedom needed for the current study. One can derive parameters of the aforesaid effective theory through the renormalization group equation and subsequently, by matching onto the complete model at some intermediate scale $\Lambda$ which we fixed at $m_t$, the top mass. The leading contribution to the tree-level potential $V_{\rm tree}$ obtained using this procedure is
\beq{\label{eq:delV}}
\Delta V=\frac{\Delta\lambda_2}{2}|H_u|^4,
\eeq
where ${\Delta\lambda_2}$ at one-loop level is given by \cite{Ellis:1991zd,Haber:1993an,Casas:1994us,Carena:1995bx},
\beq{\label{eq:dellambda2}}
\Delta\lambda_2=\frac{3}{8\pi^2}y_t^4\left[{\rm log}\left(\frac{M_{\widetilde{t}}^2}{m_t^2}\right)+\frac{A_t^2}{M_{\widetilde{t}}^2}\left(1-\frac{A_t^2}{12 M_{\widetilde{t}}^2}\right)\right].
\eeq
Here $y_t$ is the top Yukawa coupling evaluated using the running top quark mass, $M_{\wt t}= \sqrt{m_{\wt t_1} m_{\wt t_2}}$ depicts the geometric mean of two stop masses and 
$A_t$ is the soft trilinear coupling between Higgs and stops (appears within $\mathcal{L'_{\rm soft}}$ of Eq. (\ref{eq:softSUSY}) \cite{Haber:1984rc}). One can of course write down contributions like the one shown in Eq. (\ref{eq:delV}) for other scalar states, e.g., $H_d$. Such a term, however, appears due to mixing between $H_u$ and 
$H_d$ through the effective $\mu$-term and is usually sub-leading compared to the one 
shown in Eq. (\ref{eq:delV}), as long as $\mu \ll M_{\wt t}$\footnote{Such a choice helps one parameterize radiative contributions from stops effectively, even beyond the one-loop 
order \cite{Carena:1995bx,Lee:2015uza,Haber:1993an}.} and $\tan\beta$ value appears not too large. The quantity $\Delta \lambda_2$ is crucial to accommodate a 125 GeV SM-like Higgs and can be estimated using the same.

The leftover degrees of freedom also contribute to the potential (see Eq. (\ref{eq:TotScalar})) through radiative corrections. Their collective contributions are
given by Coleman-Weinberg potential \cite{Coleman:1973jx}
\beq
V_{\rm CW}^{\rm 1-loop}=\frac{1}{64\pi^2}\sum_{i=B,F}(-1)^{F_i} n_i m_i^4(\phi_\alpha)\left[{\rm log}\left(\frac{m_i^2(\phi_\alpha)}{\Lambda^2}\right)-\mathcal{C}_i\right],\label{eq:CW}
\eeq
where $i=B\,(F)$, i.e., bosons (fermions), $n_i$ represents the relevant degrees of freedom, $F_B=0\,(F_F=1)$, $\mathcal{C}_i$ is a constant with a value of $3/2\,(1/2)$ for scalars, fermions, longitudinally polarized vector bosons (transversely polarized vector bosons), $\Lambda$ is the aforesaid intermediate energy scale, fixed at $m_t$ and, 
$m^2_i(\phi_\alpha)=m^2_i (H_{\rm SM}, H_{\rm NSM}, H_S, N_R)$ denotes field-dependent masses. The latter is estimated from $V_{\rm tree} + \Delta V$ (see Eq. (\ref{eq:TotScalar}) and Eq. (\ref{eq:delV})). Contributions from  $V_{\rm tree}$ are detailed in appendix \ref{app:FM}. The set of involved $B$s are given by $S^0_{1,..,4}$, $P^0_{1,2,3}$, $H^\pm, G^0, G^\pm$, $Z^0, W^\pm$ with $n_B=4\times 1$, $3 \times 1$, $2,\,1,\,2,\,3$, $2\times 3$, depending on the nature of the concerned state, i.e., scalar or complex scalar or massless bosons or massive vector bosons. A similar approach for the fermions give $F=\widetilde{\chi}^0_{1,..,9}, \,\widetilde{\chi}^\pm_{4,5},\,t$ with $n_B=9\times 2,\,2\times 2,\,3\times 4$ considering their electric and colour charges. One should note that the presence of $G^0,\, G^\pm$ in the Coleman-Weinberg potential yields divergent contributions. However, these can be effaced by using an infrared regulator. Finally, putting all these pieces, i.e., $V_{\rm tree}$ (see Eq. (\ref{eq:TotScalar})), $\Delta V$ (see Eq. (\ref{eq:delV})) and $V_{\rm CW}^{\rm 1-loop}$ (see Eq. (\ref{eq:CW})) together, one obtains the effective scalar potential as
\beq
V_{\rm eff}=V_{\rm tree}+V_{\rm CW}^{\rm 1-loop}+\Delta V. \label{eq:T0Veff}
\eeq
Inclusion of Coleman-Weinberg contributions (see Eq. (\ref{eq:T0Veff})) to the tree-level scalar potential, however, changes the position of physical minima and masses. To restore the original position for the physical minima, keeping $\mathcal{M}^2_{S,13}, \mathcal{M}^2_{S,14} \to 0$ and maintaining the mass of the CP-even scalar state with leading $H_{\rm SM}$ composition at 125 GeV, one needs to introduce appropriate counterterms, encapsulated within another contributor $V_{\rm ct}$. The latter is normally related to a redefinition of the entries of $-\mathcal{L_{\rm soft}}$ (see Eq.\,(\ref{eq:softSUSY})) \cite{Bi:2015qva,Huber:2015znp,Bian:2017wfv} which are depicted in appendix \ref{ap:Counter}. The counterterms are, thus, not arbitrary but fixed by the aforesaid criteria. Mathematically,
\beq\label{eq:cond1cw}
\frac{\partial}{\partial \phi_i}\Big(V_{\rm eff} + V_{\rm ct}\Big)\Big|_{\phi_i=\langle\phi_i\rangle} = 0\,\,\,\, {\rm{and}} \,\,\,\,
\frac{\partial^2}{\partial \phi_i \partial \phi_j}\Big(V_{\rm eff} + V_{\rm ct}\Big)\Big|_{\phi_i=\langle\phi_j\rangle} = 0,
\eeq
with $\phi_i=\{H_{\rm SM},\,H_{\rm NSM},\,H_{\rm S},\,N_R\}$. One can figure out $\langle \phi_i \rangle$ using Eq. (\ref{eq:VEVs}) and Eq. (\ref{eq:Higgsbasis}). We note in passing that till now we have discussed modifications of the tree-level scalar potential from higher order effects at vanishing temperature, i.e., $T=0$. In reality, however, one also needs to include contributions arising from $T\neq 0$ which we will address now.

\subsection{Contributions from non-zero temperature}{\label{susec:finiteT}}
{The one-loop temperature-dependent potential is given by \cite{Dolan:1973qd}
	\beq{\label{eq:VatTnotZero}}
		V_{T\neq 0}^{\rm 1-loop}=\frac{T^4}{2\pi^2}\sum_{i=B,F}(-1)^{F_i}n_iJ_{B/F}\left(\frac{m_i^2(\phi_\alpha,T)}{T^2}\right),
	\eeq
 where $T$ represents the temperature, symbols $F_{F,B},\, n_{F,B}$ are the same as discussed in the context of Eq. (\ref{eq:CW}), $m_i^2(\phi_\alpha,T)$ depicts thermal field-dependent
 masses of the $i_{\rm th}$ degrees of freedom as:
\beq\label{Eq:Daisy_coeff_ci}
m_i^2(\phi_\alpha,T)=m_i^2(\phi_\alpha)+c_i T^2,
\eeq
 with $c_i$ representing the concerned Daisy coefficients \cite{Dolan:1973qd,Kirzhnits:1976ts,Parwani:1991gq,Espinosa:1992gq,Arnold:1992rz}\footnote{Inclusion of the next-to-leading order contributions would affect Eq. (\ref{Eq:Daisy_coeff_ci}). Such detailed thermodynamic treatment of the EWPT is beyond the scope of the current study and has already been addressed in the literature \cite{Croon:2020cgk,Schicho:2021gca,Niemi:2021qvp,Schicho:2022wty,Ekstedt:2022bff}.}. These coefficients appear non-vanishing for bosons and are given in appendix \ref{ap:Daisy}. Finally, $J_{B/F}$, i.e., the thermal function, is defined as 
 \beq\label{eq:thermalF}
J_{B/F} \Big(x^2 \equiv \frac{m_i^2(\phi_\alpha,T)}{T^2}\Big)=\pm \int_0^\infty dy~ y^2~{\rm log}\left(1\mp e^{-\sqrt{x^2+y^2}}\right),
\eeq
where $+\,(-)$ sign is for bosons (fermions). One should note that at the $m^2 \gg T^2$ limit, where $``m"$ depicts a generic mass term, $J_{B/F}$ suffers an exponential suppression from Boltzmann factor. These repressions ensure that massive degrees of freedom, e.g., squarks, gluinos, etc., that are already integrated out (see subsection \ref{susec:higher-order-pot}), do not affect $T\neq 0$ corrections.}
 
Clubbing all the pieces together, i.e., tree-level scalar potential, one-loop contributions via Coleman-Weinberg potential, and contributions from the finite temperature part, one gets  the finite temperature effective scalar potential at the one-loop order as
 \beq
V_T=V_{\rm tree}+\Delta V+{V'}_{\rm CW}^{\rm 1-loop}+ V_{\rm ct} + V_{T\neq 0}^{\rm 1-loop} \equiv V_T (\phi,\, T),\label{eq:totPotR}
\eeq
where ${V'}_{\rm CW}^{\rm 1-loop}$ has a form similar to Eq. (\ref{eq:CW}) but replacing
$m^2_i(\phi_\alpha)$ with thermal masses $m^2_i(\phi_\alpha,\,T)$, as depicted in Eq. (\ref{Eq:Daisy_coeff_ci}). We will use Eq. (\ref{eq:totPotR}) to inquire about the PT properties. 

In this study, we compute $V_T(\phi, T=0)$ and also $V_T(\phi, T)$ using Landau gauge, i.e., $\xi=0$, where the ghost fields get decoupled. The components of $V_T$ are gauge dependent \cite{Nielsen:1975fs,Patel:2011th,DiLuzio:2014bua} which, however, insufficient to affect the undertaken tasks \cite{Garny:2012cg,Kozaczuk:2013fga,Kozaczuk:2014kva,Chatterjee:2022pxf}, i.e., whether the concerned $V_T$ gives rise to an SFOPT and to study its properties. In this model,
barrier formation is possible even at the tree-level, without the loop-induced corrections, via the cubic and the quartic interactions, e.g., through couplings $\kappa$ and $A_\kappa$, respectively. The gauge dependence, appearing through the loop-induced  contributions is expected to be subdominant in our study. Gauge invariant treatment of $V_T$ and quantities related to it, however, are also 
possible and are already advocated in the literature \cite{Nielsen:1975fs,Fukuda:1975di,Baacke:1993aj,Baacke:1994ix,Laine:1994zq,Wainwright:2011qy,Wainwright:2012zn,Garny:2012cg,Profumo:2014opa,Espinosa:2016nld,Chiang:2017nmu,Ramsey-Musolf:2019lsf,Kozaczuk:2019pet,Croon:2020cgk,Arunasalam:2021zrs,Lofgren:2021ogg,Schicho:2022wty}. 
For the sake of completeness, nevertheless, we estimated the gauge dependence of our findings in subsection  \ref{sec:gaugeD} in the lines of Ref. \cite{Katz:2015uja}  and observed that the impact of gauge dependence hardly affects our conclusions, thanks to tree-level barrier formation. One should also note that ${V}_{\rm CW}^{\rm 1-loop}$ (see Eq. (\ref{eq:CW})), and hence ${V'}_{\rm CW}^{\rm 1-loop}$, depend  on the renormalization scale $\Lambda$. This effect, however, is sub-leading \cite{Kozaczuk:2014kva,Baum:2020vfl} as we are working in an effective framework after integrating the heavy states out and considering $\Lambda=m_t$. The relative dominance of the $\Lambda$ dependence over the gauge dependence \cite{Laine:2017hdk,Croon:2020cgk} is non-trivial in nature. Once again, $\Lambda$ dependence can be softened as discussed in Ref. \cite{Chiang:2017nmu}. We plan to investigate the interplay of gauge dependencies and $\Lambda$ sensitivities in a future publication.

So far we have discussed different pieces of the scalar potential needed to study the PT dynamics. Now we will address how and to which extent various model parameters can affect the same.

\section{Choice of parameters}{\label{sec:param_space}}
The set of new parameters, compared to the NMSSM, are 
\begin{equation}\label{eq:extraparam1}
 Y^i_N,\, \lambda_N,\, \,v_N,\, (A_N Y_N)^i,\,A_{\lambda_N}\lambda_N,
 \end{equation}
using Eq. (\ref{eq:sup-nmssmRhn}), Eq. (\ref{eq:TotScalar}), Eq. (\ref{eq:VEVs}), and replacing soft-SUSY breaking square mass term $M^2_N$ with the corresponding VEV. Now, as already discussed, $Y^i_N$s are associated with the neutrino mass generation through a TeV scale seesaw and thus, are constrained to be small. These $Y^i_N$ values, for TeV-scale trilinear terms, predicts $(A_N Y_N)^i \sim \mathcal{O}$ $(10^{-3}-10^{-4})$ GeV. The latter is also related to the smallness of $v_i$, i.e, the LH-sneutrino VEVs (see Eq. (\ref{eq:VEVs})), as guided by a TeV scale seesaw mechanism and neutrino data. Hence, for the PT analysis, we can neglect these tiny parameters, i.e., $v_i$,\, $Y^i_N,\, (A_N Y_N)^i$, without any loss of generality as they have negligible effects on the PT dynamics.  Now from the discussion of section \ref{sec:Model}, it is evident that relevant ``bare" parameters for the uncoloured scalar potential after trading (see appendix \ref{ap:minmcond} for details) soft-squared masses with the corresponding VEVs (see Eq. (\ref{eq:VEVs})) are, 
\beq\label{eq:IndeParameters1}
\lambda,\,\lambda_N,\,\kappa,\,\,v_u,\,v_d,\,v_S,\,v_N,\,A_\lambda,\,A_\kappa,\,A_{\lambda_N}.
\eeq
One can redefine this list further by trading $v_u,\,v_d$ with $v=\sqrt{v^2_u+v^2_d}, \, \tan\beta=v_u/v_d$ and $v_S$ with $\mu=\lambda v_S$. As $v=174$ GeV is known, Eq.(\ref{eq:IndeParameters1}) can be re-casted as
\beq\label{eq:IndeParameters2}
\lambda,\,\lambda_N,\,\kappa,\,\,\tan\beta,\,\mu,\,v_N,\,A_\lambda,\,A_\kappa,\,A_{\lambda_N}.
\eeq
One can also trade parameter $v_N$ with the RH-neutrino mass term $M_N\propto \lambda_N v_N$. Similar trading is also possible for $A_\lambda$ with $M_A$, using a relation given in subsection \ref{susec:effective pot}. We, however, do not use $M_A, M_N$ for the parameter space scanning. Parameter $\lambda$ can also be exchanged using Eq.(\ref{eq:M12zero}). The same parameter can also be constrained using an upper-bound
on the tree-level SM-like Higgs mass \cite{Drees:1988fc,Ellis:1988er,Ellwanger:2009dp}, given as $m^2_Z (\cos^22\beta + g^{-2}_2 \lambda^2 \sin^2 2\beta)$. This helps us to consider small $\tan\beta \lesssim 5$ and $\lambda \sim \mathcal{O} (0.1)$ or higher such that one gets a significant contribution to the tree-level SM-like Higgs mass\footnote{Lower $\lambda$ values suggest reduced tree-level mass and hence, needs larger corrections from the stop sector. In the NMSSM, considering the perturbative nature of $\lambda$ up to the scale of the Grand Unified Theory (GUT) one gets $\lambda \lesssim 0.7$, in the limit of $\kappa \ll \l$ \cite{Ellwanger:2009dp}.}.

The ranges of other parameters are also guided by certain aspects, e.g., in order to avoid the presence of Landau pole \cite{Landau:1954cxv,Landau:1955} below the GUT scale, i.e., $10^{16}$ GeV, one needs to consider $\lambda,\,\kappa$ values carefully at the EW scale such that $\sqrt{\lambda^2+\kappa^2} \lsim 0.7$ \cite{Ellwanger:2009dp}. Besides, smaller values of $\kappa \sim \mathcal{O} (10^{-2})$ are favoured as a stronger PT along a particular field direction prefers smaller values of the quartic coupling (e.g., $\kappa$ for PT along the $H_S$ direction) and larger values of the cubic coupling (e.g., $A_\kappa$ for a PT along the $H_S$ direction), leading to an enhanced barrier height along that specific direction. A small value of $\kappa$, together with a small $A_\kappa$ value\footnote{These ranges of $\kappa, \, A_\kappa$ are guided by the well-known $U(1)_{\rm PQ}, U(1)_{\rm R}$ limits \cite{Ellis:1988er,Dobrescu:2000jt,Dobrescu:2000yn} for the NMSSM.}, as already discussed in subsection \ref{susec:effective pot}, assure the presence of light CP-even and CP-odd states below 125 GeV. These light states help to procure a 125 GeV SM-like Higgs via the ``push-up" \cite{Escudero:2008jg,Ghosh:2014ida} effect. It is evident that one needs to consider $A_\kappa$ values carefully as for this parameter larger values are favourable for the PT dynamics while smaller ones are useful in fixing the SM-like Higgs mass around 125 GeV. Tree-level mass of the singlet-dominated CP-even state, using Eq.(\ref{eq:IndeParameters2}) and Eq. (\ref{eq:MassCPeven}), is
\beq\label{eq:masssinglettree}
\mathcal{M}^2_{S,33} \equiv \, m_{H_S}^2=\frac{-\lambda\lambda_N A_{\lambda_N} v^2_N}{2\mu}
+ \frac{\kappa A_\kappa \mu}{\lambda} + \frac{4 \kappa^2\mu^2}{\lambda^2}
+ \frac{\lambda^2 v^2 A_{\lambda} \sin 2\beta}{\mu}.
\eeq
This reduces to the known NMSSM result \cite{Ellis:1988er} at the limit $\lambda_N \to 0$ with a $\mathcal{O}(\lambda^2$) correction\footnote{This term appear to be sub-leading for small $\lambda,\,\tan\beta$ values together with $v_S \ll v$.}. It is apparent from Eq. (\ref{eq:masssinglettree}) that how different parameters appear instrumental in determining the mass of a CP-even singlet-dominated state in this framework. We consider $\kappa >0, \, A_k<0$ in this study to ensure the formation of a barrier along the $H_S$ field direction. The parameter $\mu$ plays a vital role in the PT dynamics and, as given in Eq. (\ref{eq:masssinglettree}), is also crucial for the mass and composition of a singlet-like state. Ref. \cite{Kozaczuk:2014kva} suggests that a strong EWPT favours $\mu\lesssim 300$ GeV for the $\mathbb{Z}_3$ invariant NMSSM. We consider similar ranges for $\mu$ in our analysis which also obey the ``naturalness" criteria and the LEP chargino bound \cite{L3:1999onh,ALEPH:2003acj,DELPHI:2003uqw,OPAL:2003wxm}, i.e., $|\mu|\gsim 103.5$ GeV. This range of $\mu$ values, together with the choice of $\lambda\sim \mathcal{O} (0.1)$, suggests a value for $v_S$ not too far from the EW scale as required to yield a sizable impact on the EWPT from the singlet sector. A similar observation holds true for the RH-sneutrino VEV $v_N$. The parameter $v_S$ also determines the mass term for RH-neutrino, i.e., $\propto \lambda_N v_S$ which is constrained to be around a TeV as non-zero neutrino masses in the chosen framework arise through a TeV scale seesaw. The adaptation of a TeV scale seesaw also put some bounds on the parameter $\lambda_N$ that is expected to be at most $\mathcal{O}(1)$ to avoid the existence of Landau pole below the GUT scale. The requirement of having stronger PT along the $N_R$ field direction, however, suggests smaller values of $\lambda_N$. This behaviour, is similar to $\kappa$, as addressed before. The role played by $\lambda_N$ in the PT dynamics is somewhat non-trivial and will be addressed later in detail. The remaining parameters, $A_\lambda, \, A_{\lambda_N}$ are connected to the scale of $v_S,\,v_N$ and thus, are expected to be in the ballpark of a TeV. These parameters, i.e., $A_\lambda,\, A_{\lambda_N}$ also affect tree-level masses of the CP-even and CP-odd scalar states as detailed in appendix \ref{ap:MassmMatricesEWSB}. In this analysis we consider $A_\lambda >0$ and $A_{\lambda_N} <0$. The latter choice helps to efface the possible existence of a tachyonic state in the CP-odd scalar sector (see Eq.\,(\ref{eq:MassCPodd})). We note in passing that so far we have presented a qualitative discussion in the context of the chosen independent parameters, as depicted in Eq.\,(\ref{eq:IndeParameters2}). For finding BPs through numerical analysis, one, however, also needs to consider all the relevant present and anticipated experimental bounds which we will address in the next subsection.

\subsection{Experimental Constraints}\label{susec:expconstraint}
A viable phenomenological analysis must satisfy all the concerned experimental limits, the existing and the projected ones.  The inclusion of these bounds reduces the size of the available parameter space. In this analysis, apart from considering sensitivity reaches of the existing \cite{Lentati:2015qwp,NANOGRAV:2018hou,LIGOScientific:2014qfs} and upcoming \cite{Caprini:2015zlo,Corbin:2005ny,Kudoh:2005as} GW detection setups,  we also considered constraints arising from (i) analysis of the SM-like Higgs boson properties and BSM Higgs searches at colliders, (ii) other BSM searches at the colliders, (iii) flavour-violating processes, (iv) neutrino experiments, (v) muon anomalous magnetic moment, etc. In order to employ these constraints in our numerical analysis, we first implemented the concerned model in {\tt SARAH 4.14.5} \cite{Staub:2008uz,Staub:2011dp,Staub:2012pb,Dreiner:2012dh,Staub:2013tta,Goodsell:2014bna,Staub:2015kfa,Goodsell:2018tti}. Subsequently, we use {\tt SPheno-4.0.5} \cite{Porod:2003um,Porod:2011nf,Staub:2011dp,Porod:2014xia,Goodsell:2014bna,Goodsell:2015ira,Goodsell:2016udb,Braathen:2017izn,Goodsell:2018tti,Goodsell:2020rfu} to get the mass spectrum and decay widths. The output of {\tt SPheno-4.0.5} also provides branching fractions for various flavour-violating processes, BSM contributions to the muon anomalous magnetic moment \cite{Porod:2011nf}, several LHC observables like reduced Higgs couplings, etc. We will now discuss the aforesaid constraints one by one in further detail.

(i) {\it Analysis of the SM-like Higgs boson properties and BSM Higgs searches at colliders:} Here one needs to consider two aspects: (a) SM-like Higgs analyses, and (b) the BSM Higgs searches. Concerning the first, important constraints appear from the measured mass, i.e., $\approx 125$ GeV \cite{CMS:2022dwd,ATLAS:2022net}, and couplings \cite{CMS:2022qva,CMS:2022dwd,ATLAS:2021tbi,ATLAS:2021gcn,ATLAS:2022yrq,ATLAS:2022ers,ATLAS:2022yvh,ATLAS:2022tnm,ATLAS:2022vpz}. We have used these results to assure the existence of an SM-like 125 GeV Higgs in our analysis. Besides, to assure the SM-like nature we also put a lower limit  (80\%) on the $H_u$ composition of the 125 GeV mass eigenstate. Regarding the BSM Higgs searches, i.e., for states with leading $H_{\rm NSM}, H_S$ components, and the charged Higgs, we consider the concerned experimental bounds, see for example Ref. \cite{Novak:2839763} and references therein. We used {\tt HiggsBounds} \cite{Bechtle:2008jh} {\tt 5.10.2} \cite{Bechtle:2020pkv} to implement experimental constraints from the SM and BSM Higgs searches in our numerical study.

(ii) {\it Other BSM searches at the colliders:} We already discussed in subsection \ref{susec:higher-order-pot} that we are working in an effective framework after integrating out heavy degrees of freedom like gluinos, squarks and even charged sleptons and LH-sneutrinos. We consider these states to remain heavier than 1 TeV. Such assumptions, especially for gluinos and squarks are supported by the experimental findings. In this study, we consider gluino mass $\gsim 1.8$ TeV and squark masses $\gsim 1.2$ TeV. These choices are guided by the present CMS \cite{CMS:2021beq,CMS:2022uby,CMS:2022tqr,CMS:2022bdb} and ATLAS \cite{ATLAS:2021twp,ATLAS:2021nka,ATLAS:2022rcw,ATLAS:2022ihe} observations. Experimental lower bounds on the charged slepton and LH-sneutrino masses are somewhat less \cite{ATLAS:2019gti,CMS:2021ruq,ATLAS:2022hbt}. However, we also considered them to be heavier than a TeV and integrate them out. In our numerical study, the lightest neutralino mass varies from 3 GeV to 120 GeV. However, this does not contradict any experimental bounds, e.g., SM-like Higgs decaying to a pair of neutralinos, (see for example Refs. \cite{CMS:2021cox,CMS:2021few,ATLAS:2021yqv,CMS:2021ruq,ATLAS:2021kob,ATLAS:2022rcw}) as its predominant composition ($\gtrsim 90\%$) is from the singlino and the RH-neutrino. For charginos, we used a lower bound of $103.5$ GeV \cite{L3:1999onh,ALEPH:2003acj,DELPHI:2003uqw,OPAL:2003wxm} in our analysis. It is important to note that experimental lower bounds are often interpreted in the context of simplified models and hence, they may not directly restrict the concerned model parameter space.

(iii) {\it Flavour-violating processes:} The presence of BSM states can significantly enhance branching fractions (BR) of certain flavour-violating processes, e.g., $B\to X_s \gamma$, $B^0_s \to \mu^+ \mu^-$ (see Refs. \cite{Bertolini:1986tg,Bertolini:1990if,Ciuchini:1998xy,Borzumati:2003rr,Degrassi:2006eh,Domingo:2007dx,Arbey:2012ax} and references therein), etc., compared to the SM predictions. One can minimize these new contributions by taking  $\tan\beta \lsim 5$ and fixing squarks, gluinos, sleptons, etc., masses to be heavier than a TeV. However, finite BSM contributions to these processes still appear through the EW scale uncoloured neutral scalars, neutral pseudoscalars, charged scalars, charginos and neutralinos, as required for the EWPT. Thus, we consider the following $2\sigma$ bounds 
\begin{itemize}
\centering
    \item[] ${\rm BR}(B\rightarrow X_s \gamma) =(3.49 \pm 0.38) \times 10^{-4}$ \cite{ParticleDataGroup:2022pth,hflavichep2022},
    \item[] ${\rm BR}(B^0_s\rightarrow \mu^{+}\mu^{-}) = (3.45 \pm 0.58) \times 10^{-9}$ \cite{ParticleDataGroup:2022pth,hflavichep2022}.
\end{itemize}
We note in passing that ${\rm BR}(B\rightarrow X_s \gamma),\,{\rm BR}(B^0_s\rightarrow \mu^{+}\mu^{-})$ also receive extra contributions due to R-parity breaking \cite{Kong:2004bg,Xu:2006vk}. However, given the framework of a TeV scale seesaw, the size of R-parity violating couplings, i.e., $Y^i_N v_N$, appears to be $\sim \mathcal{O} (10^{-3}-10^{-4})$ GeV and hence, hardly yield any significant contributions.

We consider charged Yukawa couplings to be diagonal for this work which helps to bypass constraints from the flavour-violating Higgs decays \cite{ATLAS:2019pmk,CMS:2021rsq}. One can also consider slepton soft squared masses to be diagonal to minimize  mixing among sleptons (both charged and neutral). With these choices, the effective bilinear R-parity violating couplings, i.e., $Y^i_N v_N$, and the LH-sneutrino VEVs appear to be main sources for the various charged lepton flavour violating (cLFV) processes like $\mu\rightarrow e \gamma$, $\mu \rightarrow eee$, etc. However, the scale of these couplings, i.e., $\sim \mathcal{O} (10^{-3}-10^{-4})$ GeV, as required for a TeV scale seesaw, can easily evade these bounds. This behaviour is very similar to the SUSY models with bilinear R-parity violation \cite{Frank:2000gs,Cheung:2001sb,Carvalho:2002bq}. We note in passing that in our numerical studies we emphasized on the cLFV processes for the $\mu$ over the similar ones from $\tau$ as the concerned existing and upcoming experimental sensitivities are much more stringent for $\mu$. Nevertheless, we also include constraints for cLFV processes involving a $\tau$ in our analysis, e.g., BR$(\tau \to \mu \gamma) < 4.4 \times 10^{-8}$ \cite{BaBar:2009hkt}. The $\mu$-based cLFV bounds included in the current analysis are given by 
\begin{itemize}
\centering
    \item[] ${\rm BR}(\mu\rightarrow e \gamma) < 4.2 \times 10^{-13}$ ~\cite{MEG:2016leq},
    \item[] ${\rm BR}(\mu \rightarrow eee) < 1.0 \times 10^{-12}$~\cite{SINDRUM:1987nra},
    \item[] ${\rm CR}(\mu N \rightarrow e N^*) < 7\times 10^{-13}$~\cite{SINDRUMII:2006dvw},
\end{itemize}
where CR$(\mu N \rightarrow e N^*)$ represents muon to electron conversion ratio in atomic nuclei with $N\,(N^*)$ representing the nucleus in the normal (excited) state. The given number, i.e., $7\times 10^{-13}$ is for the gold nuclei.

(iv) {\it Neutrino experiments:} With one generation of RH-neutrino, as already stated in section \ref{sec:Model}, it is not possible to accommodate the experimentally observed three-flavour neutrino masses and mixing \cite{deSalas:2020pgw,Esteban:2020cvm,Gonzalez-Garcia:2021dve}, even with the inclusion of loop effects \cite{Ibarra:2003up}. Thus, one will get one massive and two nearly massless neutrinos in this model. Nevertheless, even in such a scenario, we used constraints from the atmospheric mass squared difference $\Delta m^2_{\rm atm}$, i.e., $2.430 (- 2.574) \times 10^{-3} - 2.593 (- 2.410) \times 10^{-3}$ eV$^2$  for normal (inverted) hierarchy \cite{deSalas:2020pgw,Esteban:2020cvm,Gonzalez-Garcia:2021dve}, and the sum of three neutrino masses in the range $0.06$ eV - $0.12$ eV \cite{Vagnozzi:2017ovm,Planck:2018vyg,Giusarma:2018jei,Esteban:2018azc,ParticleDataGroup:2018ovx,Ivanov:2019hqk,ParticleDataGroup:2022pth,Tanseri:2022zfe}.

(v) {\it Muon anomalous magnetic moment:} Just like the flavour violating processes, the anomalous magnetic moment of muon also receives extra contributions over the SM from new parameters and the BSM states (see Refs.\,\cite{Kobakhidze:2016mdx,Hundi:2011si} and references therein). The recent comprehensive SM prediction of the muon anomaly is
$116591810$ $(43) \times 10^{-11}$ (0.37 ppm) \cite{Aoyama:2020ynm} while 
the experimental average\footnote{Here we have used combined experimental average obtained from the FNAL \cite{Muong-2:2021ojo} and the BNL E821 \cite{Muong-2:2006rrc} results.} is $116592061 (41) \times 10^{-11}$ (0.35 ppm). These numbers, adding errors in quadrature, gives $\Delta a_\mu = (251 \pm 59) \times 10^{-11}$ which is arising from the BSM sources. This, in $4\sigma$ span, gives $(1.5-48.7)\times 10^{-10}$. The BSM contributions, especially involving charged sleptons states below a TeV \cite{Chakraborti:2021dli,Badziak:2019gaf,Domingo:2008bb}, can affect this process significantly and can easily accommodate the latest experimental observation \cite{Muong-2:2021ojo}. In our analysis, as already discussed in subsection \ref{susec:higher-order-pot}, we kept charged slepton masses around a TeV. Nevertheless, by playing with the other concerned parameters we checked that the aforesaid $\Delta a_\mu$ range, i.e., $(1.5-48.7)\times 10^{-10}$ is not violated in our BPs. In fact, the choice of slepton, squark masses around a TeV or more yields suppressed cLFV processes and smaller BSM contributions to the anomalous magnetic moment of the muon. All the chosen BPs respect all the five aforesaid classes of constraints. We now discuss this study's key objectives in detail, i.e., PT properties and GW production.

\section{The EWPT and its Properties}\label{sec:EWPTnu}
	
As we already discussed, understanding the EWPT properties in the early Universe in a Particle Physics model has twofold advantages. Firstly, it can be confirmed whether the model carries the prospect to explain the origin of EWBG at some corner of the parameter space. Secondly, it provides scope to test the model at GW detectors beyond the conventional BSM searches. One of the prerequisites of EWBG is the FOPT with sufficient strength along the $SU(2)_L$ field directions so that it can suppress the processes which wash out the baryon asymmetry after it is produced, namely $SU(2)_L$ sphalerons \cite{Cline:2006ts}. The same FOPT may yield a detectable amount of GWs that could be accessible by future GW interferometers.

The structure of the thermal effective potential for a PT reveals that at very high temperatures the Universe would be in a symmetric phase with the relevant field (say $\phi_i$) being located at zero. As the Universe cools down, the symmetric vacuum may disappear and the corresponding field values could be finite. Additionally, a second minimum can be formed at some higher field value which becomes degenerate with the previous one at $T=T_c$, known as critical temperature. At temperature below $T_c$, the transition from the high-T VEVs to the low-T VEVs can take place. We should note here that a high-T (low-T) phase means an unstable (stable) vacuum below $T_c$ or above nucleation temperature. Therefore, to have an in-depth understanding of PT dynamics, an estimate of critical temperature $T_c$ and the strength of PT are enormously important.

Theoretically, the critical temperature can be obtained from the following equality:
\beq\label{degenerate at Tc}
V_T (v'_X, T_c) = V_T (v_X, T_c),
\eeq
where $v'_X$ and $v_X$ represent the high-T and low-T VEVs, respectively, along a particular field direction\footnote{As LH-sneutrinos hardly affect the PT dynamics, we do not consider the possibility $v_X=v_i$.}. We also need to ensure the existence of high- and low-T vacua which can be confirmed by the following equalities,
\beq\label{low and high vacua at Tc} 
\partial_{\phi_\alpha} V_T (v'_X, T_c) = 0,\, \partial_{\phi_\alpha} V_T (v_X, T_c) = 0,
\eeq
where $\phi_\alpha=\{H_{\rm SM}, H_{\rm NSM}, H_{\rm S}, N_R\}$. In many cases, including ours, analytical solutions of Eq. (\ref{degenerate at Tc}) and Eq. (\ref{low and high vacua at Tc}) are almost impossible to derive in order to obtain the estimates of the relevant parameters to study the PT properties. We have used the publicly available package {\tt cosmoTransitions} \cite{Wainwright:2011kj} to carry out the numerical calculation for our model in consideration.
	
A FOPT proceeds via bubble nucleation and the nucleation rate $(\Gamma)$ per unit volume $(V)$ at finite temperature is given by $\frac{\Gamma}{V} \propto T^4 e^{-S_E/T}$, where $S_E$ is the three-dimensional effective Euclidean action known as bounce action. The criterion which set the condition for the onset of bubble nucleation is given by \cite{Linde:1981zj,Mazumdar:2018dfl},
\beq\label{eq:SEdefn}
\frac{S_E (T_n)}{T_n} \simeq 140,
\eeq
where $T_n$ is the nucleation temperature. If it happens that the quantity $\frac{S_E(T_n)}{T_n} > 140$, then the transition does not occur due to low tunnelling probability.

As mentioned earlier, we use {\tt cosmoTransitions} \cite{Wainwright:2011kj} to compute 
$S_E$ and $T_n$, which also allows for estimating the probability of a transition taking place. Since we have four-dimensional field space, relevant to EWPT, a detailed scan of the model parameter space is challenging and numerically expensive as well. Therefore in the present work, we first provide a representative BP-based study which will be detailed
subsequently.  We will see that such BPs are sufficient to understand the parameter space of NMSSM + one RHN framework that can potentially give rise to an SFOPT and can also be interesting from the viewpoint of EWBG. Subsequently, we discuss the impact of new parameters in the present setup compared to the NMSSM on PT strength along different field directions by providing a scan of the relevant parameter spaces.
	
Before we proceed further, let us now define different criteria to consider a PT to be a strong one. Conventionally, in the critical temperature analysis, the order parameter that decides the fate of PT is given by,
\beq\label{SFOPT in Tc}
\gamma_c \equiv \frac{v_c (T_c)}{T_c} = \frac{\sqrt{\left\langle H_{\rm SM} \right\rangle^2   + \left\langle H_{\rm NSM} \right\rangle^2}}{T_c} \gtrsim 1.0,
\eeq
where $v_c(T_c)$ denotes VEVs of the $SU(2)_L$ Higgs fields, i.e., $H_{\rm SM},\,H_{\rm NSM}$, at $T_c$. For the nucleation temperature calculation, we define an SFOPT along the respective field directions as follows:
\begin{itemize}
\item Along $SU(2)_L$ doublet Higgs direction:
\beq\label{SFOPT condn}
\frac{\Delta \phi_{SU(2)}}{T_n} = \frac{\sqrt{(\left\langle H^{\rm lT}_{\rm SM}\right\rangle-\left\langle H^{\rm hT}_{\rm SM}\right\rangle)^2+(\left\langle H^{\rm lT}_{\rm NSM}\right\rangle-\left\langle H^{\rm hT}_{\rm NSM}\right\rangle)^2}}{T_n} \gtrsim 1.0
\eeq
\item Along the $SU(2)_L$ singlet Higgs and the RH-sneutrino direction:
\beq
\frac{\Delta \phi_{S}}{T_n} =  \frac{\sqrt{(\left\langle H^{\rm lT}_{S} \right\rangle-\left\langle H^{\rm hT}_{S} \right\rangle)^2}}{T_n} \gtrsim 1.0\;; \quad \frac{\Delta \phi_{\widetilde{N}}}{T_n} =  \frac{\sqrt{(\left\langle {N}^{\rm lT}_R \right\rangle-\left\langle {N}^{\rm hT}_R \right\rangle)^2}}{T_n} \gtrsim 1.0,
\eeq
\end{itemize}
where $\frac{\Delta \phi_{SU(2)}}{T_n},\, \frac{\Delta \phi_{S}}{T_n}$ and $\frac{\Delta \phi_{\widetilde{N}}}{T_n}$ represent PT strength along the $SU(2)_L$-doublet, $SU(2)_L$-singlet and the RH-sneutrino field direction, respectively. The notation, $\left\langle \Phi^{\rm lT} \right\rangle$ denotes the low temperature minimum while $\left\langle \Phi^{\rm hT} \right\rangle$ is the high temperature minimum of a scalar field ($\Phi$) before nucleation. A favourable condition to yield the observed baryon asymmetry of the Universe via the EWBG is $(\left\langle H^{hT}_{\rm SM}\right\rangle, \left\langle H^{hT}_{\rm NSM}\right\rangle) = (0, 0)$ with $\frac{\Delta \phi_{SU(2)}}{T_n}\gtrsim 1$. In contrast, when $(\left\langle H^{hT}_{\rm SM}\right\rangle, \left\langle H^{hT}_{\rm NSM}\right\rangle) \neq (0, 0)$, the sphaleron processes outside the bubble gets substantially suppressed which lead to inefficient production of the baryon asymmetry of the Universe from the EWBG.


\subsection{PT in the NMSSM + one RHN  model}\label{sec:EWPTnusubsec2}
As we already specified, the field space relevant to the PT analysis is four-dimensional in the present framework. This opens up the possibility of obtaining a richer PT pattern compared to the case of the NMSSM. We define the high-temperature symmetric vacuum of the scalar potential as $\Omega_0$. In principle, one can have many distinct PT patterns in the whole parameter region of the NMSSM + one RHN framework. Here we summarise a few such possibilities that advocate some unique PT patterns along the various field directions:

\begin{itemize}
\item  {\bf Type-{\tt I}:} As already stated, at $T\gg T_c$, the Universe remains in the symmetric phase where each of the four fields has zero VEV. The simplest possibility for a PT is that at critical temperature the symmetry-breaking minimum of the total scalar potential appears only along the $H_{\rm SM}$ direction. Then the PT happens from symmetric to the broken phase directly in that direction. We denote this by $\Omega_0 \xrightarrow{\text{PT}}$ $\Omega_{H_{\rm SM}}$ where $\Omega_{H_{\rm SM}}$ represents the vacuum along SM Higgs direction.

\item {\bf Type-{\tt IIa}:}  This pattern involves displacement of the $H_S$ field VEV (at $T>T_c$) from the initial zero value as the Universe cools down. We label it as Type II. Below $T_c$, the PT occurs along both the $H_{\rm SM}$ and $H_S$ field directions. We denote this particular pattern (IIa) as $\Omega_0 \rightarrow$ $\Omega_{H_S}^\prime \xrightarrow{\text{PT}} \Omega_{H_{\rm SM}}+\Omega_{H_S}$.
		
\item {\bf Type-{\tt IIb}:} This is similar to the earlier case where for $T>T_c$, a shift of the $H_S$ field value from zero vacuum appears. Below the critical temperature, the transition also takes place along the $H_S$ direction only and is represented by $\Omega_0 \rightarrow$ $\Omega_{H_S}^\prime$ $\xrightarrow{\text{PT}} \Omega_{H_S}$.
		
\item {\bf Type-{\tt IIc}:} This case also falls under the Type II category. However, below critical temperature, the PT happens along both $H_S$ and $N_R$ field directions as indicated by $\Omega_0 \rightarrow$ $\Omega_{H_S}^\prime$ $\xrightarrow{\text{PT}}$ $\Omega_{H_S}+\Omega_{N_R}$.
		
\item {\bf Type-{\tt IIIa}:} In this category, for $T > T_c$, the shifts of $H_{\rm SM}$ and $H_{S}$ VEVs from the initial zero values take place. When $T<T_c$, PT also occurs along the same field directions. This pattern is represented by $\Omega_0 \rightarrow$ $\Omega_{H_{SM}}^{\prime}+ \Omega_{H_S}^{\prime}$ $\xrightarrow{\text{PT}}$ $\Omega_{H_{\rm SM}}$+$\Omega_{H_{S}}$.
		
\item {\bf Type-{\tt IIIb}:} In this category, at $T>T_c$, the behaviour of the scalar potential is similar to the last one. However, at $T < T_c$, the PT occurs along $H_{\rm SM}$, $H_{\rm S}$ and $N_R$ directions as indicated by $\Omega_0 \rightarrow$ $\Omega_{H_{SM}}^\prime+ \Omega_{H_S}^\prime$ $\xrightarrow{\text{PT}}$ $\Omega_{H_{SM}}$+$\Omega_{H_{S}}$+$\Omega_{N_R}$.
		
\item {\bf Type-{\tt IV}}: This category is defined to indicate a particular PT pattern where at a $T>T_c$, the symmetric vacuum of the total scalar potential gets displaced along the $S$ and $N_R$ field directions. The PT occurs below $T_c$ along any of the four field directions.
		
\end{itemize}

As described earlier, any BP showing either of the type-I or type-IIa PT pattern is preferred in view of efficient EWBG, provided the corresponding PT strength satisfies the condition $\frac{\Delta\phi_{SU(2)}}{T_n}\gtrsim 1$. Whereas, the rest of the types as listed above may not lead to EWBG due to non-satisfaction of either of the conditions, $\left(\left\langle H^{hT}_{\rm SM}\right\rangle, \left\langle H^{hT}_{\rm NSM}\right\rangle\right) \neq (0, 0)$ or $\frac{\Delta \phi_{SU(2)}}{T_n}\gtrsim 1$. The PT types that do not favour EWBG, can be still interesting if it triggers an SFOPT along the $SU(2)_L$ doublet or singlet field directions and subsequently radiates GW at a detectable amount.

\subsection{A simplified model to understand EWPT in NMSSM+ one RHN model}
\label{sec:semiana}
It is always perceptive to investigate a chosen phenomenological task in the light of analytical or semi-analytical calculations. In the present framework, however, given the structure of the scalar potential $V_T$ (see Eq. \ref{eq:totPotR})), it is hardly possible to perform an exact analytical computation of vacuum structure at the critical temperature $T_c$. The exact analytical expression of the nucleation probability is also rather non-trivial and requires to be investigated numerically. However, one can consider some reasonable simplifications in the scalar potential to compute the quantity $\frac{\phi_c}{T_c}$ analytically and gain some insight into the behaviour of the new parameters that leave a non-negligible impact on the PT dynamics along the $SU(2)_L$ field directions. The quantity $\frac{\phi_c}{T_c}$, in this simplified setup, represents the PT strength along the $SU(2)_L$ field direction and is similar to what is depicted in Eq. (\ref{SFOPT in Tc}). The quantity $\phi_c$ represents the value of $\phi$ field at $T=T_c$. The semi-analytical approach adopted here closely follows \cite{Menon:2004wv, Carena:2011jy}. For this analysis, we have used the basis $\{H_u, H_d, S, \wt{N}\}$ which represents the relevant field space used to investigate the PT patterns.

It has already been advocated in subsection \ref{susec:effective pot} that the presence of lighter states below 125 GeV favours the SFOEWPT. One way to assure the same is by considering the limit $\kappa\to 0$ (known as Peccei Quinn limit, see Ref. \cite{Ellwanger:2009dp} for details). In this limit, following Eqs. (\ref{eq:softSUSY}) - (\ref{eq:Dterm}), the tree-level CP-even electrically neutral scalar potential (named as $ V_0^{\rm toy}(H_u^0, H_d^0, S, N_R)\equiv V_0^{\rm toy}$) for the NMSSM + one RHN model is given by
\bea\label{eq:potV_0}
V_0^{\rm toy}&=& \frac{1}{8}\, (g^2_1+g^2_2) \left[(H_u^0)^2 - (H_d^0)^2\right]^2 
+ \l^2 \left[(H_u^0)^2 (H_d^0)^2 + S^2 (H_u^0)^2+ S^2 (H_d^0)^2\right]\nn\\ 
&-& \l\, \l_N H_u^0 H_d^0 N_R^2 + \l_N^2 S^2 N_R^2 + \frac{\l_N^2}{4} N_R^4-2 \l A_\l S H_u^0 H_d^0 + \l_N A_{\l_N}  S N_R^2 \nn\\
&+& m_u^2 (H_u^0)^2+ m_d^2 (H_d^0)^2 + m_S^2 S^2 + M_N^2 N^2_R\,\,.
\eea

Now, we assume that the ratio of $H_u^0$ and $H_d^0$ field values at the broken phase minimum is constant up to $T = T_c$. Therefore, by keeping $\tan\b$ fixed, we define $\phi = \sqrt{(H_u^0)^2 + (H_d^0)^2}$. Since one-loop correction to the scalar potential is subdominant compared to the tree-level potential, we include only the leading one-loop effects in our semi-analytic calculation. We also consider the thermal corrections (predominantly by gauge bosons and top quark) to the effective potential, by including terms proportional to $T$ and $T^2$. Further, we neglect
contributions from charginos and neutralinos in this approximate analysis. With these contributions, the simplified effective potential Eq. (\ref{eq:potV_0}}) appears as,

\bea\label{eq:potVT_toy1}
V^{\rm toy}_T &&= V_0^{\rm toy}+c\,T^2\phi^2-E\,T\phi^3,\nn\\
&&=M^2 \phi^2 + c T^2 \phi^2 -E T \phi^3 + m_S^2 S^2 + M_N^2 N_R^2 + \frac{\wt{\l}}{2} \phi^4 + \l_N^2 S^2 N_R^2 \nn \\
&&  ~~+ \frac{\l^2_N}{4} N_R^4 + \l^2 S^2 \phi^2 - 2 \wt{a} \phi^2 S - \wt{\l}_N \phi^2 N_R^2  + a_n S N_R^2,
\eea
where
\bea\label{eq:potVT_toy2}
&& M^2 = m^2_u \sin^2\b + m^2_d \cos^2\b, \,\,
c =\frac{1}{8} \left( g^2_2 + \frac{g^2_1+g^2_2}{2} + 2 y^2_t \sin^2\b \right), \nn \\
&& \wt{a} = \l A_\l \sin\b \cos\b, ~~~ a_n = \l_N A_{\l_N},\,\, 
\wt{\l}_N = \l \l_N \sin\b \cos\b, \nn\\
&& \frac{\wt{\l}}{2} = \frac{g^2}{8} \cos^2 2\b + \l^2 \sin^2\b \cos^2\b + \frac{\Delta \l_2}{2},
\eea
with $m^2_u,\,m^2_d$ and $y_t$ denote soft square mass terms for the up, down-type Higgses and top Yukawa coupling, respectively. The quantity $\Delta {\l_2}$
is defined in Eq. (\ref{eq:delV}). In Eq. (\ref{eq:potVT_toy1}), while including the thermal corrections to the scalar potential we have only considered the leading contributions originating from gauge bosons and the top quark. The value of $E$ is roughly $\sim 0.02$ for such contributions.

Different field space trajectories render distinctive PT patterns. Given the three-field configuration (i.e., $\phi,\,S$ and $N_R$) of $V^{\rm toy}_T$ (see Eq. (\ref{eq:potVT_toy1})), a number of different combinations of PT patterns are possible depending on which of the field(s) develop(s) zero or non-zero VEV(s) during the PT. Subsequently, we explore three such possibilities analytically that can guide us to understand the role of the new parameters in the chosen model. As already stated, the estimation of $\frac{\phi_c}{T_c}$ in this semi-analytic analysis is always associated with the $SU(2)_L$ Higgs field. Therefore, throughout our calculation, we shall assume $\langle \phi \rangle \neq 0$ during the PT.\\

\noindent {\bf \underline{Case-I}:} To begin with, we consider a transition pattern where, $\langle \phi \rangle \neq 0, \langle S \rangle \neq 0$  and $\langle N_R \rangle = 0$ during the PT. To realize this, we assume the field space trajectory $N_R = 0$ and set$\frac{\partial V^{\rm toy}_T}{\partial S } = 0$ with $\langle S\rangle \neq 0$ during the PT. In this case, we can safely decouple the $N_R$ field from 
$V^{\rm toy}_T$ which now reduces to 
 \bea
V^{\rm toy}_T\Big|_{N_R=0} &=& M^2 \phi^2 + c\, T^2 \phi^2 -E T \phi^3 + \frac{\wt{\l}}{2} \phi^4 + m^2_S S^2 \nn \\
&&+ \l^2 S^2 \phi^2 - 2\, \wt{a} \, \phi^2 S.
\label{eq:potVT_toy_withoutPhiN}
\eea
Since our focus is to find $\frac{\phi_c}{T_c}$, we would like to express Eq. (\ref{eq:potVT_toy_withoutPhiN}) only in terms of $\phi$ after replacing $S$ with its respective $\phi$ field dependent VEV obtained through $\frac{\partial V^{\rm toy}_T}{\partial S}\Big|_{N_R=0} = 0$ as
\bea
\label{eq:phiSextrema}
S = \frac{\wt{a} \phi^2}{\left( m^2_S + \l^2 \phi^2 \right)}\,.\label{eq:SvevA}
\eea
Inserting Eq. (\ref{eq:phiSextrema}) into  Eq. (\ref{eq:potVT_toy_withoutPhiN}) at $T=T_c$, with $M^2(T_c)=M^2 + c T_c^2$, gives 
\bea
\label{eq:potVT_toyCrit}
V^{\rm toy}_T(\phi_c,T_c)\Big{|}_{N_R = 0} =  M^2(T_c) \phi_c^2 -E T_c \phi_c^3 + \frac{\wt{\l}}{2} \phi_c^4 - \frac{\wt{a}^2 \phi_c^4}{\left( m^2_{S} + \l^2 \phi_c^2 \right)}.
\eea
The quantities $M^2(T_c),\,\, E\gamma \, (\gamma = T_c/\phi_c)$ can be estimated from Eq. (\ref{eq:potVT_toyCrit}) using $V_T^{\rm toy}(0,T_c)=V_T^{\rm toy}(\phi_c,T_c)$ and $\partial_{\phi_c} V_T^{\rm toy}(\phi_c,T_c)=0 $, similar to Eq. (\ref{degenerate at Tc}) and Eq. (\ref{low and high vacua at Tc}). After some rearrangements, one ends up with
\beq\label{eq:CII1steq}
M^2 + c T_c^2=\frac{\phi_c^2}{2}\left(\tilde{\lambda}-\frac{2\tilde{a}^2m_S^2}{(m_S^2+\lambda^2\phi_c^2)^2}\right)+\frac{\lambda^2\tilde{a}^2\phi_c^4}{(m_S^2+\lambda^2\phi_c^2)^2},
\eeq
\beq\label{eq:CII2ndeq}
E \gamma=\tilde{\lambda}-\frac{2\tilde{a}^2m_S^2}{(m_S^2+\lambda^2\phi_c^2)^2}. 
\eeq
The quantity for $M^2\equiv M^2(T = 0)$ in Eq. (\ref{eq:CII1steq}) is obtained 
through $\partial_{\phi} V_T^{\rm toy}=0$ at $T=0,\,\phi=v$ and is given by
\beq\label{eq:MatT0case1}
-M^2=G(v) = v^2 \left(  \wt{\l} - \frac{\wt{a}^2 (2 m^2_S + \l^2 v^2)}{(m^2_S + \l^2 v^2)^2}  \right).
\eeq
Eq. (\ref{eq:CII1steq}), using Eq. (\ref{eq:MatT0case1}), translates to 
\beq\label{eq:PhicByTc2}
c \g^2 = \frac{G(v)}{\phi^2_c} + \frac{E \g}{2} + \frac{\wt{\l} \wt{a}^2 \phi^2_c}{(m^2_{S} + \l^2 \phi^2_c)^2}.
\eeq
One can now use Eq. (\ref{eq:CII2ndeq}) to get $\phi_c^2=\frac{1}{\lambda^2}\left(-m_S^2+\sqrt{\frac{2\tilde{a}^2m_S^2}{\wt{\l}-\gamma E}}\right)$ which turns Eq. (\ref{eq:CII1steq}) as 
\beq\label{eq:caseIIgamma}
-\frac{\tilde{\lambda}}{2}+\gamma E-c \gamma^2+\frac{\sqrt{\tilde{a}^2(\tilde{\lambda}-\gamma E)}}{\sqrt{2}m_S}+\frac{\lambda^2 G (v)}{-m_S^2+\sqrt{\frac{2\tilde{a}^2m_S^2}{\tilde{\lambda}-\gamma E}}}=0.
\eeq
We solve Eq. (\ref{eq:caseIIgamma}) considering some representative values (these are of similar order to our benchmarks as detailed later) of the relevant parameters to obtain $\phi_c/T_c$. We consider
\beq\label{eq:BPre}
\lambda=0.3,\,\lambda_N=0.1,\,\tan\b=5,\, A_\lambda =600\,{\rm GeV},
\eeq
together with $\Delta \l_2=0.146$ which assures (see Eq. (\ref{eq:delV}) for details) an SM-like Higgs around $125$ GeV.
\begin{center}
\begin{figure*}[h!]
\centering
\includegraphics[height=6.5cm,width=8.5cm]{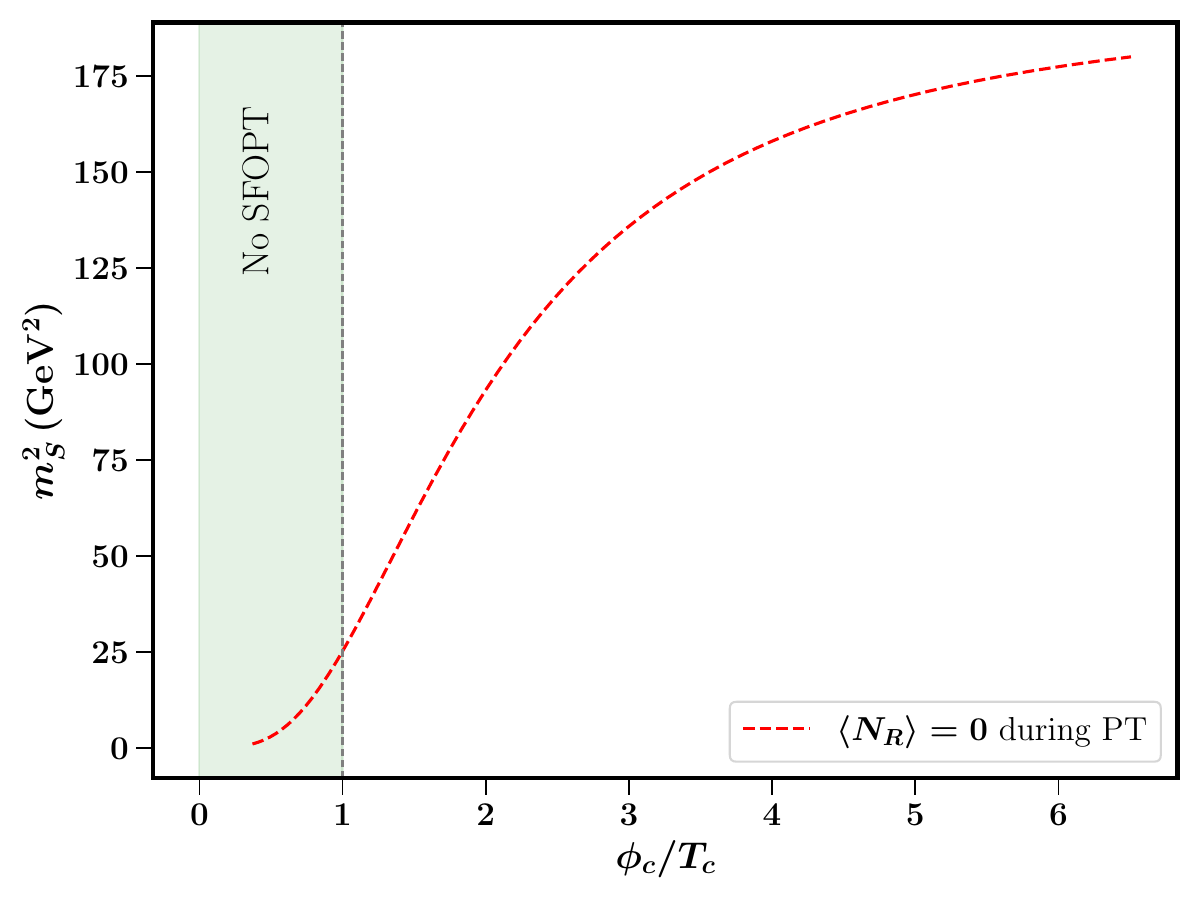}
\caption{ 
\label{fig:strengthPlot1}
The variation of $\phi_c/T_c$ with $m^2_S$, as evaluated from 
Eq. (\ref{eq:caseIIgamma}) using parameters given by Eq. (\ref{eq:BPre}). This 
corresponds to the field trajectory $\langle N_R\rangle = 0$, $\langle S\rangle\neq 0,\,
\langle \phi\rangle\neq 0$ during the PT.}
\end{figure*}
\end{center}

The variation of the PT strength $\phi_c/T_c$ with the relevant model parameter $m^2_S$, in the light of Eq. (\ref{eq:caseIIgamma}) and Eq. (\ref{eq:BPre})
is depicted in Figure \ref{fig:strengthPlot1}. It is evident from the plot that a larger value of $m^2_S$ enhances the PT strength. In addition, for this particular field trajectory, the $\frac{\phi_c}{T_c}$ has no dependence on $A_{\lambda_N}$ which is also clear from Eq. (\ref{eq:caseIIgamma}). One should note from Eq. (\ref{eq:minmcond4}), together with (\ref{eq:MassCPeven}), that $m^2_S$ is connected to the sum of $\mathcal{M}^2_{S, 33}$, $\mathcal{M}^2_{S, 44}$ and $\l^2 v^2$ up to a good approximation, neglecting terms that are $\propto \mathcal{O} (\kappa)$. A larger value of $m^2_S$, as shown in Figure \ref{fig:strengthPlot1}, favours stronger PT in the $SU(2)_L$ field direction which indicates lighter values of $\mathcal{M}^2_{S, 33}$ and $\mathcal{M}^2_{S, 44}$. This, in turn, implies that a light $S$ or $N_R$ state below 125 GeV is perhaps preferred in order to enhance the PT strength along the $SU(2)_L$ doublet field direction. We note in passing that our toy model is only suggestive up to a good approximation, and a complete numerical treatment is needed to study the exact dependence of PT strength on the new parameters.\\


\noindent {\bf \underline{Case-II}:} Now we investigate a similar PT pattern but with $\langle \phi \rangle \neq 0, \langle N_R \rangle \neq 0$ and $\langle S \rangle = 0$ during the transition. This time we decouple the $S$ field from $V^{\rm toy}_T$ in Eq. (\ref{eq:potVT_toy1}) and get,
\beq\label{eq:caseII-pot}
V^{\rm toy}_T\Big{|}_{S = 0} = M^2 (T) \phi^2 - E T \phi^3 + \frac{\wt{\l}}{2} \phi^4 +M_N^2 N_R^2  + \frac{\l_N^2}{4} N_R^2 - \wt{\l}_N \phi^2 N_R^2.
\eeq
The field space trajectory $\frac{\partial V^{\rm toy}_T}{\partial N_R}\Big{|}_{S = 0} = 0$ gives,
\beq
\label{eq:phiNextrema}
N_R^2 = - \frac{2 (M_N^2 - \wt{\l}_N \phi^2)}{\l^2_N}.
\eeq
Using Eq. (\ref{eq:phiNextrema}) one can replace $N^2_R$ in Eq. (\ref{eq:caseII-pot}).
The latter at $T=T_c,\,\phi=\phi_c$ yields
\beq
V^{\rm toy}_T(\phi_c,T_c)\Big{|}_{S = 0} = \wt{M}^2 (T_c) \phi_c^2 - E T \phi_c^3 + \frac{\wt{\l}^{'}}{2} \phi_c^4  - \frac{2 M_N^4}{\l_N^2},
\eeq
where, $\wt{M}^2(T=T_c) = M^2 (T=T_c) + \frac{2 \wt{\l}_N}{\l_N^2} M_N^2$ and $\frac{\wt{\l}^{'}}{2} = \frac{\wt{\l}}{2} - \frac{2 \wt{\l}_N^2}{\l_N^2}$. 
Following the similar approach as of Case-I, we find the following equalities to hold at $T=T_c$,
\beq\label{eq:case-II-main-eqns1}
c\g^2 = \frac{\wt{\l}^{'}}{2} - \frac{1}{\phi^2_c}\left( M^2 + \frac{2 \wt{\l}_N M_N^2}{\l_N^2}\right), 
\eeq
\beq\label{eq:case-II-main-eqns2}
E \gamma = \wt{\l}^{'}, \,{\rm ~~~where~~~}\, \gamma=\frac{T_c}{\phi_c}.
\eeq
\begin{center}
\begin{figure*}
\centering
\includegraphics[height=6.5cm,width=8.5cm]{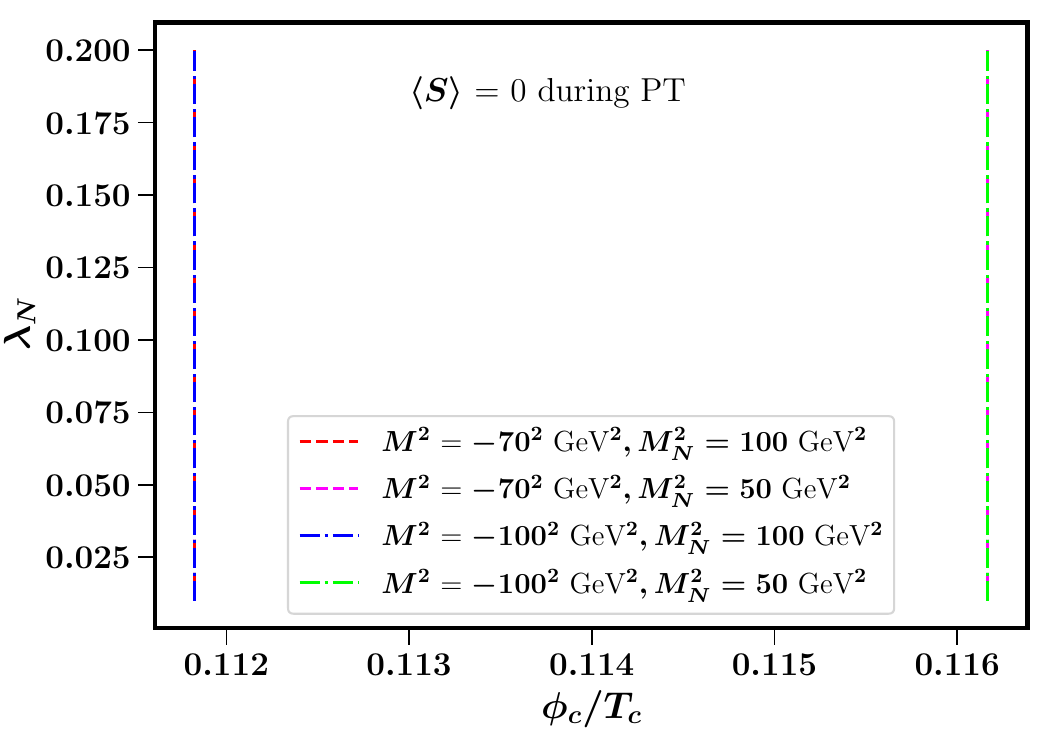}
\caption{\label{fig:strengthPlot-caseII}The variation of $\phi_c/T_c$ with $\l_N$ as evaluated from Eq. (\ref{eq:case-II-main-eqns1}) and Eq. (\ref{eq:case-II-main-eqns2}), for two values of $M^2$ and $M_N^2$, using parameters given by Eq. (\ref{eq:BPre}).	This plot is the result of a semi-analytic computation considering the field trajectory $\langle S \rangle = 0$, $\langle N_R \rangle\neq 0,\, 
\langle \phi \rangle \neq 0$ during the PT.
}
\end{figure*}
\end{center}

Just like the Case-I, one can solve Eq. (\ref{eq:case-II-main-eqns1}) and Eq. (\ref{eq:case-II-main-eqns2}) simultaneously to obtain $\frac{\phi_c}{T_c}$. The equalities in Eqs. (\ref{eq:case-II-main-eqns1}) and (\ref{eq:case-II-main-eqns2}) also evince the dependence of $\frac{\phi_c}{T_c}$ on $\lambda,\l_N,\tan\beta,\,M^2_N,\,M$ (see Eq. (\ref{eq:potVT_toy2})) in the present case. The variation of $\phi_c/T_c$ with
$\l_N$ is shown in Figure \ref{fig:strengthPlot-caseII} for two different $M^2$ values which is related to Higgs soft mass parameters $m^2_u,\,m^2_d$ and $\tan\beta$ (see Eq. (\ref{eq:potVT_toy2})). We set the other relevant parameters as specified in Eq. (\ref{eq:BPre}) together with $M^2_N$ value as $50\,\,{\rm GeV}^2$ and $100\,\,{\rm GeV}^2$. It is evident from this figure that $\phi_c/T_c$ 
is hardly sensitive to sector specific parameters, i.e., $\l_N,\,M^2_N$ values. Nevertheless,
marginal enhancement appears in $\phi_c/T_c$ values
for lower $M^2_N$ values. Thus, any PT with the transition pattern $\langle \phi \rangle \neq 0, \langle N_R \rangle \neq 0$ and $\langle S \rangle = 0$ hardly shows any sensitivity on the relevant new parameters. Hence, in our full numerical investigation of the total NMSSM + one RHN effective potential, we shall omit all the transitions which involve $\langle S \rangle = 0$ during the PT.
We note in passing that $\phi_c/T_c$ varies with the parameter $\wt{\l}'$. However, it is independent of $\l_N$ as the expression of $\wt{\l}'$ contains $\wt{\l}^2_N/\l^2_N$ which removes the $\l_N$ dependence (see Eq. (\ref{eq:potVT_toy2})).\\

\noindent {\bf \underline{Case-III}:} As the  final case, we explore the most general possibility, i.e., $\langle \phi \rangle \neq 0, \langle N_R \rangle \neq 0$ and $\langle S \rangle \neq 0$ during the PT. For this case also we start from Eq. (\ref{eq:potVT_toy1}) and first,
derive the trajectory along  $N_R$ field direction by setting $\frac{\partial V_T^{\rm toy}}{\partial N_R} = 0$. The latter gives
\beq
\label{eq:phiNextrema-caseIII}
N_R^2 = - \frac{2 (M_N^2 + a_n S - \wt{\l}_N \phi^2 + \l_N^2 S^2) }{\l^2_N}.
\eeq
One can use Eq. (\ref{eq:phiNextrema-caseIII}) to remove $N^2_R$ dependence of Eq. (\ref{eq:potVT_toy1}) which yields
\bea
\label{eq:potVT_toy3}
V^{\rm toy}_T (\phi, S) &&= m^2 \phi^2 + c T^2 \phi^2 -E T \phi^3 +\frac{\wt{\l}^{'}}{2} \phi^4 + \l^{' 2} \phi^2 S^2 -2 \wt{a}^{\prime} \phi^2 S \nn \\
&&\hspace{0.3cm}+ \wt{m}^2_S S^2 - \l_N^2 S^4 - 2 a_n S^3 - \frac{1}{\l^2_N} (M^2_N + a_n S)^2,
\eea
where $\wt{\l}^{'}/2 = \wt{\l}/2 - \l^2 \sin^2\b \cos^2\b$, $\l^{' 2} = \l^2 + 2 \wt{\l}_N$, $\wt{a}^{\prime} = \wt{a} - \l A_{\l_N} \sin\b \cos\b$, $m^2 = M^2 +  \frac{2 \wt{\l}_N}{\l^2_N} M^2_N$ and $\wt{m}^2_S = m_S^2 - 2 M_N^2$. Subsequently, we consider the field-space trajectory along the $S$ field direction, i.e., $\frac{\partial V_T^{\rm toy}}{\partial S} = 0$ which gives
\beq\label{eq:case-III-S-trajectory}
2 \l_N^2 S^3 + 3 a_n S^2 - S \left( \wt{m}_S^2 + \l^{' 2} \phi^2 - A_{\l_N}^2 \right) + \wt{a}^{'} \phi^2 + \frac{M_N^2}{\l_N} A_{\l_N} = 0,
\eeq
which, without any assumptions, does not give a simple analytical form of $S$. Therefore, to simplify our calculation in obtaining an approximate relation for $\frac{\phi_c}{T_c}$, we shall assume $\l_N \ll 1$. Accordingly, we can safely discard $\mathcal{O}(\l_N^2)$ or any higher order terms or terms that are a combination of the type $\mathcal{O}(\l \l_N^2)$ and so on compared to $\mathcal{O}(\l_N)$ term. With this assumption, Eq. (\ref{eq:potVT_toy1}) can be re-written as
\bea
\label{eq:pot-caseIII-1}
V^{\rm toy}_T &&=M^2 \phi^2 + c T^2 \phi^2 -E T \phi^3  + \frac{\wt{\l}}{2} \phi^4 + m_S^2 S^2 + M_N^2 N_R^2 \nn \\
&&  ~~ + \l^2 S^2 \phi^2 - 2 \wt{a} \phi^2 S - \wt{\l}_N \phi^2 N_R^2  + a_n S N_R^2,
\eea
where $\frac{\partial V_T^{\rm toy}}{\partial S} = 0$ gives
\beq
\label{eq:case-III-S-extrema}
S = \frac{\wt{a} \phi^2 - \frac{a_n}{2} N_R^2}{(m_S^2 + \l^2 \phi^2)}\,.
\eeq
Putting the above expression for $S$ in Eq. (\ref{eq:pot-caseIII-1}) one gets
\beq
\label{eq:pot-case-III-2}
V^{\rm toy}_T = M^2(T) \phi^2 -E T \phi^3 + \frac{\wt{\l}}{2} \phi^4 - \frac{(\wt{a} \phi^2 - \frac{a_n}{2} N_R^2 )^2 }{\left( m^2_{S} + \l^2 \phi^2 \right)} + M_N^2 N_R^2 - \wt{\l}_N \phi^2 N_R^2.
\eeq
From Eq. (\ref{eq:pot-case-III-2}), the condition $\frac{\partial V^{\rm toy}_T}{\partial N_R} = 0$ implies
\beq
\label{eq:case-III-NR-minima}
N_R^2 = \frac{2}{a_n^2} (m_N^2 - \wt{\l}_N \phi^2) (m_S^2 + \l^2 \phi^2) + \frac{2 \wt{a}}{a_n} \phi^2
\eeq.
Putting this value of $N^2_R$ in 
Eq. (\ref{eq:pot-case-III-2}) one finds
\bea
\label{eq:pot-case-III-3}
V^{\rm toy}_T (\phi) &&= \wt{M}^2(T) \phi^2 - E T \phi^3 + \frac{\wt{\l}^{'}}{2} \phi^4 -\frac{1}{a_n^2} (M_N^4 - 2 M_N^2 \wt{\l}_N \phi^2) (m_S^2 + \l^2 \phi^2) \nn \\ 
&& + \frac{2 M_N^4 m_S^2}{a_n^2}, 
\eea
where $\wt{M}^2(T) \approx M^2(T) + \frac{2 \wt{a}}{a_n} M_N^2 - \frac{4 \wt{\l}_N {M_N^2 m_S^2}}{a^2_n} + \frac{\l^2 M_N^4}{a_n^2}$ and $\frac{\wt{\l}^{'}}{2} \approx \frac{\wt{\l}}{2} - \frac{2 \wt{\l}_N \wt{a}}{a_n}$ dropping $\mathcal{O}(\wt{\l}_N \l^2)$ term. With the help of 
Eq. (\ref{eq:pot-case-III-3}), similar to Case-I and Case-II,
one can derive the following relations with a similar approach.
These are written as
\beq\label{eq:case-III-solution1}
c \g^2 = \frac{\wt{\l}^{'}}{2} + \frac{\wt{G}(v)}{\phi_c^2} - \frac{\wt{\l}_N M_N^2}{a_n^2} (m_S^2 +\l^2 \phi^2),
\eeq
\beq\label{eq:case-III-solution2}
E \g = \wt{\l}^{'},
\eeq
with
\beq
\wt{G}(v) = \wt{\l}^{'} v^2 + \frac{\wt{\l}_N M_N^2}{a_n^2} (m_S^2 +\l^2 v^2). 
\eeq
\begin{center}
\begin{figure}[t!]
\centering	
{\includegraphics[height=6.5cm,width=8.5cm]{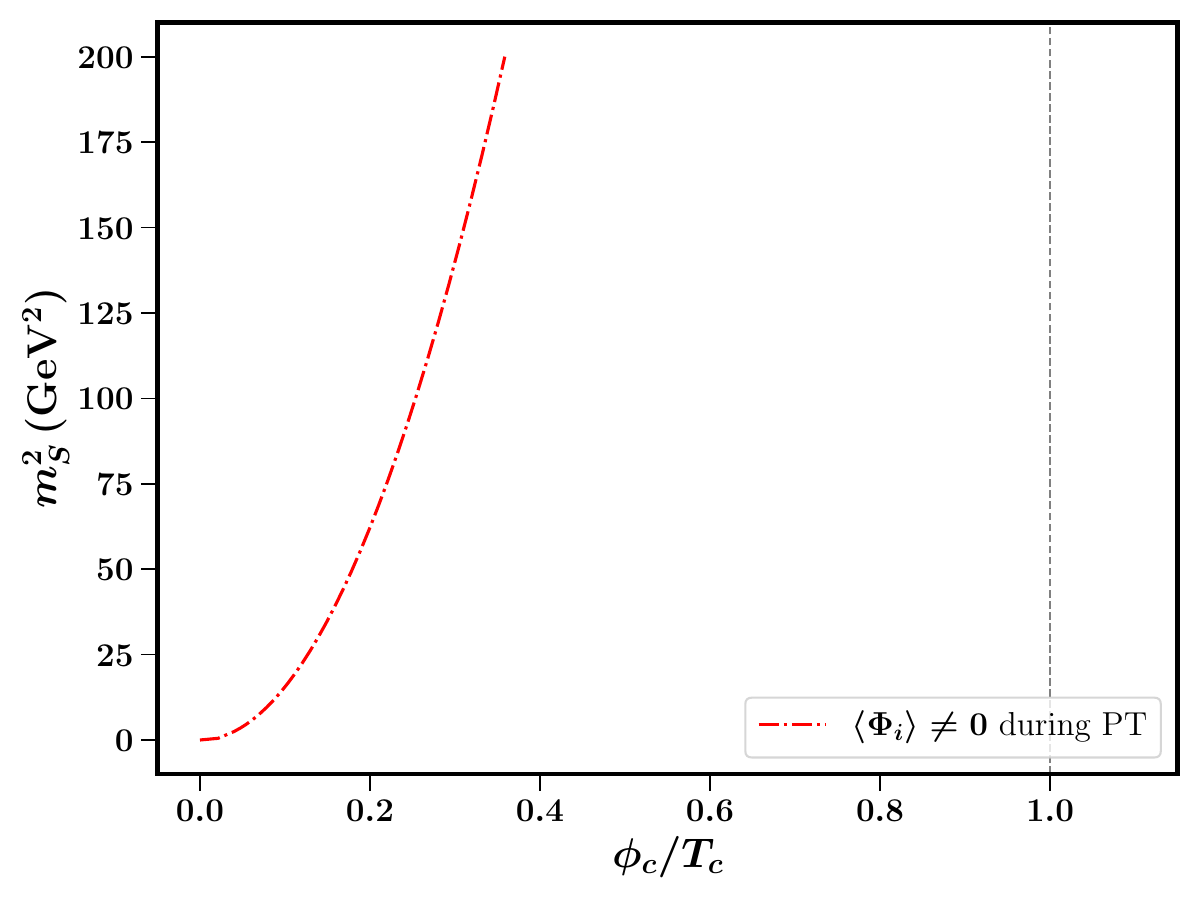}}
\caption{\label{fig:strengthPlot-case-III} The variation of $\phi_c/T_c$ with of $m^2_S$, as evaluated from our semi-analytic calculation considering the field trajectory $
\langle \Phi_i \rangle\neq 0 \equiv
\langle \phi \rangle\neq 0,\,
\langle S \rangle\neq 0\,$ and, $
\langle N_R \rangle\neq 0,\,
$ during the PT.}
\end{figure}
\end{center}

Once again, like previous cases, we solve for $\frac{\phi_c}{T_c}$ numerically using Eq. (\ref{eq:case-III-solution1}) and Eq. (\ref{eq:case-III-solution2}) considering the representative point given by Eq. (\ref{eq:BPre}), together
with $M_N=150$ GeV (as of the Case-II) and, $|A_{\l_N}| = 300\,{\rm GeV}$. We plot the result in Figure \ref{fig:strengthPlot-case-III} in the $m_S^2$ - $\frac{\phi_c}{T_c}$ plane. It is evident from Figure \ref{fig:strengthPlot-case-III} that, in the phase space of our toy model if we have a transition pattern with $\langle \phi \rangle \neq 0, \langle S \rangle \neq 0, \langle N_R \rangle \neq 0$ simultaneously during the PT, the PT strength increases in the $SU(2)_L$ filed direction with increasing $m_S^2$. This leads to the same conclusion that we have observed in Case-I, indicating the presence of light singlet or RH-sneutrino-dominated states. However, it never fulfills the criteria of an SFOPT along the $SU(2)_L$ filed direction. Therefore, we can interpret that Case-III is less favourable from the viewpoint of an SFOPT along $SU(2)_L$ filed direction.

In summary, our semi-analytic calculation suits perfectly for the $SU(2)_L$ direction as $\frac{\phi_c}{T_c}$ calculation is always associated with the $SU(2)_L$ Higgs field. Therefore, to explore the influence of the new parameters on the PT strength along the different field directions, we need to solve the bounce action. This helps us to obtain the nucleation probability of a successful PT, and hence, to determine the nucleation temperature of the transition. For this purpose, and given the complexity of our model framework, we must rely on numerical routines and we shall use {\tt{cosmoTransitions}} \cite{Wainwright:2011kj} to obtain the necessary observables. We note 
in passing that our toy model is only suggestive up to a good approximation and we will analyse the total effective potential numerically in the following subsection.

\subsection{Numerical results}\label{subsec:numerical results}
As earlier mentioned, we would like to begin with a benchmark-based study of EWPT in the present work. In the later part, we will be discussing explicitly the dependence of new parameters in the current setup compared to the NMSSM. We first tabulate six BPs in Table\,\ref{tab:BPs} that are consistent with all relevant theoretical and experimental constraints, as discussed in subsection \ref{susec:expconstraint}. We select the BPs in such a way that they show distinct PT characteristics with some of them favouring EWBG and carrying good to moderate detection prospects at GW detectors. Note that we have four soft-SUSY breaking parameters (i.e., $A_\lambda,\, A_\kappa, \, A_{\lambda_N},\, A_N$) in our model. We discuss the possible role of all the $A-$ parameters in section\,\ref{sec:param_space}. Recall that one of the soft parameters $A_N$ does not contribute much to the PT dynamics since it is always associated with the tiny neutrino Yukawa coupling $Y^i_N$ as earlier clarified. We  keep $A_N$ above the TeV scale for all BPs, which ensures slepton masses $\gtrsim \mathcal{O}(1$ TeV). In Table\,\ref{tab:BPs}, we provide the eigenvalues of the four CP-even mass eigenstates, i.e., $m_{h_{125}}, m_H, m_{H_S}, m_{\wt{N}}$, corresponding to each BPs. The leading composition in these states are coming from the $H_{\rm SM}$, $H_{\rm NSM}$, $H_S$ and $N_R$ fields, respectively. We have explicitly checked that all the BPs evade the relevant experimental bounds as detailed in subsection \ref{susec:expconstraint}. Nevertheless, we have explicitly shown values of the various flavour-violating processes $\Delta a_\mu$ and $\Delta m_{\rm atm}^2$ for the sake of completeness. In Table\,\ref{tab:PTproperties1} and Table \ref{tab:PTproperties2}, we have summarised the PT outputs of the BPs as obtained from the {\tt cosmoTransitions} \cite{Wainwright:2011kj} package. Below we discuss the PT characteristics for each of the BPs in detail. \\
\begin{table}[!ht]
\begin{center}
\scriptsize
\begin{tabular}{| c | c | c | c | c | c | c |}
\hline
&BP-I  &BP-II & BP-III & BP-IV  & BP-V & BP-VI \\
\hline
$\tan \beta$ & 2.90 & 2.74  & 2.90 & 5.77 & 4.79  & 5.86\\
$\lambda$ & 0.416 & 0.412  & 0.416 & 0.384 &  0.118  & 0.111\\
$\kappa$ & 0.022 & 0.019  &0.022  & 0.012 & 0.013  &  0.051 \\
$\lambda_N$ & 0.146 & 0.142  &0.146 & 0.130 &  0.260  & 0.238 \\
$Y^1_N \times 10^7$ & 0.9 & 0.65  & 1.1 & 1.0 &  3.6  & 4.3 \\
$Y^2_N\times 10^7$ & 0.9 & 0.65  & 1.1 & 1.0 &  3.6 & 4.3 \\
$Y^3_N\times 10^7$ & 0.9 & 0.65  & 1.1 & 1.0 &  3.6  & 4.3 \\
$A_{\lambda}$ [GeV] & 775.48 &  705.32  & 775.48 & 1184.87 & 988.08 &  920.08  \\
$A_{\kappa}$ [GeV] & -62.75 & -25.37  & -95.61 & -107.08 & -11.70 & -41.61 \\
$A_{\lambda_N}$ [GeV] & -349.68 & -337.77  &-326.60 &-363.16 & -1358.30 & -1528.57  \\
$A_N$ [GeV] &  -16000.0 & -12000.0 & -8500.0 & -12000.0 & -6500.0 & -5000.0  \\
$\mu$ [GeV]  & 224.56 & 220.86  & 224.56 & 203.12 & 153.59 & 162.64 \\
$v_{N}$ [GeV] & 308.80 &  325.21  & 284.50 & 386.45 & 136.57 & 355.66   \\
$v_1\times 10^4$ [GeV] & 1.0 &  0.55  & 1.0 & 1.2 & 1.0 & 1.0   \\
$v_2\times 10^4$ [GeV] & 1.0 &  0.55  & 1.0 & 1.2 & 1.0 & 1.0   \\
$v_3\times 10^4$ [GeV] & 1.0 &  0.55  & 1.0 & 1.2 & 1.0 & 1.0   \\
\hline
\hline
$m_{{h_{125}}}$ [GeV] & 126.02 &  124.80  & 125.64 & 125.63 & 126.28    & 124.05\\
$m_{H}$ [GeV] & 772.36 &  718.07  & 772.73 & 1213.76 & 897.40 & 1012.14 \\
$m_{H_{S}}$ [GeV] & 83.60 & 88.98 & 69.48 & 109.54 & 97.31  & 195.41  \\
$m_{\widetilde{N}}$ [GeV] & 48.60 & 51.65  & 51.89 & 27.65 & 65.18 & 115.63 \\
\hline
\hline
${\rm BR}(B\rightarrow X_s \gamma) \times 10^4$ & $3.61 $ &  $3.70$  & $3.60$ & $3.47$ & $3.59$ & $3.55$   \\
${\rm BR}(B^0_s\rightarrow \mu^{+}\mu^{-})\times 10^9$ & $3.24$ &  $3.26$  & $3.24$ & $3.19$ & $3.20$ & $3.19$ \\
${\rm BR}(\mu\rightarrow e \gamma)\times 10^{30}$ & $394$ & $0.61$  & $4.98$ & $51.4$  & $404$ & $173$ \\
${\rm BR}(\mu\rightarrow eee)\times 10^{29}$ & $113.0$ & $363.7$ & $44.6$ & $53.9$ & $2.04$ & $2.04$ \\
${\rm CR}(\mu N \rightarrow e N^*)\times 10^{28}$ & $1.81$ & $0.11$ & $2.49$ & $4.43$ & $4.85$ & $7.31$ \\
$\Delta m^2_{\rm atm} \times 10^{3}\,{\rm eV}^2$ & $2.51$ & $2.57$ & $2.58$ & 2.54 & $2.58$ & $2.46$ \\
$\Delta a_\mu \times 10^{10}$ & $3.88$ & $0.75$ & $3.42$ & $1.94$ & $1.54$  & $3.24$ \\
\hline
\end{tabular}
\end{center}
\caption{The representative BPs that we will use to study the PT patterns in the present framework. Apart from the parameters mentioned above, we fix the gaugino mass parameters $M_1 = 300 \,{\rm GeV}, M_2 = 2 M_1, M_3 = 6 M_1$, trilinear soft coupling $A_t$ around $2 \,{\rm TeV}$. We also consider RH-slepton soft masses above 1 TeV and squarks soft masses $M_{\widetilde{Q}_i}, M_{\widetilde{u}^c_i}, M_{\widetilde{d}^c_i}$ all above 1.2\, TeV. 
With the chosen values of parameters $Y^i_N,\, v_i$ and $A_N$, the LH-sneutrino
and LH-slepton masses also appear in the ballpark of a TeV. As already stated in subsection \ref{susec:expconstraint}, suppressed cLFV processes and smaller BSM contributions to the anomalous magnetic moment of muon are evident now due to slepton, squark masses around a TeV or more. In fact, for BP-II, $\Delta a_\mu$ remains below the
aforesaid $4\sigma$ range. ${\rm CR}(\mu\, N \rightarrow e\, N^*)$ value is estimated for the gold nuclei.}
\label{tab:BPs} 
\end{table}
\begin{table}[!ht]
\begin{center}
{\footnotesize
\begin{tabular}{| c | c | c | c | c | c | c |}
\hline	
&	 BP-I			&	BP-II				&	{BP-III}	  \\	
\hline
\hline	
Transition Type	&	Type-{\tt IIa}		&	Type-{\tt IIa}	&	Type-{\tt IIIa}						\\
\hline
$v_c/T_c$	& 1.30 ({\tt In}); 0 ({\tt Out})	&	0.73 ({\tt I}); 0 ({\tt O})		& 1.83 ({\tt I}); 0.61 ({\tt O})						\\
$\Delta \phi_{SU(2)}/T_n$	& 1.58	&	0.81				&	1.28					\\
$\Delta \phi_S/T_n$		&	4.70	&	1.16				& 7.61						\\
$\Delta \phi_{\widetilde{N}}/T_n$		&	0			&	0				&0					\\
\hline
$T_c$ (GeV)		&	117.8 &		127.2	&	101.6		 	 \\
$T_n$ (GeV)		&	109.9 &		126.7		& 82.9					\\
\hline
high-$T_n$ VEVs			&	(0, 0, 113.8, 0)			&	(0, 0, 341.6, 0)			&	(105.8, 32.5, 88.8, 0)					\\
low-$T_n$ VEVs			&\hspace{.08cm} 	(173.1, 9.5, 631.3, 0)\hspace{.08cm}		&\hspace{.08cm} 	(102.3, 11.3, 488.7, 0)\hspace{.08cm}  	&\hspace{.08cm} 	(208.1, 4.8, 719.7, 0)	\hspace{.08cm} 	 		\\
\hline
high-$T_c$ VEVs			&	(0, 0, 72.6, 0)		&	(0, 0, 333.1, 0)			&	(62.4, 20.9, 35.6, 0)					\\
low-$T_c$ VEVs			&\hspace{.08cm} 	(152.9, 11.8, 572.5, 0)\hspace{.08cm} 		&\hspace{.08cm} 	(92.5, 10.1, 467.7, 0)\hspace{.08cm} 	&\hspace{.08cm} 	(186.4, 10.6, 625.4, 0)	\hspace{.08cm} 	
\\ \hline
\end{tabular}}
\end{center}
\caption{The PT properties for first three BPs as tabulated in Table\,\ref{tab:BPs}.}
\label{tab:PTproperties1}
\end{table}
	
\begin{figure*}[!ht]
\centering
\subfigure{\includegraphics[width=7cm,height=5.5cm]{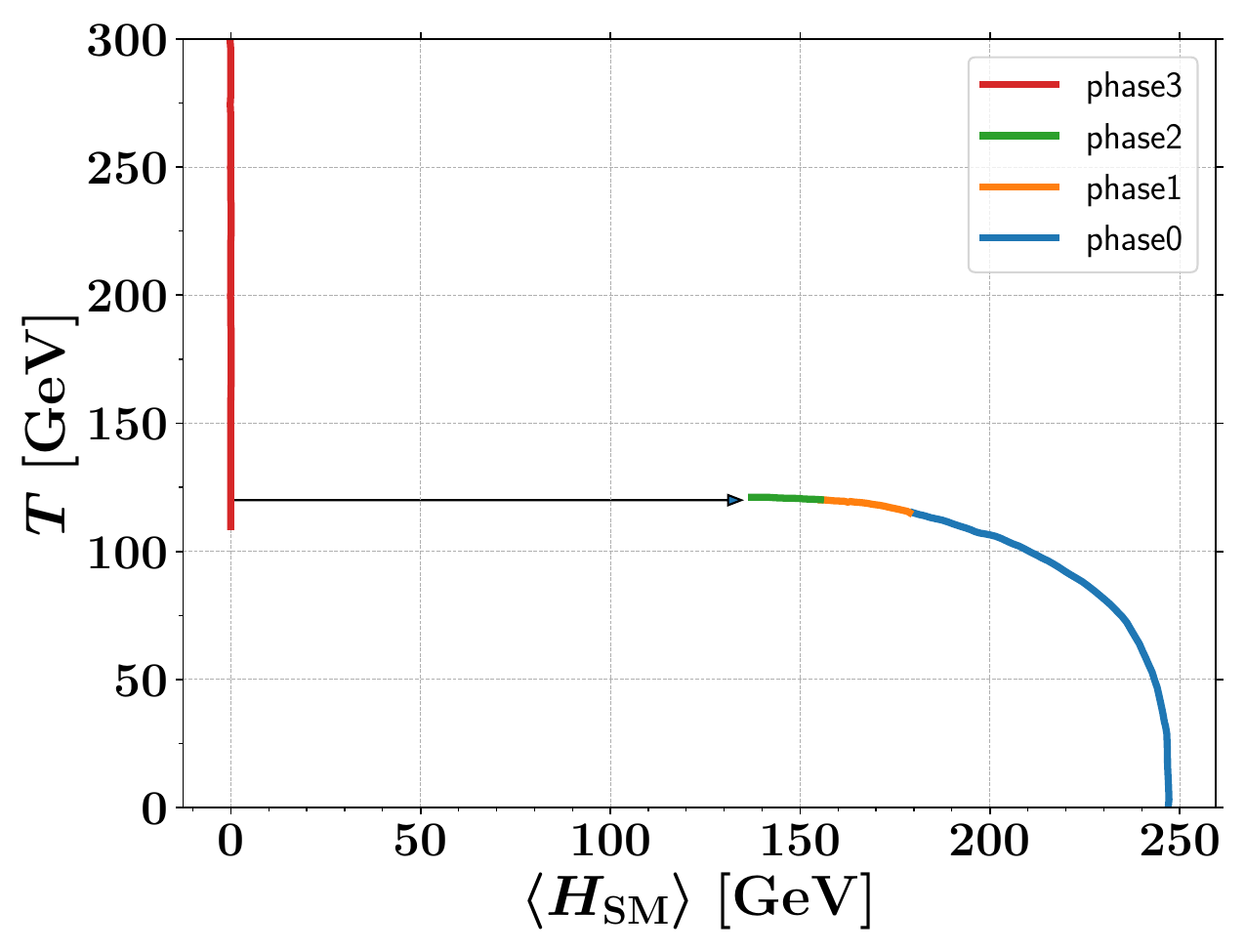}}
\subfigure{\includegraphics[width=7cm,height=5.5cm]{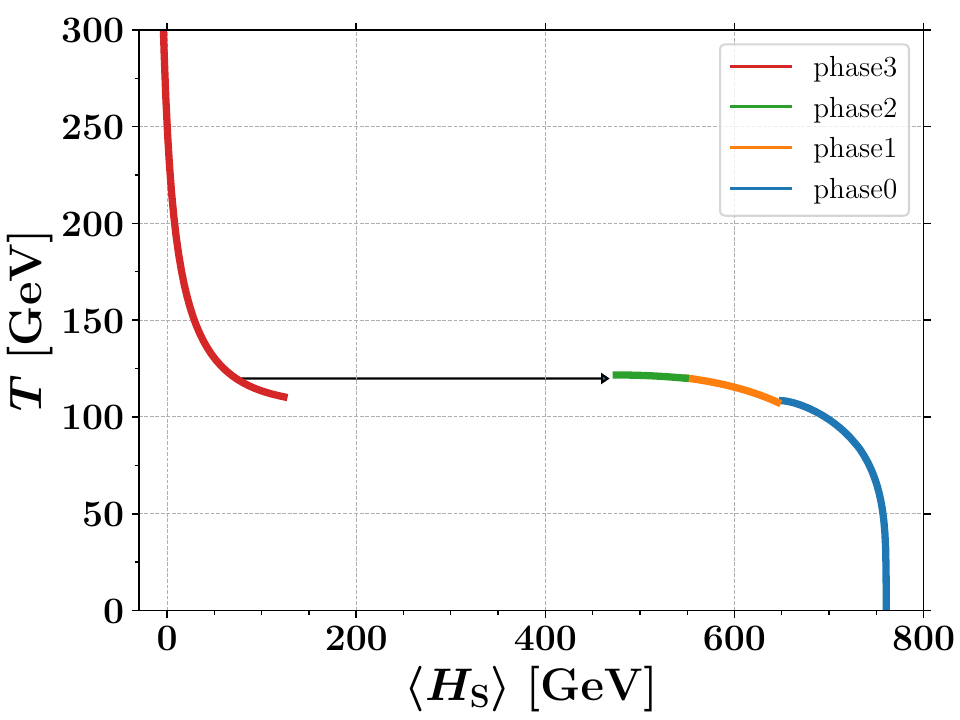}} 
\caption{\label{fig:resultsPT-BPI}
Phase structures as a function of temperature along the $H_{\rm SM}$ and $H_S$ field directions for BP-I. Different colours represent the locations of a particular field as a function of temperature. The black coloured line with the arrow connects two degenerate phases at $T=T_c$ and the direction of the arrow indicates a possible FOPT.}
\end{figure*}

\vspace{0.1cm}	
\noindent $\bullet$ \textbf{BP-I and BP-II :} Out of these two representative BPs, BP-I shows an SFOPT along both the $SU(2)_L$-doublet and singlet field directions. On the other hand, we obtain a weaker FOPT for BP-II in the $SU(2)_L$ doublet directions whereas a stronger one along the $SU(2)_L$ singlet direction. In Figure\,\ref{fig:resultsPT-BPI}, we have shown the evolution of the phase structures along the $H_{\rm SM}$ (left) and the $H_S$ (right) field directions as a function of temperature for BP-I. The critical temperature for BP-I is 117.8 GeV as noted in Table\,\ref{tab:PTproperties1}. Above the critical temperature $H_{\rm SM}$ is located at zero (as pointed by the legend phase 3, red coloured, in Figure\,\ref{fig:resultsPT-BPI}). At $T=T_c$, we find another degenerate minimum along the same field direction, which is $\langle H_{\rm SM}\rangle = 152.9\, {\rm GeV}$ (as marked by phase 2, green coloured). The black coloured line with the arrow connects the high-T and low-T VEVs indicating a possible FOPT. The bubble nucleation occurs afterwards and it ends at 109.9 GeV which we have highlighted in orange colour (also labelled as phase 1). A similar pattern can be observed along $H_S$ direction too as shown in the right panel of Figure\,\ref{fig:resultsPT-BPI}. The interesting point to mention here is that the $\langle H_S \rangle$ starts to get displaced from zero value even at a temperature above $T_c$. This is in contrast to the evolution of phase structure along $H_{\rm SM}$ direction for this particular BP. The BP-II shows similar characteristics although the strong PT occurs only along the $H_{\rm S}$ direction. The high-temperature behaviour of the total scalar potential leads us to identify the PT properties for both BP-I and BP-II as Type-IIa. For BP-I, we observe from Table\,\ref{tab:PTproperties1}, that the PT strength at $T=T_c$ is greater than one inside the bubble and zero outside the bubble. Therefore a baryon number may be generated in the broken phase and the wash-out effects are likely to be suppressed. In view of this, BP-I is favoured in order to address EWBG. However, BP-II shows a weaker FOPT in the $SU(2)_L$ doublet directions and hence is not suitable to address the question of EWBG. In subsection\,\ref{sec:GW} we will discuss the strength of emitted GW spectrum during bubble nucleation for both BP-I and BP-II in view of the proposed sensitivities of a few forthcoming GW experiments.

\vspace{0.1cm}
\noindent $\bullet$ \textbf{BP-III:} The BP-III falls into Type-IIIa category. It implies that at a temperature above $T_c$,  both $H_{\rm SM}$ and $H_{\rm S}$ attain non-zero VEVs. The critical temperature for this BP comes out to be $101.6$ GeV. At this temperature, the presence of two degenerate vacua is noticed having nonzero field values for both $SU(2)_L$ doublet and singlet fields, which set the possibility of a PT. We obtain SFOPT along both the $SU(2)_L$-doublet and singlet field directions where the PT strength turns out to be larger than one. However, the quantity $\frac{\phi_c}{T_c}$ becomes non-zero both inside and outside the bubble. This gives rise to a stronger wash-out effect which is likely to suppress the yield of baryon asymmetry and hence seemingly disfavored in view of EWBG. Nevertheless, it carries good detection prospects in the GW detectors due to relatively larger PT strength $\frac{\Delta\phi_S}{T_n}$ compared to BP-I.
	
\begin{figure*}[!ht]
\centering
\subfigure{\includegraphics[width=7cm,height=5.5cm]{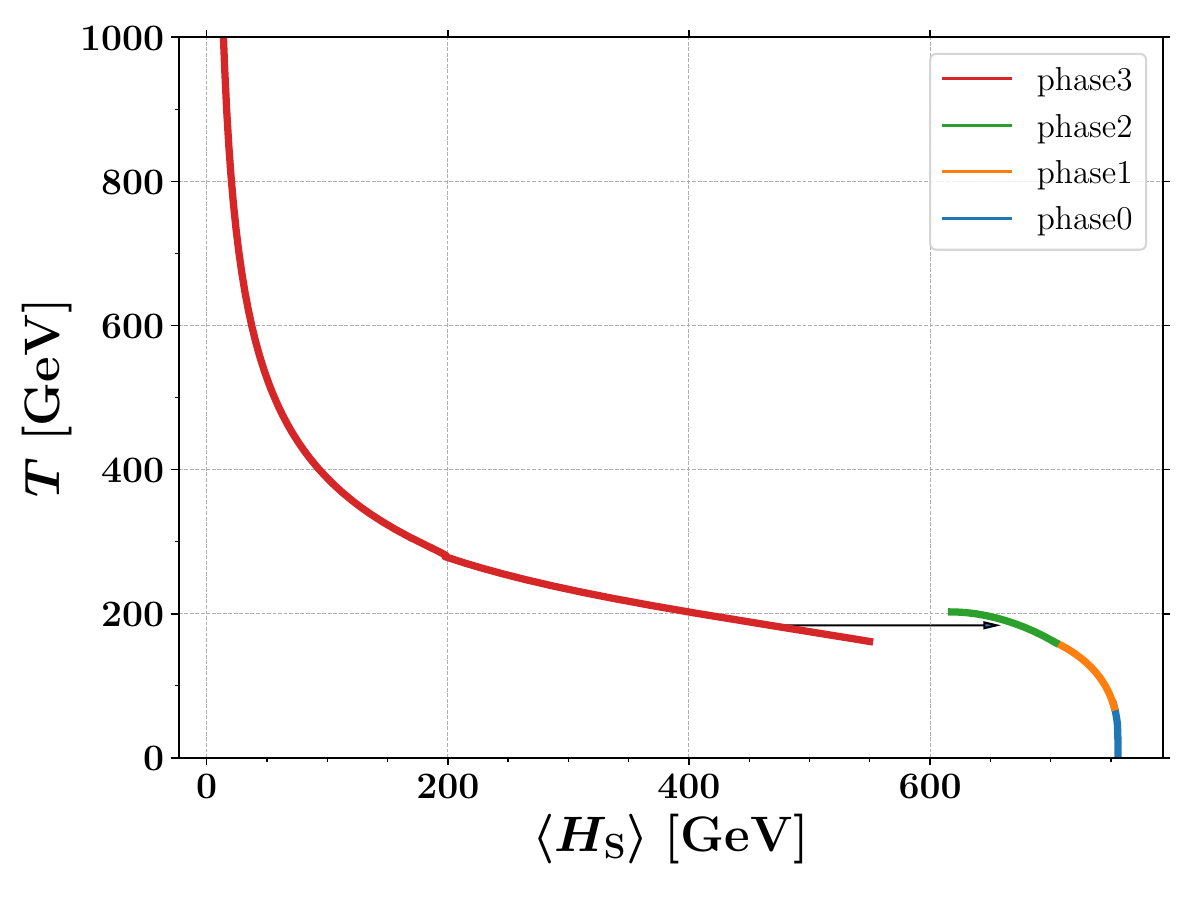}}
\subfigure{\includegraphics[width=7cm,height=5.5cm]{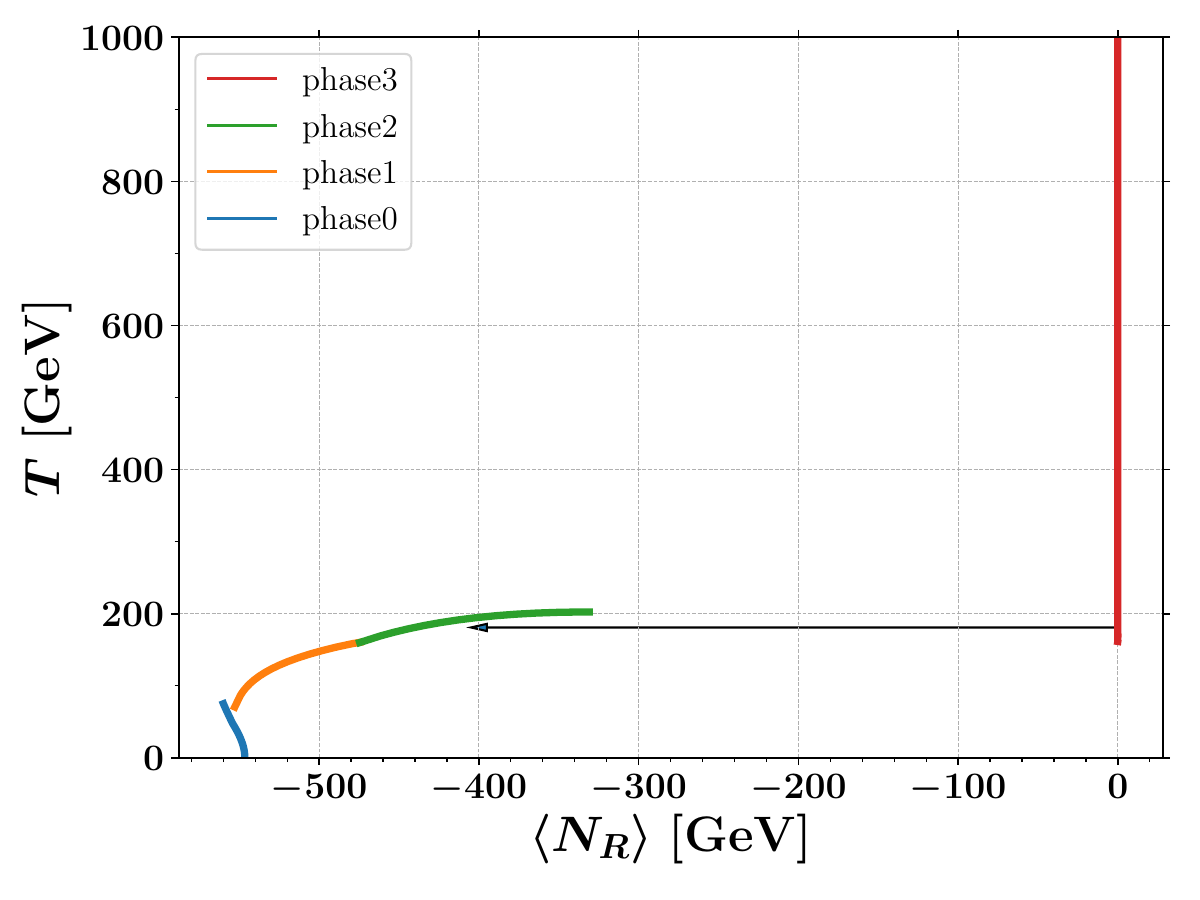}} 
\caption{ 
\label{fig:resultsPT-BPIV}\small Phase structures as function of temperature along $H_{\rm S}$ and $N_R$ field directions for BP-IV. Different colours show the evolution of minimum along a particular field direction with temperature. The line with the arrow connects two degenerate phases at $T=T_c$ and the direction of the arrow indicates a possible FOPT.}
\end{figure*}
	
\vspace{0.1cm}
\noindent $\bullet$ \textbf{BP-IV:} The BP-IV in Table\,\ref{tab:BPs} shows type-IIc PT pattern. The numerical estimates of the relevant parameters that govern the PT dynamics for BP-IV are listed in Table\,\ref{tab:PTproperties2}. We find SOFPT along both the $H_S$ and $N_R$ directions. Clearly, this BP is not preferred to address EWBG. In Figure\,\ref{fig:resultsPT-BPIV}, we show the phase structure along $H_S$ and $N_R$ directions for BP-IV as a function of temperature. At temperature above $T_c=184.5$ GeV, $H_S$ takes a non-zero field value which is the typical type-II feature. The black coloured line with arrow in Figure\,\ref{fig:resultsPT-BPIV} connects two degenerate phases at the critical temperature and paves the way for the PTs in the respective singlet field directions. 
	
\vspace{0.1cm}
\noindent $\bullet$ \textbf{BP-V:} This BP is unique in the sense that we obtain FOPT  below the critical temperature along the directions of $SU(2)_L$ fields, $H_S$ and $N_R$ at the same time. This BP falls into the type-III category since at temperature above $T_c$, we find high-T VEV to be non-zero for both $H_{\rm SM}$ and $H_S$ fields. Although this particular BP shows FOPT along $H_{\rm SM}$ direction, the strength is relatively weaker as can be seen from Table\,\ref{tab:PTproperties2}. Therefore, the possibility of EWBG remains unlikely for this BP. Nevertheless, we obtain SFOPT along $H_S$ and $\tilde{N}$ directions in contrast to weaker FOPT in the $H_{\rm SM}$ direction.  
	
\begin{table}[!ht]
\begin{center}
{\footnotesize
\begin{tabular}{| c | c | c | c  | c | c |}
\hline	
&	BP-IV				&	BP-V		&	
BP-VI 	\\
\hline
\hline	
Transition Type		& Type-{\tt IIc}		&		Type-{\tt IIIb}		&	 Type-{\tt IV}			\\
\hline
$v_c/ T_c$	&   	0.0 ({\tt In})~; 0.0 ({\tt Out})			&	0.0 ({\tt I}); 0.0 ({\tt O})	&	\Centerstack{{\tt 1st:} 0.0 ({\tt In})~; 0.0 ({\tt Out}) \\ {\tt 2nd:} 0.54 ({\tt In}); 0.0 ({\tt Out})}		\\
$\Delta \phi_{SU(2)}/T_n$	&   	0			&	0.04		&	{\tt 1st:} 0 ; {\tt 2nd:} 0.57			\\
$\Delta \phi_S/ T_n$  &	 	1.01				&	1.56		&	{\tt 1st:} 0; {\tt 2nd:} 0		\\
$\Delta \phi_{\widetilde{N}}/ T_n$		&  		2.81		&	1.71	&	{\tt 1st:} 0.2; {\tt 2nd:} 0.13	\\
\hline
$T_c$ (GeV)	&	 184.5	 		&	177.9		& 	\Centerstack{{\tt 1st:} 232.8 \\ {\tt 2nd:} 206.3} \\
$T_n$ (GeV)	&		165.8 				& 	144.3		&	\Centerstack{{\tt 1st:} 232.6 \\ {\tt 2nd:} 204.6}		\\
\hline
high-$T_n$ VEVs	&	(0, 0, 529.9, 0)	&	(137.9, 3.5, 1606.9, 0)	&	\Centerstack{{\tt 1st:} (0, 0, 2087.7, $-845.4$) \\ {\tt 2nd:} (0, 0, 2087.9, $-720.9$)}	\\
low-$T_n$ VEVs	&	 		(0, 0, 696.6, $-465.28$)	 		&  		(143.2, 0, 1832.7, 247.2)		& \Centerstack{{\tt 1st:} (0, 0, 2087.7, -807.2) \\ {\tt 2nd:} (117.2, 0, 2088.1, $-747.6$)}	\\
\hline
high-$T_c$ VEVs			&	(0, 0, 459.9, 0)		& (0, 0, 1484.6, 0)		&	\Centerstack{{\tt 1st:} (0, 0, 2087.7, $-846.2$) \\ {\tt 2nd:} (0, 0, 2087.9, $-724.6$)}			\\
low-$T_c$ VEVs		&  		(0, 0, 671.2, $-429.5$)	 		&  		(0, 0, 1827.6, 275.9)		& \Centerstack{{\tt 1st:} (0, 0, 2087.4, $-808.5$) \\ {\tt 2nd:} (112.3, 0, 2088.1, $-749.3$)}
\\ \hline
\end{tabular}}
\end{center}
\caption{The PT properties for the last three BPs as tabulated in Table\,\ref{tab:BPs}.}
\label{tab:PTproperties2}
\end{table}
	
\vspace{0.1cm}
\noindent $\bullet$ \textbf{BP-VI:} So far, for all the BPs we have obtained single-step FOPT. In contrast, BP-VI shows a two-step FOPT. The outputs are tabulated in Table\,\ref{tab:PTproperties2}. In both steps, the high-temperature behaviour of the scalar potential closely follows the Type-IV pattern. On the other hand, in the first step FOPT occurs along the $N_R$ direction only, while in the second step, we find FOPT in both the $N_R$ and $H_{\rm SM}$ directions. Note that, this BP shows a weaker FOPT and hence, is not suitable for the EWBG.

\begin{figure*}[!ht]
\centering
\subfigure{\includegraphics[height=6.1cm,width=7.5cm]{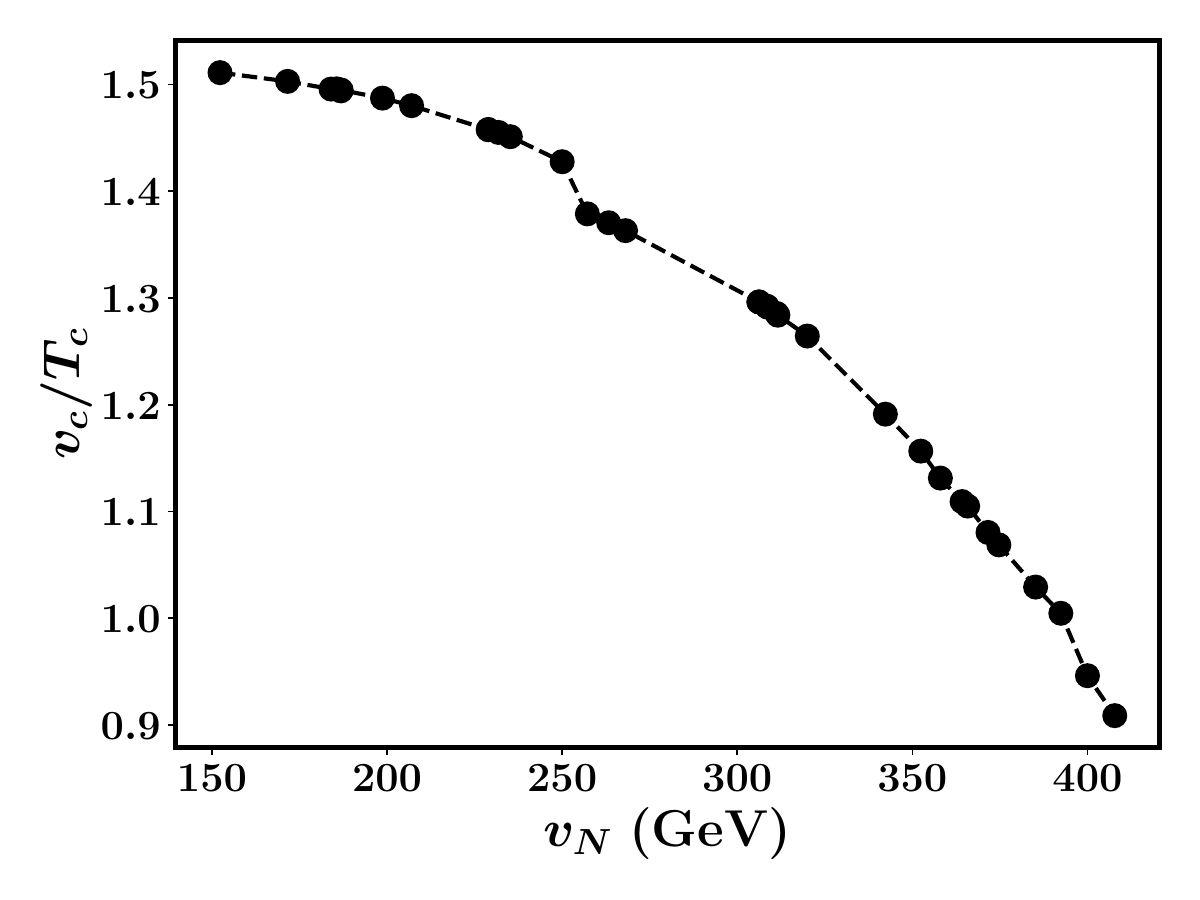}}
\subfigure{\includegraphics[height=6cm,width=7.2cm]{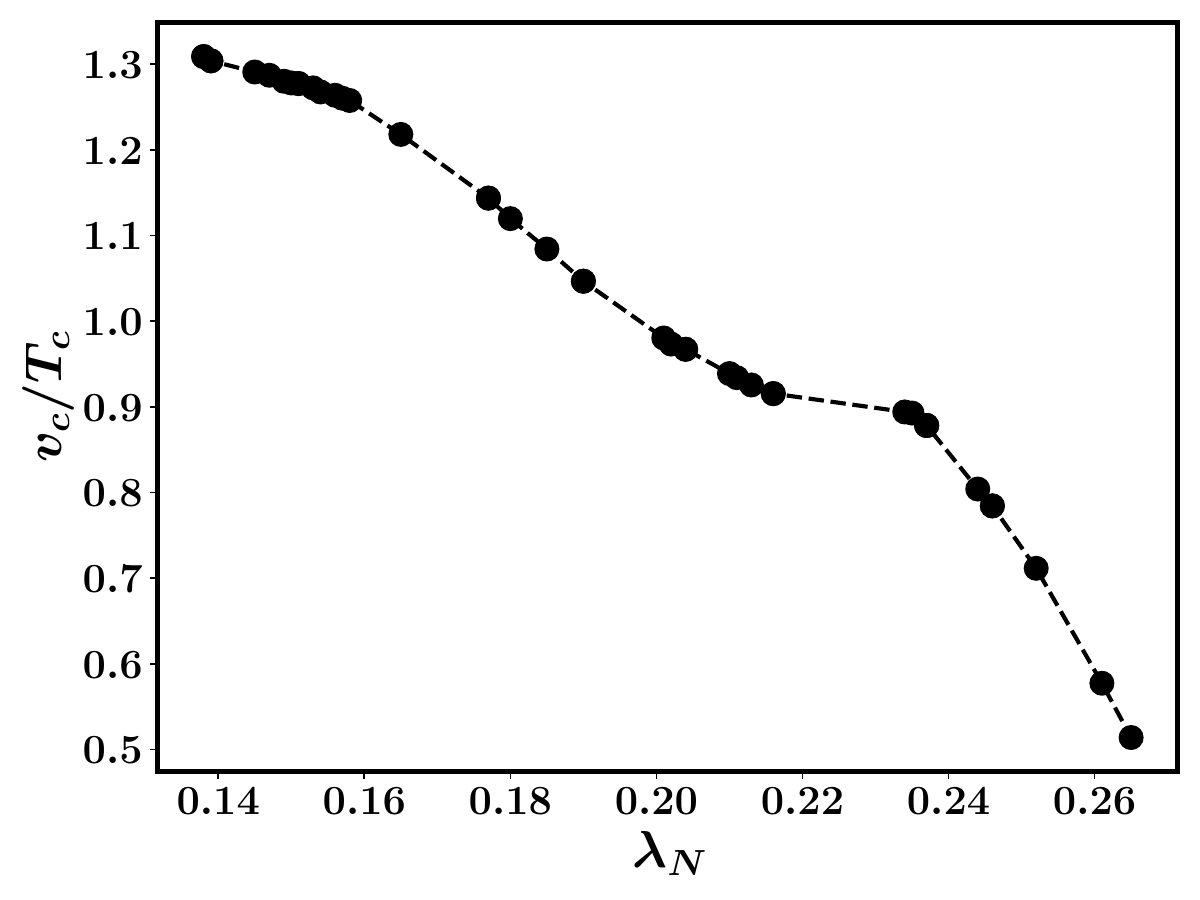}}
\caption{These plots show the dependence of PT strength on $v_{N}$ (left) and $\lambda_N$ (right) in the $T_c$ calculation.  Parameters $Y^i_N,\,v_i$ and $A_N$ have no significant effect in PT dynamics and thus, we keep their values $\sim\mathcal{O}(10^{-7})$, $\sim\mathcal{O}(10^{-4}$ GeV), $\sim\mathcal{O}(1$ TeV), respectively. Other relevant parameters are fixed as in BP-I of Table\,\ref{tab:BPs}, except $v_{N}$ and $\lambda_N$.}
\label{fig:lightsneu1}
\end{figure*} 
	
\begin{figure*}[!ht]
\centering
\subfigure{\includegraphics[height=6.1cm,width=7.5cm]{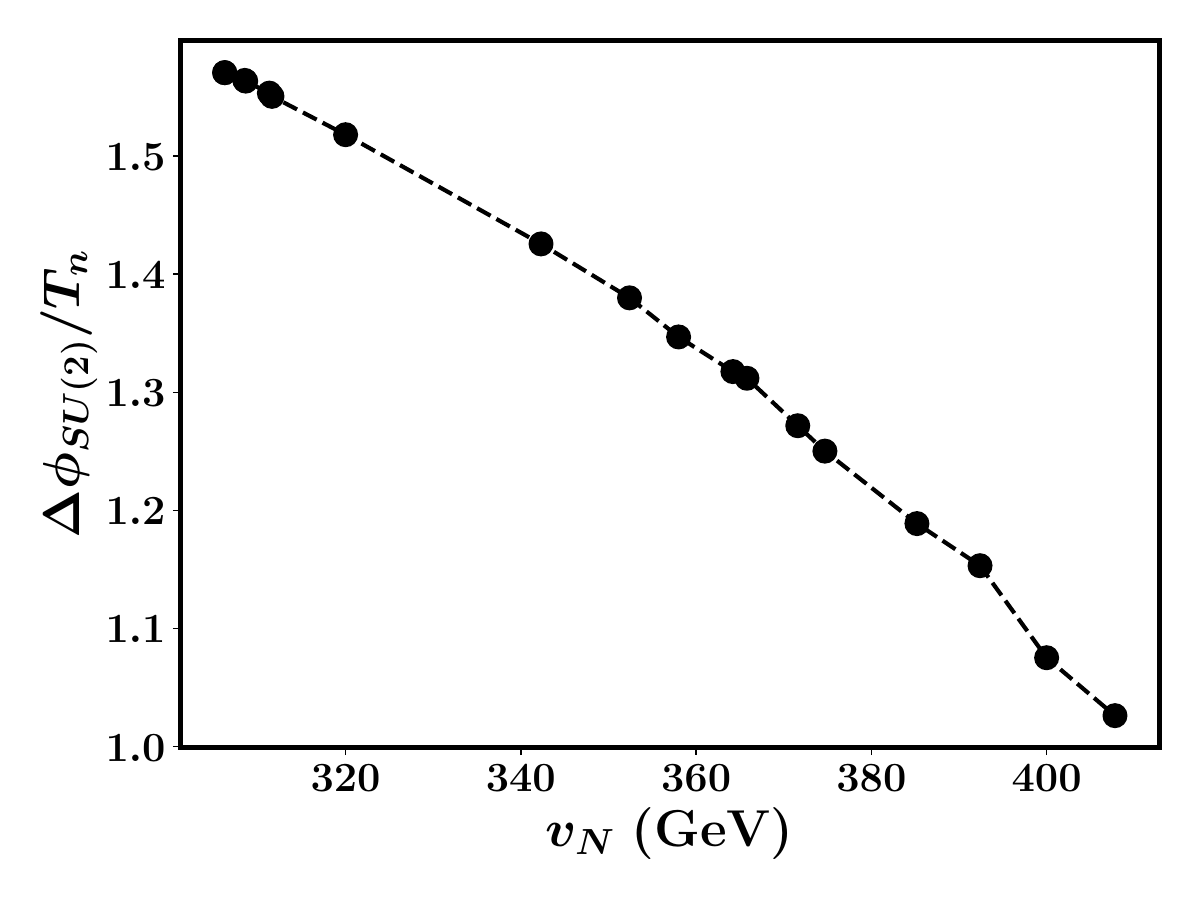}}
\subfigure{\includegraphics[height=6cm,width=7.5cm]{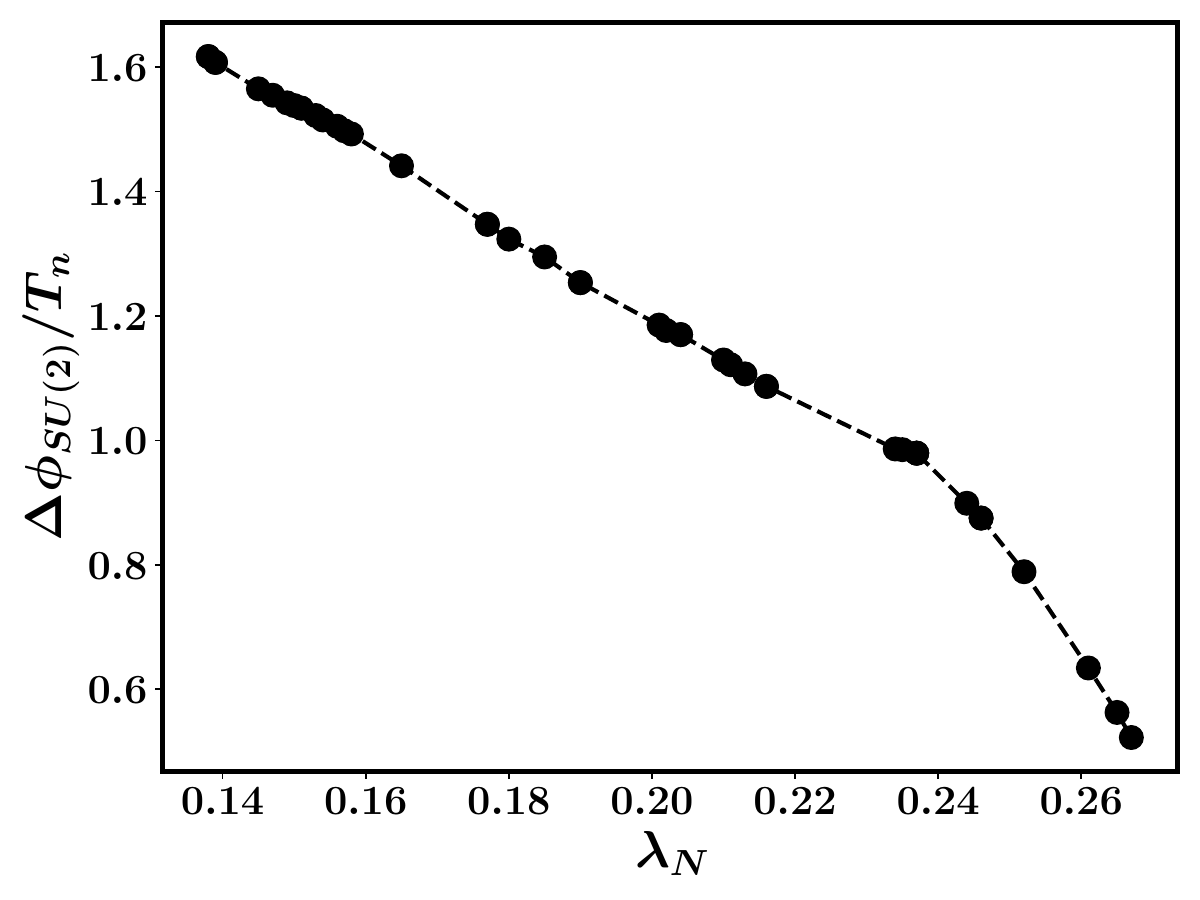}}
\subfigure{\includegraphics[height=6cm,width=7.7cm]{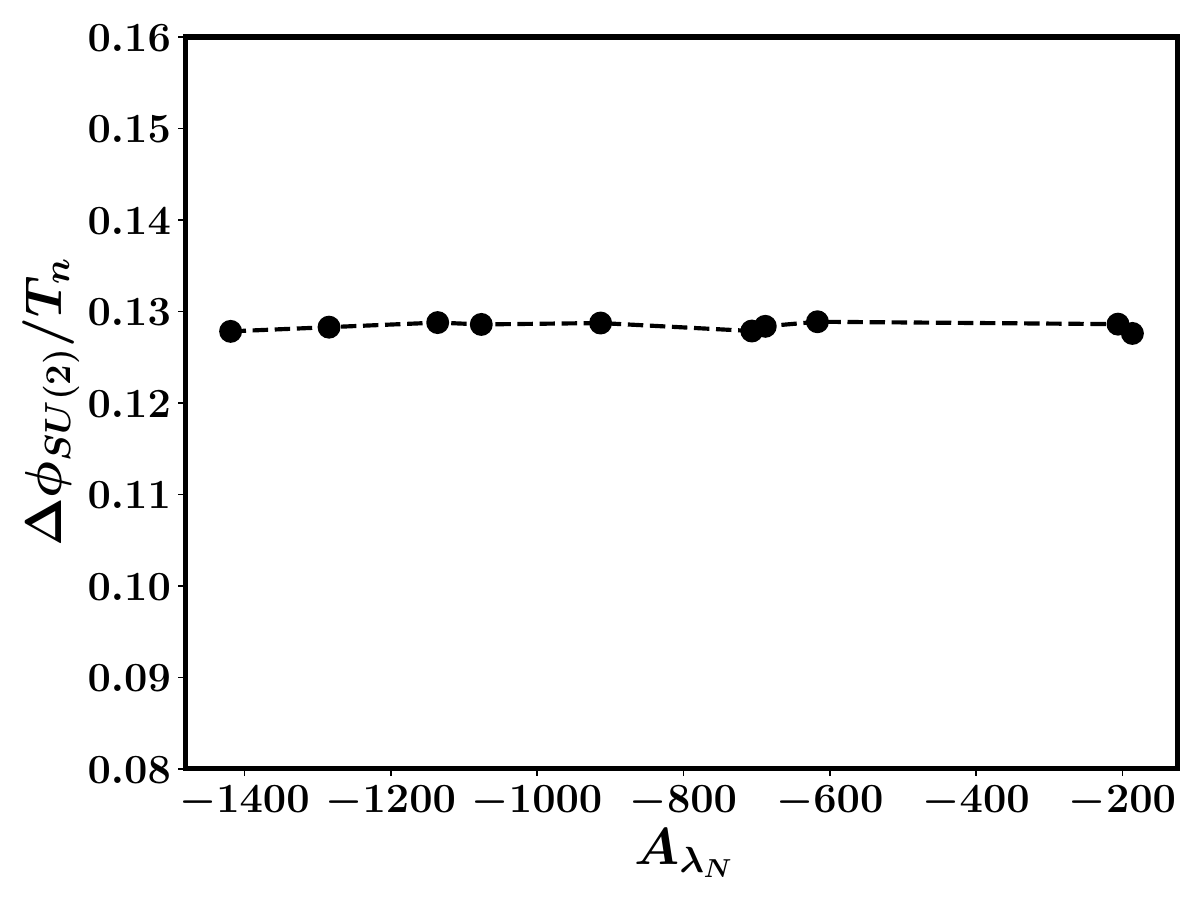}}\caption{These plots show the dependence of PT strength $\Delta \phi_{SU(2)}/T_n$ on $v_{N}$ (top left), $\lambda_N$ (top right) and $A_{\lambda_N}$ (bottom) along the $SU(2)_L$ field direction, in the $T_n$ calculation. Here, orders of parameters $Y^i_N,\, v_i$ and $A_N$ are chosen as in Figure \ref{fig:lightsneu1} and the other relevant parameters are fixed as in BP-I of Table\,\ref{tab:BPs}, except $v_{N}$, $\lambda_N$ and $A_{\lambda_N}$.}
\label{fig:lightsneu2}
\end{figure*} 

Recall from section \ref{sec:param_space} that the new physics parameters, relevant for the study of PT in the current framework are $\{\lambda_N, A_{\lambda_N},v_{N}\}$ compared to the $\mathbb{Z}_3$ symmetric NMSSM. In the subsequent analysis, we like to inquire about the impact of these new parameters on the PT strength along different field directions. Also, note that a FOPT apparently favours a lighter RH-sneutrino-like state below 125 GeV as we observe from the BP-based study of PT and their outcomes. This characteristic is likely to be further confirmed while we vary the new parameters and obtain the sensitivity of PT strength on these parameters. 
	
First, in Figure\,\ref{fig:lightsneu1} we show the impact of $v_{N}$ (left) and $\lambda_N$ (right) on the PT strength $\frac{v_c}{T_c}$. In each of the sub-figures, we have fixed the other relevant parameters as in BP-I of Table \ref{tab:BPs}. We find the PT strength decreases with the rise of both $v_{N}$ and $\lambda_N$. We repeat the analysis for the same BP as shown in the top panel of Figure\,\ref{fig:lightsneu2} considering nucleation temperature calculation. In particular, we estimate the PT strength in the $SU(2)_L$ field directions, i.e., $\Delta \phi_{SU(2)}/T_n$ as function of $v_{N}$ and $\lambda_N$ and notice similar trends as in Figure\,\ref{fig:lightsneu1}. Now a smaller $\lambda_{N}$ or $v_{N}$ implies lighter sneutrino following the CP-even mass matrices mentioned in Appendix  \ref{ap:MassmMatricesEWSB}. Hence, Figures \,\ref{fig:lightsneu1} and \ref{fig:lightsneu2} further reinforce the fact that a comparatively lighter RH-snuetrino is indeed preferred to trigger a possible FOPT along the $SU(2)_L$ doublet field directions in the present framework. Now the remaining new parameter $A_{\lambda_N}$ is expected to show a minor impact on the $\Delta \phi_{SU(2)}/T_n$. This is because it is not directly connected to the relevant terms at the tree-level in the Lagrangian involving the ${SU(2)}$ doublet Higgs fields. Indeed, in our analysis,  we have found that the 
$\Delta \phi_{SU(2)}/T_n$ remains more or less unaltered upon varying $A_{\lambda_N}$ as shown in the bottom panel of Figure \,\ref{fig:lightsneu2}. These important findings, as explained in Figure \ref{fig:lightsneu1} and Figure\,\ref{fig:lightsneu2}, are well supported by our semi-analytical calculation as demonstrated in the subsection\,\ref{sec:semiana}.

\begin{figure*}[!ht]
\centering
\subfigure{\includegraphics[height=6.1cm,width=7.5cm]{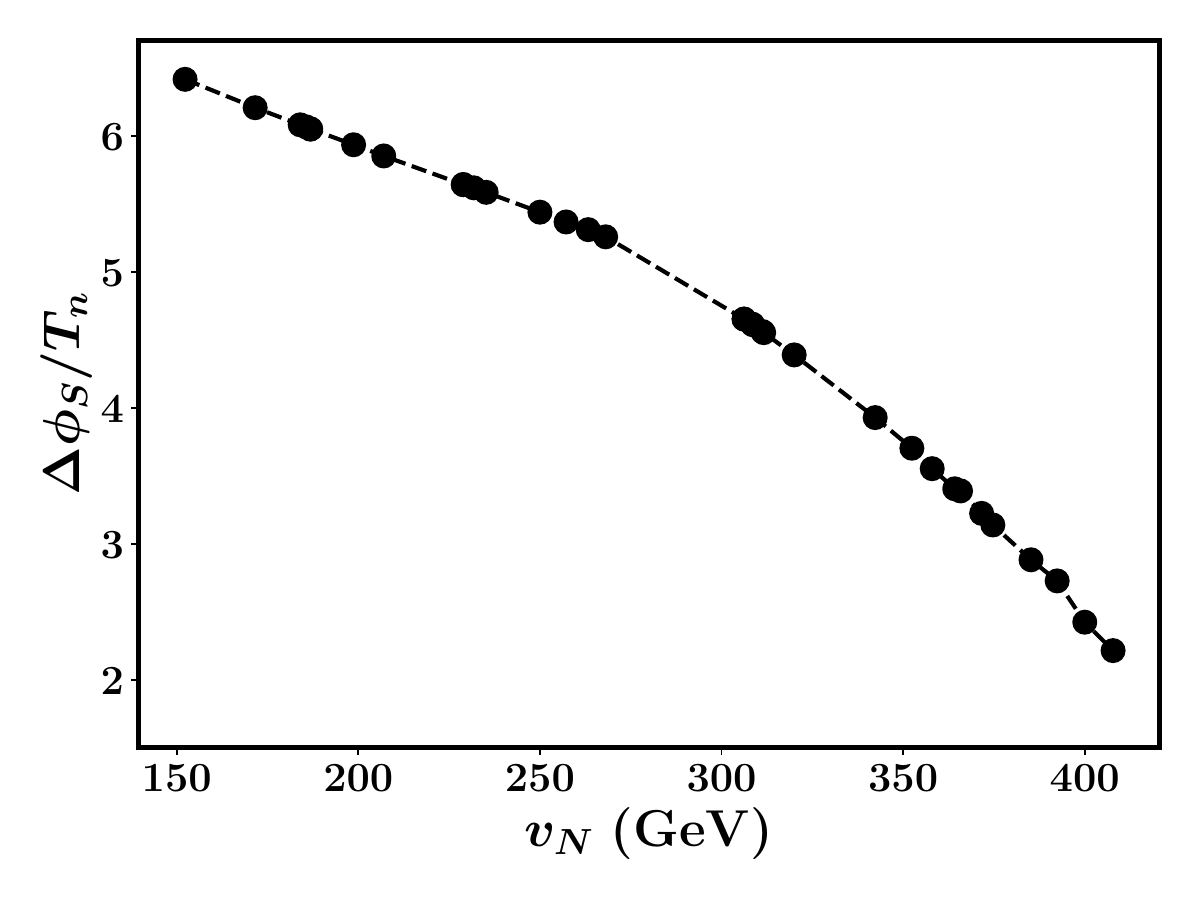}}
\subfigure{\includegraphics[height=6cm,width=7.4cm]{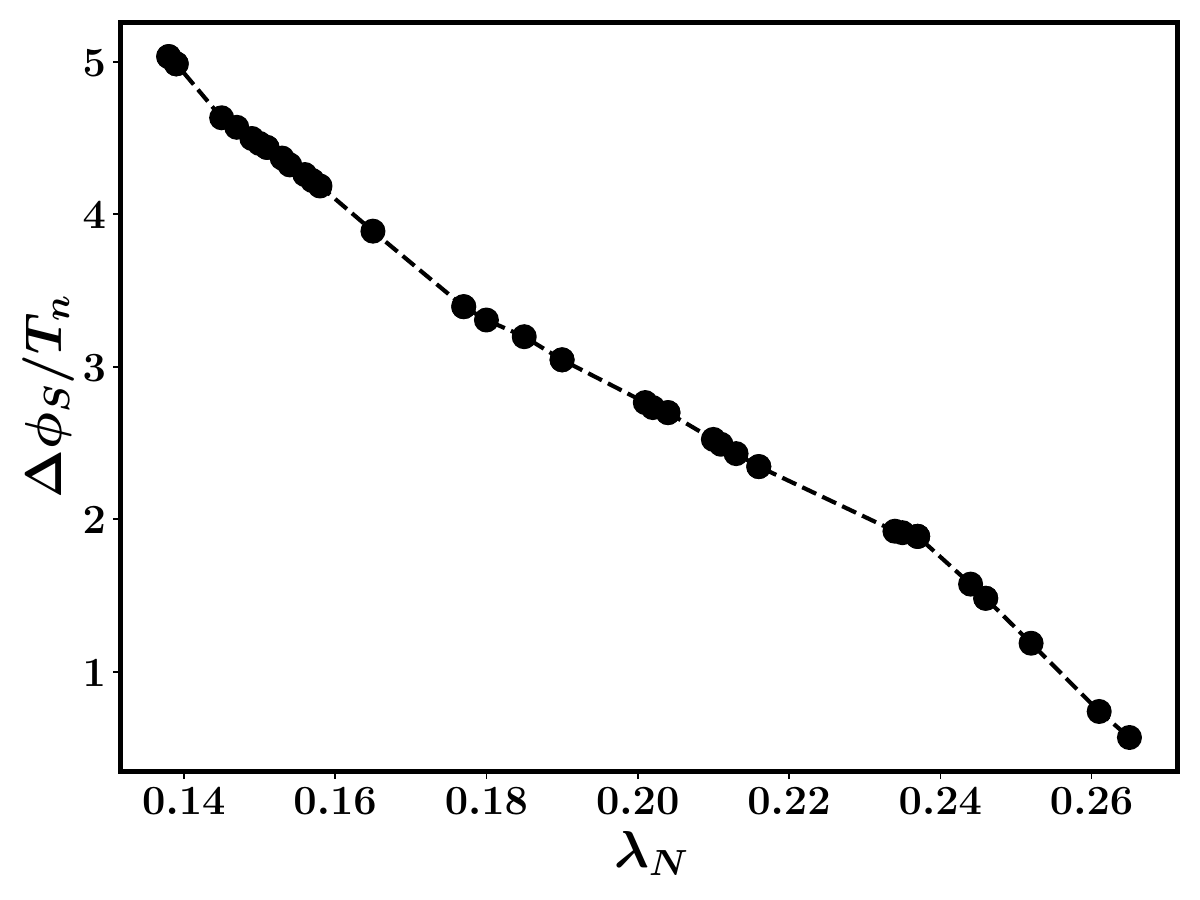}}
\subfigure{\includegraphics[height=6cm,width=7.5cm]{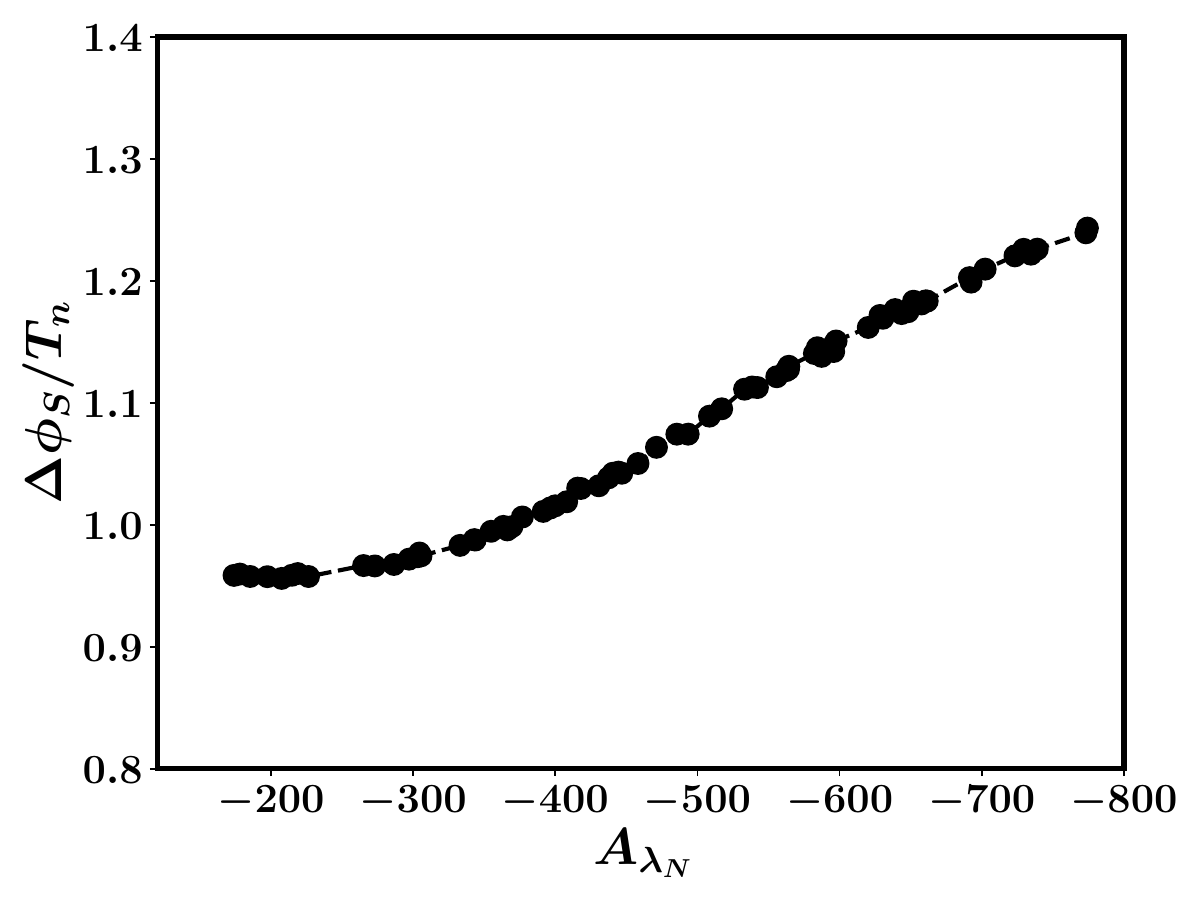}}
\caption{These plots show the dependence of PT strength $\Delta \phi_S/ T_n$ on $v_{N}$ (top left), $\lambda_N$ (top right) and $A_{\lambda_N}$ (bottom) along the $SU(2)_L$-singlet field direction, in the $T_n$ calculation. Here, orders of parameters $Y^i_N,\, v_i$ and $A_N$ are chosen as in Figure \ref{fig:lightsneu1} and the other relevant parameters are fixed as in BP-III of Table\,\ref{tab:BPs}, except $v_N, \, \l_N$ and $A_{\l_N}$.}
\label{fig:VeVlamNScanHs}
\end{figure*} 
	
\begin{figure*}[!ht]
\centering
\subfigure{\includegraphics[height=6.1cm,width=7.5cm]{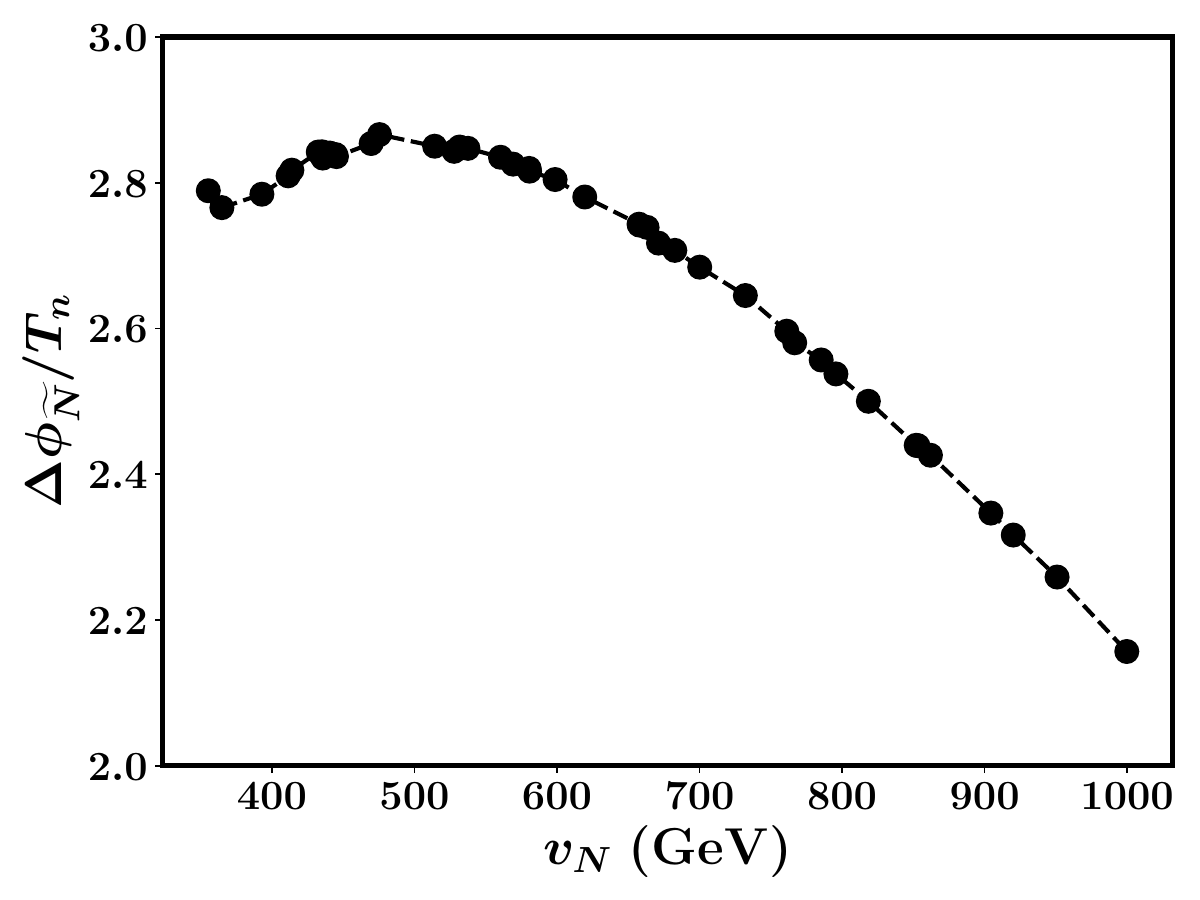}}
\subfigure{\includegraphics[height=6cm,width=7.4cm]{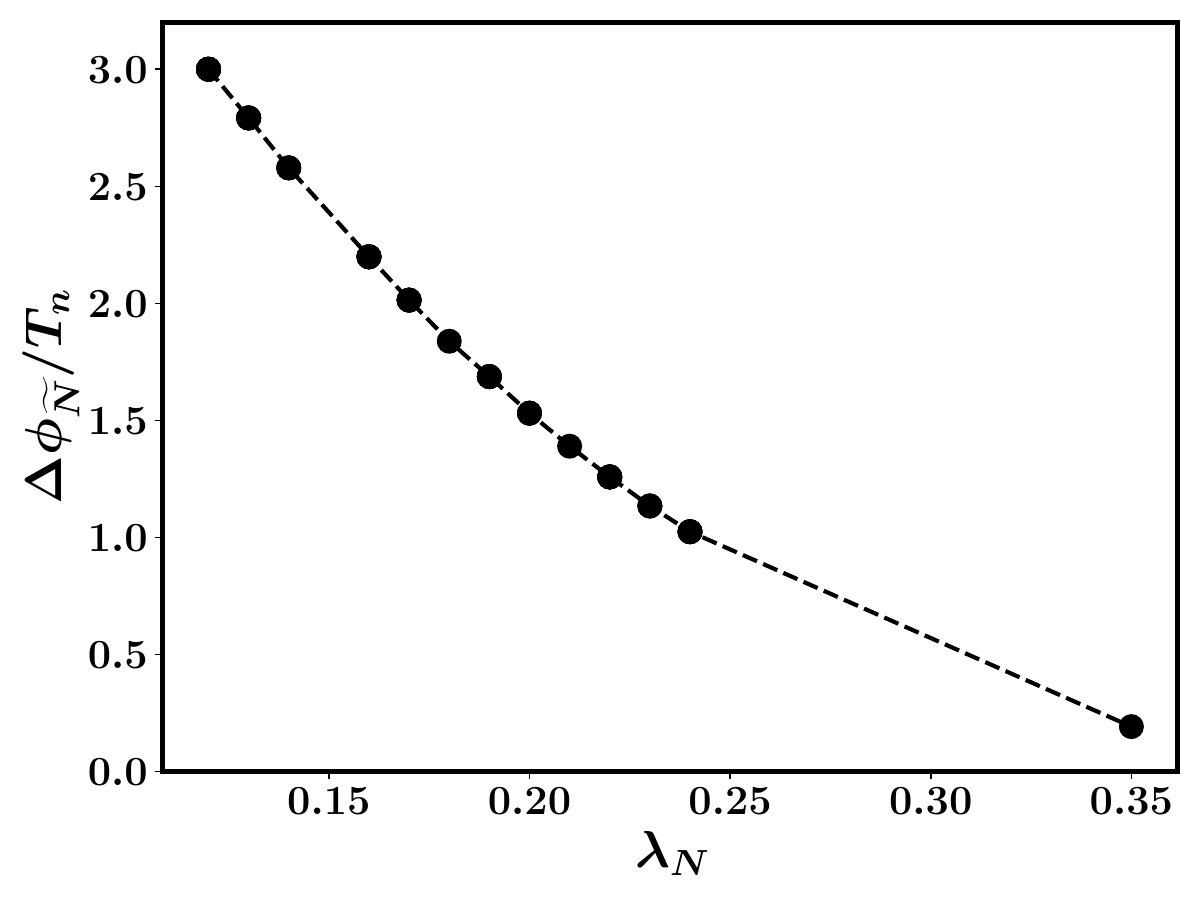}}
\subfigure{\includegraphics[height=6cm,width=7.7cm]{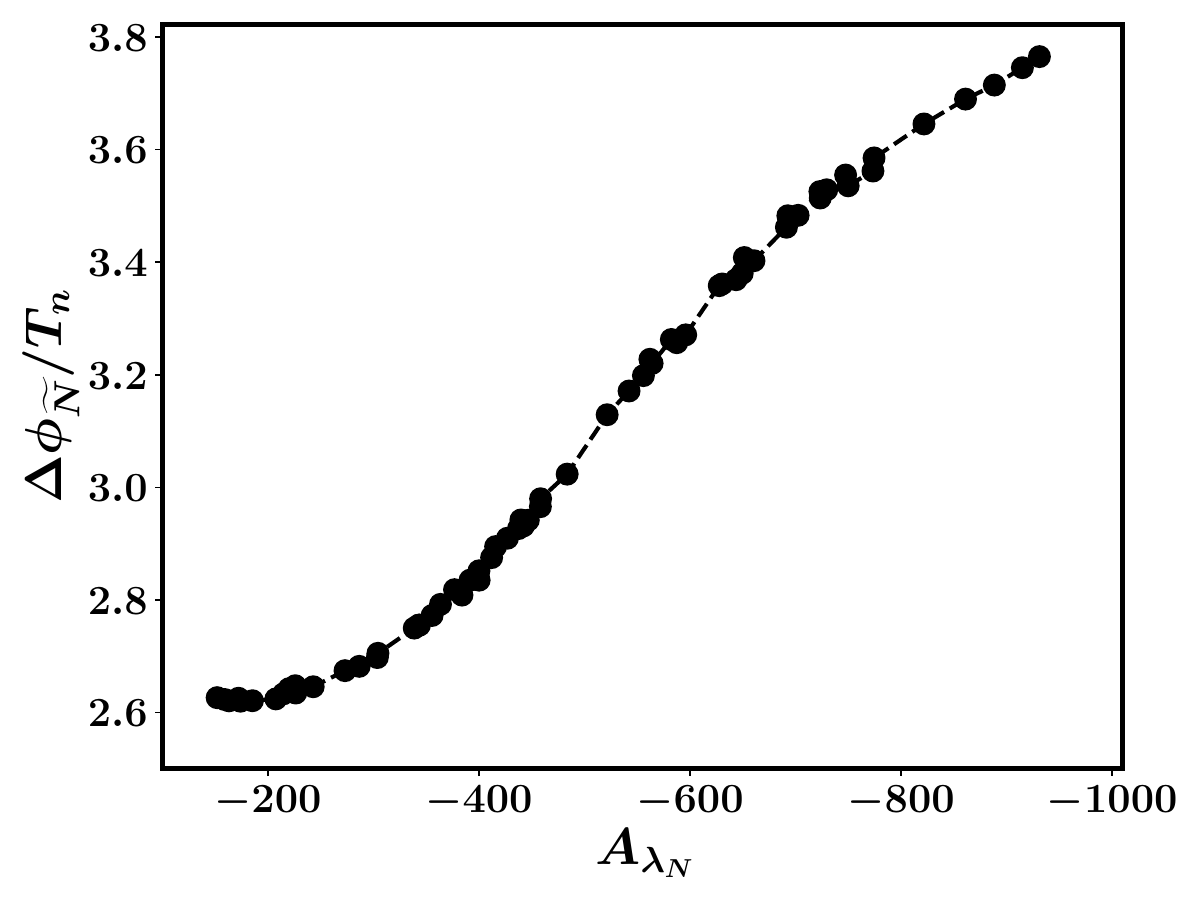}}
\caption{These plots show the dependence of PT strength $\Delta \phi_{\widetilde{N}}/ T_n$ on $v_{N}$ (top left), on $\lambda_N$ (top right) and on $A_{\lambda_N}$ (bottom) along the $N_R$ direction, in the $T_n$ calculation. Here, orders of parameters $Y^i_N,\, v_i$ and $A_N$ are chosen as in Figure \ref{fig:lightsneu1} and the other relevant parameters are fixed as in 
BP-IV of Table\,\ref{tab:BPs}, except $v_N, \, \l_N$ and $A_{\l_N}$.}
\label{fig:VeVlamNScanNs}
\end{figure*} 

Next, we like to examine the impact of the new physics parameters as earlier specified on the PT strength along $SU(2)_L$-singlet field direction $\Delta \phi_S/ T_n$ while the other parameters are set according to BP-III of Table\,\ref{tab:BPs}. In top panel of Figure\,\ref{fig:VeVlamNScanHs} we depict the variation of $\Delta \phi_S/ T_n$ as function of $v_{N}$ (left) and $\lambda_N$ (right). We observe that the quantity $\Delta \phi_S/ T_n$ increases upon lowering $\lambda_N$ when  $v_{N}$ is fixed. In the other case when we fix $\lambda_N$ and vary $v_{N}$, the $\Delta \phi_S/ T_n$ gets enhanced for a smaller $v_{N}$. Once again, these observations further strengthen our earlier finding that a lighter RH-sneutrino below 125 GeV is favoured for the occurrence of an SFOPT in the $SU(2)_L$-singlet, i.e., $H_S$ direction as well. On the other hand, we also notice that the $\Delta \phi_S/ T_n$ increases with the rise of $A_{\lambda_N}$ as shown in the bottom panel of Figure\,\ref{fig:VeVlamNScanHs}. Note that $A_{\lambda_N}$ is appearing as the coefficient of the cubic interaction $S\widetilde{N}\widetilde{N}$ (see Eq. (\ref{eq:softSUSY})). Hence a larger $A_{\lambda_N}$ is expected to increase the barrier height which results in a stronger {$\Delta \phi_S/ T_n$.
	
Previously, we have found that BP-IV provides us with a SOFPT along the $N_R$ field direction $\Delta \phi_{\widetilde{N}}/T_n$. We would like to utilize this particular BP to enquire about the dependence of new parameters on $\Delta \phi_{\widetilde{N}}/ T_n$. In top left of Figure\,\ref{fig:VeVlamNScanNs}, we show the dependence of $\Delta \phi_{\widetilde{N}}/ T_n$ on $v_{N}$. We find that for $v_{N} \lesssim 500$ GeV, the $\Delta \phi_{\widetilde{N}}/ T_n$ remains more or less constant, however, decreases while we increase $v_{N}$ further. Additionally, from top right of Figure\,\ref{fig:VeVlamNScanNs} the $\Delta \phi_{\widetilde{N}}/ T_n$ gets reduced as well upon increasing $\lambda_N$. The reason for this is twofold. As we mentioned earlier, a smaller $\lambda_N$ leads to lighter RH-sneutrino states below 125 GeV which in turn enhances the $\Delta \phi_{\widetilde{N}}/ T_n$. Moreover, a smaller $\lambda_N$ also assists in increasing the barrier height and hence results in enhanced $\Delta \phi_{\widetilde{N}}/ T_n$. In bottom panel of Figure\,\ref{fig:VeVlamNScanNs}, we have shown the $\Delta \phi_{\widetilde{N}}/ T_n$ strength gets enhanced upon increasing $A_{\lambda_N}$. This is once again caused by the enhanced barrier height for a larger $A_{\lambda_N}$ similar to the earlier case.

After examining the individual dependence of new parameters on PT strength, we now give a random scan on new physics parameters highlighting the region allowed by the experimental constraints and favouring an SFOPT along $SU(2)_L$ field directions. We vary $(\lambda_N,\,v_{N})$ and fix the other relevant parameters in Eq. (\ref{eq:IndeParameters2}) following BP-I. However, orders of parameters $Y^i_N,\, v_i$ and $A_N$ are chosen 
as $\sim\mathcal{O}(10^{-7})$, $\sim\mathcal{O}(10^{-4}$ GeV), $\sim\mathcal{O}(1$ TeV), respectively, as they hardly affect the PT dynamics. We have randomly generated pairs of $(\lambda_N,v_{N})$ and pass through all the experimental bounds mentioned in subsection \ref{susec:expconstraint}. We first sort out the points that pass all the experimental constraints as shown in green colour in Figure \ref{fig:resultsPT-scan2}. Next, we apply the condition of SFOPT along the $SU(2)_L$ field direction and pin down the points that favour SFOPT only and SFOPT with possible EWBG having minimal wash-out effects. We have marked them in Figure \ref{fig:resultsPT-scan2} by coloured `$\blacktriangle$' and `$\blacksquare$', respectively. These points depict the variation of $\Delta \phi_{SU(2)}/T_n$ in the $v_N$ - $\l_N$ plane.

\begin{figure*}[!ht]
\centering
\includegraphics[height=9cm,width=13cm]{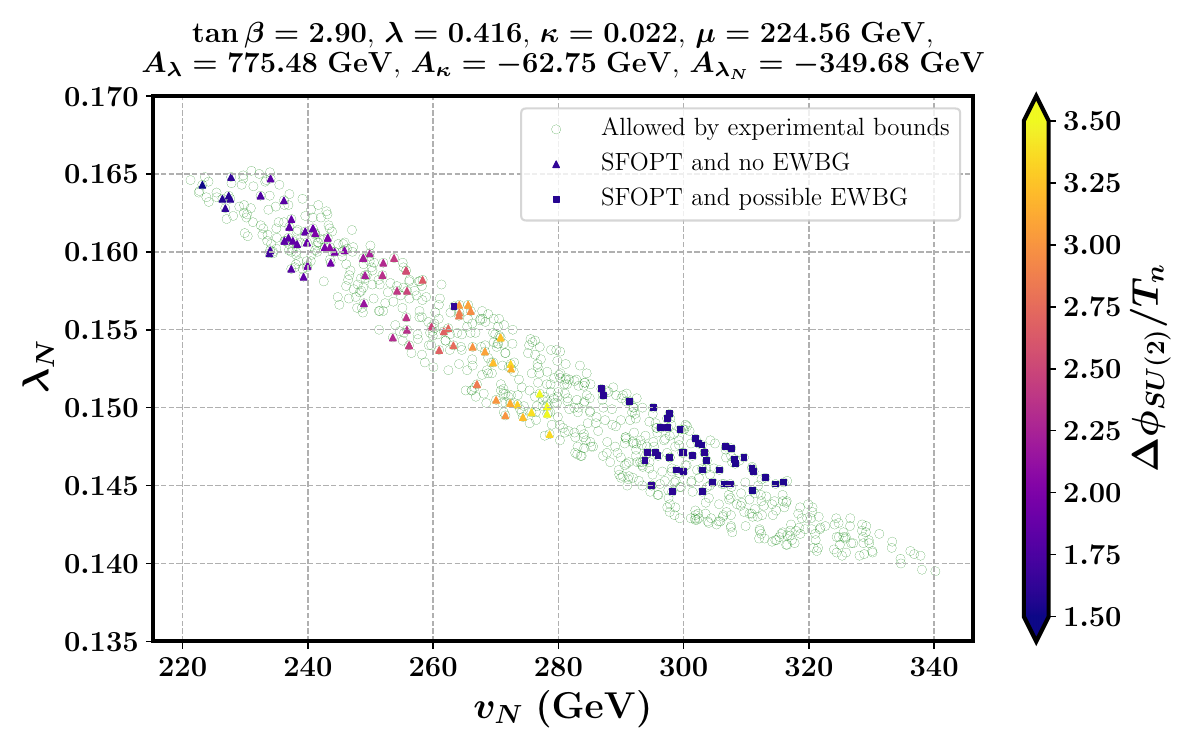}
\caption{This figure shows variations of $\Delta \phi_{SU(2)}/T_n$ in the $v_N$ - $\l_N$ plane. The green-coloured points pass all the experimental constraints as discussed in subsection \ref{susec:expconstraint}.  The points favoured for SFOPT along the $SU(2)_L$ field direction without and with EWBG are marked by coloured `$\blacktriangle$' and `$\blacksquare$' symbols, respectively. Orders of parameters $Y^i_N,\, v_i$ and $A_N$ are chosen as in Figure \ref{fig:lightsneu1} and the other relevant parameters are fixed as in BP-I of Table\,\ref{tab:BPs}.}
\label{fig:resultsPT-scan2}
\end{figure*} 
	
\begin{figure*}[!ht]
\centering
\subfigure{\includegraphics[height=9cm,width=13cm]{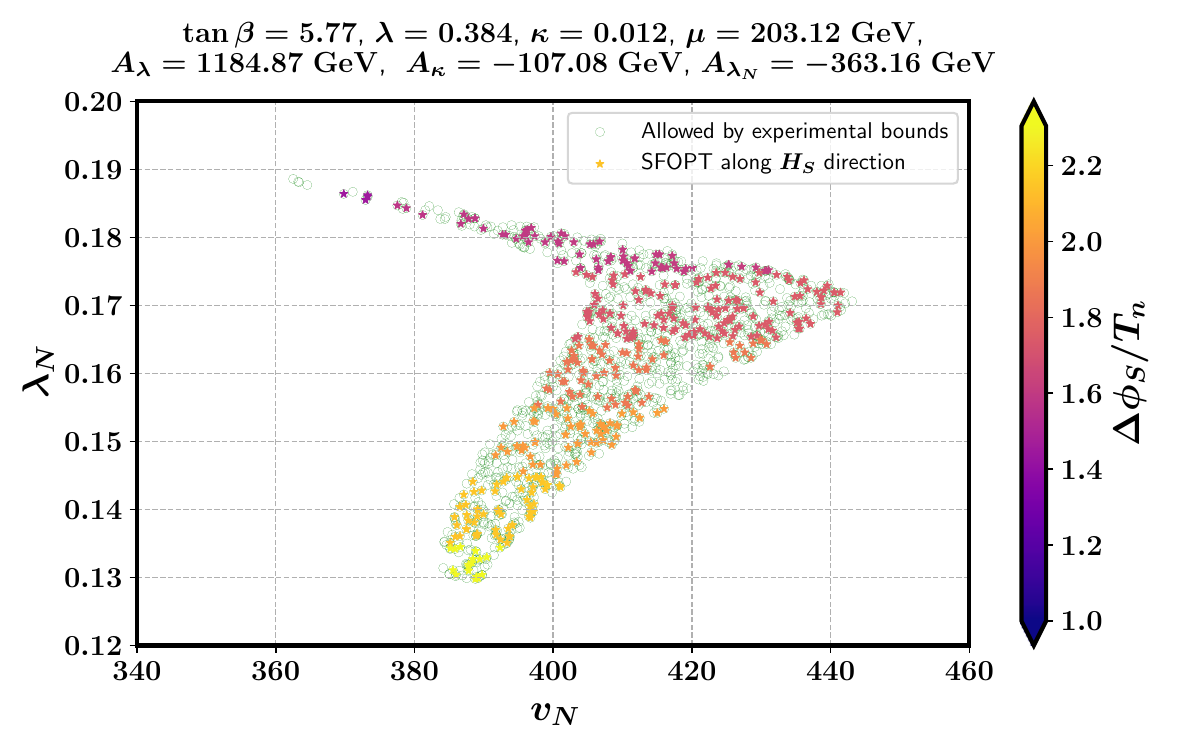}}
\caption{This figure shows variations of $\Delta \phi_S/T_n$ in the $v_N$ - $\l_N$ plane. The green-coloured points pass all the experimental constraints as discussed in subsection \ref{susec:expconstraint}. The points favoured for SFOPT along the $H_S$ field direction are marked by coloured `$\bigstar$'. Orders of parameters $Y^i_N,\, v_i$ and $A_N$ are chosen as in Figure \ref{fig:lightsneu1} and the other relevant parameters are fixed following BP-IV of Table\,\ref{tab:BPs}. }
\label{fig:resultsPT-scan4}
\end{figure*} 

\begin{figure*}[!ht]
\centering
\includegraphics[height=9cm,width=13cm]{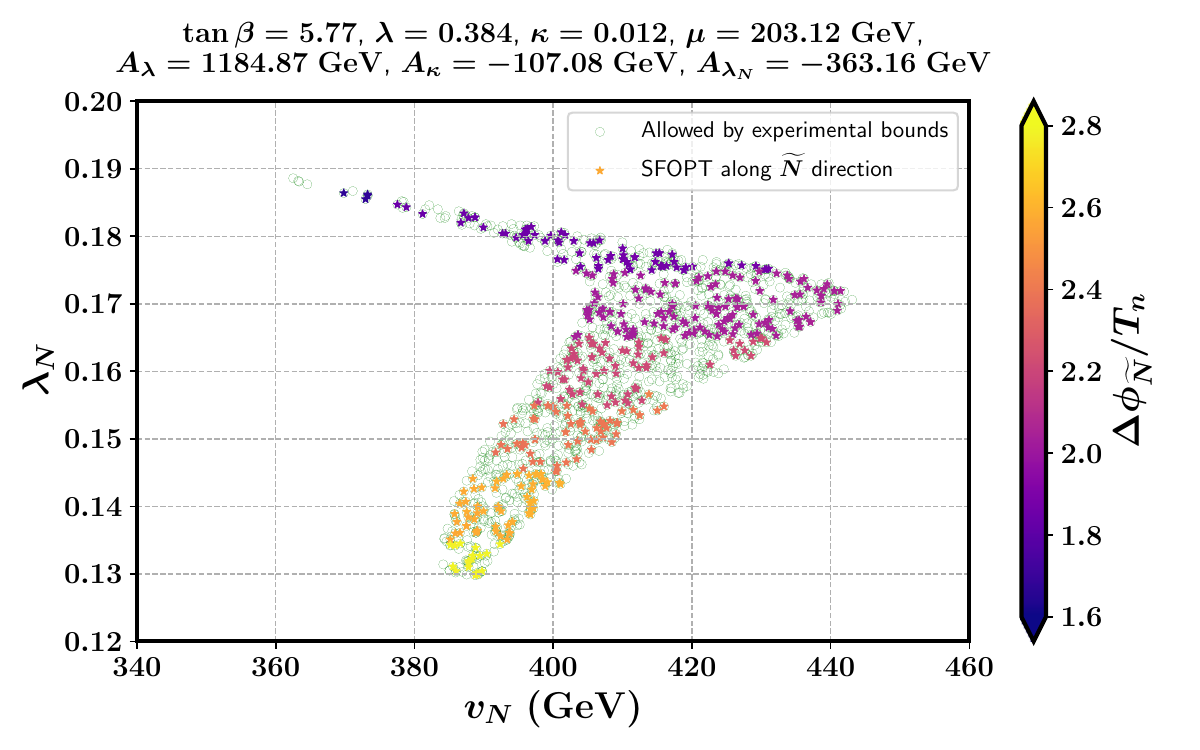}
\caption{This figure shows variations of $\Delta \phi_{\widetilde{N}}/T_n$ in the $v_N$ - $\l_N$ plane. The green-coloured points pass all the experimental constraints as discussed in subsection \ref{susec:expconstraint}.  The points favoured for SFOPT along the $N_R$ field direction are marked by coloured `$\bigstar$'. Orders of parameters $Y^i_N,\, v_i$ and $A_N$ are chosen as in Figure \ref{fig:lightsneu1} and the other relevant parameters are fixed following BP-IV of Table\,\ref{tab:BPs}.}
\label{fig:resultsPT-scan3}
\end{figure*} 
	
Next in Figure \ref{fig:resultsPT-scan4}, we made a scenario similar to that of Figure \ref{fig:resultsPT-scan2}, however, in the $H_S$ field direction in the context of BP-IV,
as shown in Table\,\ref{tab:PTproperties2}. Here, points which undergo SFOPT are marked by `$\bigstar$'. We also compute the $\Delta \phi_S/ T_n$ strength and find that the $\Delta \phi_S/ T_n$ strength is maximum when both $\lambda_N$ and $v_{N}$ are small, which is in agreement with our earlier observations. 
 
Finally, in Figure\,\ref{fig:resultsPT-scan3} we perform an analogous exercise to show the variation of $\Delta \phi_{\widetilde{N}}/ T_n$ in the $v_N$ - $\l_N$ plane. In this case, we have utilized the BP-IV of  Table\,\ref{tab:PTproperties2} once again to fix the other relevant parameters, except $v_N$ and $\l_N$. The green-coloured points are allowed by the various experimental constraints as stated in subsection \ref{susec:expconstraint}. We mark the points that favour SFOPT in the $N_R$ direction by coloured `$\bigstar$'. Once again, we find that the $\Delta \phi_{\widetilde{N}}/ T_n$ is maximum for simultaneous lower values of $v_{N}$ and $\lambda_N$, consistent with our earlier findings.

\subsection{A brief note on gauge dependency of the effective potential}\label{sec:gaugeD}
It has been discussed earlier that gauge dependence in $V_T$ (see Eq. (\ref{eq:totPotR})) may arise through the one-loop induced corrections and thermal corrections to the masses of the relevant particles. Therefore it seems that the results for PT analysis may change upon switching from one gauge choice to another. Note that, gauge independent treatments of the effective potential, relevant for the PT are already proposed in the literature (see, for e.g., some recent works \cite{Chiang:2017nmu,Kozaczuk:2019pet,Croon:2020cgk,Lofgren:2021ogg,Arunasalam:2021zrs,Schicho:2022wty} and references therein). We follow the approach of Ref. \cite{Katz:2015uja} to remove gauge dependency from $V_T$, in an attempt to cross-check our previously obtained results (with the original $V_T$ as shown in Eq. (\ref{eq:totPotR})). Note that the gauge dependency first appears at $\mathcal{O}(g^3)$ in the high-temperature expansion of $V_T$, where $g$ denotes the generic gauge coupling. Hence it is possible to obtain a gauge invariant potential by retaining the high-temperature expansion up to $\mathcal{O}(g^2)$. This approach of eliminating gauge dependency is pertinent where the gauge degrees of freedom play a sub-dominant role in the generation of the potential barrier between the symmetric and broken vacua as noted in Ref. \cite{Katz:2015uja}. This is exactly the case in our framework where a potential barrier is formed even at $T=0$, as already mentioned in section \ref{sec:Intro} and subsection \ref{susec:finiteT}. Hence, to evaluate $v_c$ and $T_c$ in a gauge invariant manner we truncate the one-loop effective potential at $\mathcal{O}(g^2)$ and repeat the numerical analysis.  

Our investigation reveals that the previously obtained results do not change drastically even after considering the gauge-invariant approach as earlier described. We have checked the estimate of $v_c/T_c$ and $T_c$ in the gauge invariant approach and found that the estimation of $T_c$ gets reduced by $\sim 3\%$ at most whereas $v_c/T_c$ shows an enhancement of $\lesssim 0.1\%$ only from earlier results, as obtained using Landau gauge. To demonstrate the impact of gauge dependence further, in the top panel of Figure \,\ref{fig:histogram} we provide a histogram that explicitly shows the comparison in the estimate of $v_c/T_c$, obtained by gauge dependent and independent approaches separately. For this purpose, we have utilised the points that satisfy the criterion of SFOEWPT in Figure\,\ref{fig:resultsPT-scan2}. The blue histogram in the top panel of Figure\,\ref{fig:histogram} corresponds to distributions that were derived by taking into account the gauge-independent effective potential, whereas the red histograms in the figure reflect distributions obtained in Landau gauge. We observe that the blue histograms are slightly shifted toward the higher $v_c/T_c$ directions, indicating a small increment in the PT strength. The bottom left and right panels of Figure \,\ref{fig:histogram}, portray a comparison of results obtained by using the gauge dependent and independent approaches in the $\frac{v_c}{T_c}-T_c$ and $\frac{\Delta\phi_{SU(2)}}{T_n}-T_n$ plane, respectively, where one can notice minor differences in the $T_c$ and $T_n$ calculation whereas $\frac{v_c}{T_c}$ and $\Delta\phi_{SU(2)}/T_n$ remain more or less the same. Hence, we draw the conclusion that if an SFOEWPT occurs for a sample point considering Landau gauge, then it also does in the gauge-independent potential in our framework.

\begin{center}
\begin{figure*}[h!]
\centering
\subfigure{\includegraphics[height=6cm,width=7.4cm]{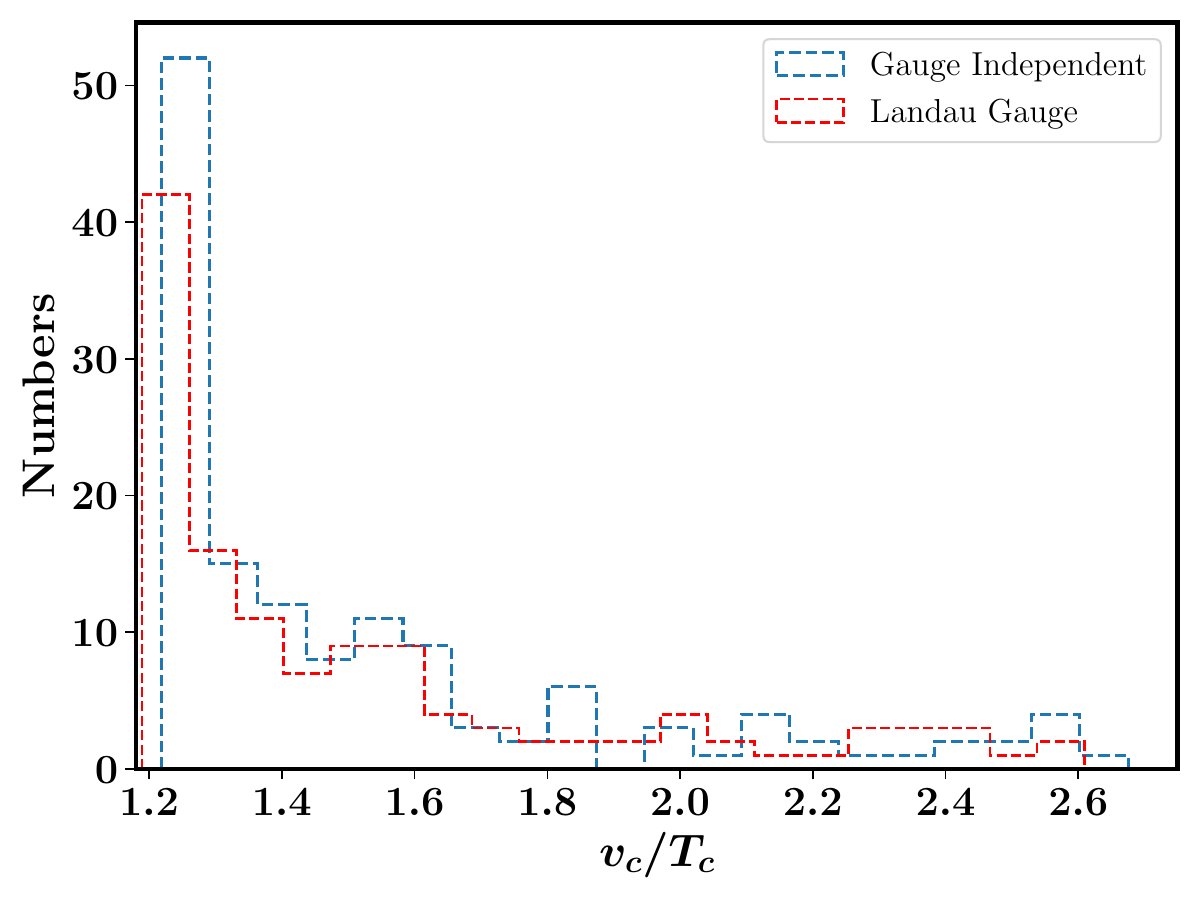}}\\
\subfigure{\includegraphics[height=6cm,width=7.4cm]{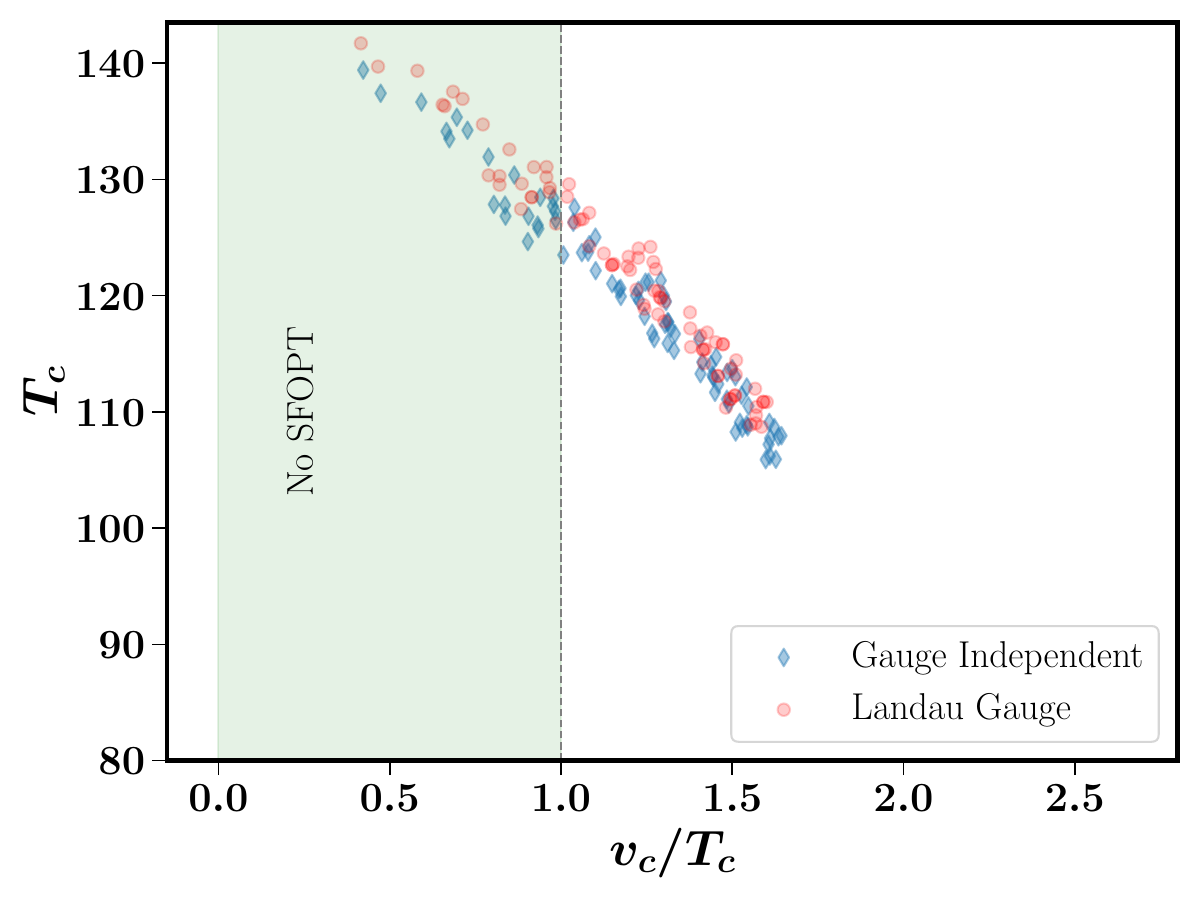}}
\subfigure{\includegraphics[height=6cm,width=7.3cm]{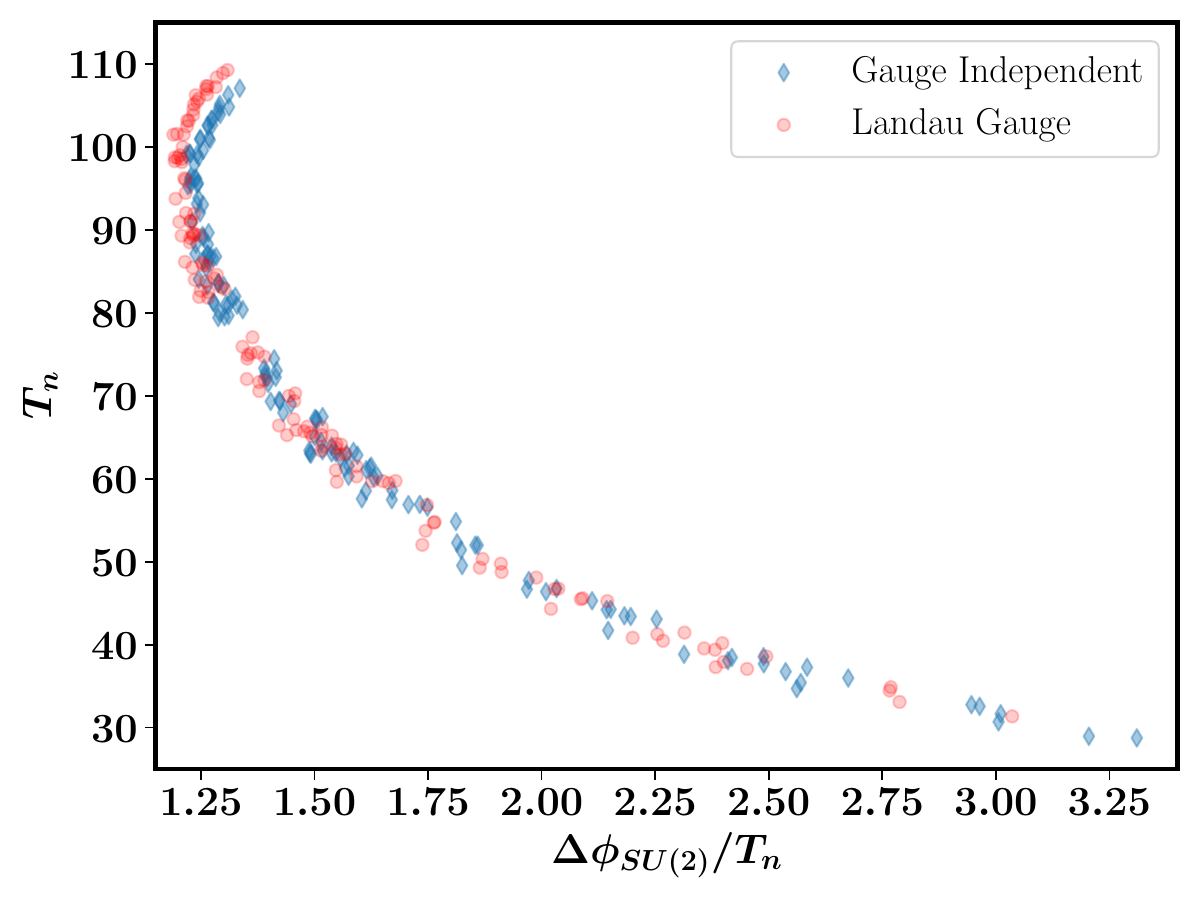}}
\caption{The comparison of results obtained using the gauge dependent and the gauge invariant approaches is shown by utilising the sample points that exhibit the SFOEWPT in Figure \ref{fig:resultsPT-scan2}. See subsection \ref{sec:gaugeD} for details.}
\label{fig:histogram}
\end{figure*}
\end{center}

\subsection{GW spectrum from SFOPT in the NMSSM + one RHN model}\label{sec:GW}
A cosmological FOPT can produce GWs in the early Universe that contains information about the strength of different model parameters. In the preceding section, we have discussed different PT characteristics in the proposed framework and computed the relevant quantities that determine the strength of a PT. In the current section, we will be talking about the production of GW and its detection prospects within our model setup.
	
As we have mentioned earlier, a FOPT is characterized by critical temperature $T_c$, and nucleation temperature $T_n$. The critical temperature indicates the moment when the location of the global minimum changes from one vacuum phase to another. However, the critical temperature analysis does not assure that the associated PT is indeed taking place. On the other hand, FOPT proceeds via bubble nucleation, and hence calculation of nucleation temperature is very crucial in order to obtain the phenomenological parameters that are important from the standpoint of estimating GW spectra. When the nucleation happens, at a temperature below $T_c$, the probability of tunneling $\Gamma (T)$ from the false vacuum to the true one is given by \cite{Grojean:2006bp},
\beq\label{eq:tunnelling-prob}
\Gamma (T) \approx T^4 \left( \frac{S_E}{2 \pi T} \right)^{3/2}\; e^{-\frac{S_E}{T}},
\eeq
where $S_E$ is the bounce action corresponding to the critical bubble and can be written as \cite{Linde:1981zj},
\beq\label{eq:bounce-action}
S_E = \int_{0}^{\infty} 4\pi r^2 dr \left( V_{T} (\phi, T) + \frac{1}{2}\;\left( \frac{d\phi(r)}{dr} \right)^2  \right),
\eeq
with $r$ being the radial coordinate and $\phi$ corresponding to the scalar dynamical fields present in a model framework. The scalar field solution $\phi$ can be derived by solving the classical field equation \cite{Affleck:1980ac,Linde:1977mm,Linde:1981zj}
\beq\label{eq:phivariationr}
\frac{d^2 \phi}{dr^2} + \frac{2}{r} \frac{d\phi}{dr} = \frac{dV_{T} (\phi, T)}{dr},
\eeq
and subsequently applying proper boundary conditions: $\frac{d\phi}{dr} = 0$ when $r \rightarrow 0$ and $\phi(r) \rightarrow \phi_{\rm false}$ when $r \rightarrow \infty$, where $\phi_{\rm false}$ represents the four-dimensional field values at the false vacua. We reiterate here that in order to solve the differential equation and the bounce action numerically, we have implemented our model in the {\tt cosmoTransitions} \cite{Wainwright:2011kj} package.
		
The essential parameters that are required for the estimation of GW spectra from FOPT are relative change in energy density during the PT ($\alpha$), and the inverse of the duration of the PT ($\beta$). Both the parameters, $\alpha$, and $\beta$, are defined at the nucleation temperature $T_n$. The first parameter, $\alpha$, is computed from \cite{Kamionkowski:1993fg},
\beq\label{eq:alphaP}
\alpha = \frac{\Delta \rho}{\rho_{\rm rad}},
\eeq
where $\Delta \rho$ is the released latent heat and it is expressed as \cite{Kehayias:2009tn},
\beq\label{eq:delrho}
{\Delta \rho = \left[ V_T(\phi_0, T)-T\frac{dV_T(\phi_0, T)}{dT}  \right]_{T = T_n} - \left[ V_T(\phi_n, T)-T\frac{dV_T(\phi_n, T)}{dT}  \right]_{T = T_n},}
\eeq
with $\phi_0$ and $\phi_n$ represent, in our case, the four-dimensional field values at the false and true vacua, respectively, and $V_T(\phi, T)$ is the finite-temperature effective potential as mentioned in Eq. (\ref{eq:totPotR}). We should note here that the quantity $\Delta \rho$ measures the strength of a PT, the larger value of the same corresponds to a stronger FOPT. In Eq.\,(\ref{eq:alphaP}), $\rho_{\rm rad}$ corresponds to the radiation energy in the plasma and it is expressed as, $\rho_{\rm rad} = \frac{\pi^2g_{*} }{30} T^4_n$, with  $g_{*}$ being a temperature-dependent quantity that counts the total number of relativistic energy degrees of freedom.
	
The parameter $\beta$ is defined as \cite{Nicolis:2003tg},
\beq\label{eq:betaP}
\frac{\beta}{H_{\ast}} = T \frac{d}{dT} \left( \frac{S_E}{T} \right) \Bigg|_{T = T_{\ast}} \equiv \, T \frac{d}{dT} \left( \frac{S_E}{T} \right) \Bigg|_{T = T_n},
\eeq
where $H_{*}$ is the expansion rate of the Universe during the PT and $T_*$ stands for the PT temperature. We have considered $T_{\ast} \simeq T_n$ in the present work. We have tabulated the obtained values of $\alpha$ and $\beta$ in Table \ref{tab:GW_obs} for different BPs shown in Table \ref{tab:BPs}. As stated earlier, the quantity $\alpha$ is proportional to the energy released during the PT and hence a larger PT strength should lead to a larger $\alpha$ value. In fact, this is exactly the case where we find the largest $\alpha$ for the BP-III (see Table\,\ref{tab:GW_obs}) having $\Delta \phi_S/ T_n = 7.61$ (see Table \ref{tab:PTproperties1}.) We obtain the lowest $\alpha$ for the first-step PT of BP-VI since the corresponding $\Delta \phi_{\widetilde{N}}/ T_n$ is weakest among all as can be seen from Tables\,\ref{tab:PTproperties1} and \ref{tab:PTproperties2}. 
	
\begin{table}[h!]
\centering
\begin{tabular}{| c | c | c |}
\hline 
~BPs~ & ~$\alpha$~ & ~$\beta/H_{*}$~  \\
\hline
\hline
~BP-I~ & ~0.0456~ & ~37535.2~ \\
\hline
~BP-II~ & ~0.0121~& ~143931.0~ \\\hline
~BP-III~ & ~0.0870~& ~11729.8~  \\ \hline
~BP-IV~ & ~0.0101~ & ~7596.0~  \\ \hline
~BP-V~ & ~ 0.0027 ~ & ~ 4611.3 ~  \\ \hline
~BP-VI-I~ & ~0.0002 ~& ~516911.0~\\ \hline
~BP-VI-II & ~ 0.0017 ~&   ~63837.8 ~ \\
\hline
\end{tabular}
\caption{Estimates of the parameters $\alpha$ and $\beta$ as defined in Eq. (\ref{eq:alphaP}) and Eq. (\ref{eq:betaP}), respectively for the six BPs listed in Table\,\ref{tab:BPs}. Note that the BP-VI-I shows two-step PT patterns and we have made the estimates of $\alpha$ and $\beta$ in both steps.}
\label{tab:GW_obs}
\end{table}
There are mainly three different processes that trigger the emission of GWs in a FOPT: (i) bubble wall collisions, (ii) sound waves, and (iii) magneto-hydrodynamic (MHD) turbulence in the plasma. Therefore, the total energy spectrum of the emitted GW can approximately be given as a sum of these three contributions \cite{Caprini:2015zlo,Ellis:2020awk},
\beq\label{eq:GWTotal}
\Omega_{\rm GW}h^2 \approx \Omega_{\rm col} h^2 + \Omega_{\rm sw} h^2 + \Omega_{\rm tur} h^2, \,\, {\rm respectively,}
\eeq
where, $h = H_0/(100 \text{\,km}\cdot \text{sec}^{-1} \cdot \text{Mpc}^{-1})$ \cite{DES:2017txv} with $H_0$ corresponding to Hubble's constant at the present epoch. The contribution to the total GW energy density from the bubble wall collision can be computed using the envelope approximation and it can be estimated as a function of frequency ``$f$" as \cite{Jinno:2016vai},
\beq\label{eq:GWcoldetails}
\Omega_{\rm col} h^2 = 1.67 \times 10^{-5} {\left(\frac{\beta }{H_{\ast}} \right)^{-2}} \left( \frac{\kappa_c \alpha}{1 + \alpha} \right)^2 \left(  \frac{100}{g^{\ast}} \right)^{1/3} \left( \frac{0.11 v^3_w}{0.42 + v^2_w} \right) \frac{3.8 \left( f/f_{\rm col} \right)^{2.8}}{1 + 2.8 \left( f/f_{\rm col} \right)^{3.8}}\;,
\eeq
where $v_w$ is the bubble wall velocity and $\kappa_c$ is the efficiency factor of bubble collision, given as,
\beq\label{eq:kcfac}
\kappa_c = \frac{0.715 \alpha + \frac{4}{27} \sqrt{\frac{3 \alpha}{2}}}{1 + 0.715 \alpha}.
\eeq
The red-shifted peak frequency $f_{\rm col}$ \cite{Jinno:2016vai} is expressed as (with the approximation $T_{\ast} \approx T_n$),
\beq\label{eq:PF1}
f_{\rm col} = 16.5 \times 10^{-6} \left( \frac{f_{\ast}}{\beta} \right) \left( \frac{\beta}{H_{\ast}} \right) \left( \frac{T_n}{100 {\rm \,GeV}} \right) \left( \frac{g^{\ast}}{100} \right)^{1/6} {\rm Hz},
\eeq
where the fitting function, $f_{\ast}/\b$, at the time of the PT is given by,
\beq\label{eq:fastbybetadetails}
\frac{f_{\ast}}{\b} = \frac{0.62}{1.8 - 0.1 v_w +v^2_w}.
\eeq
In order to obtain a GW spectrum with higher strength, it is generally assumed that the expanding bubbles attain a relativistic terminal velocity in the plasma and we consider $v_w \simeq 1$ in our calculations \footnote{A precise determination of bubble wall velocity is non-trivial \cite{Dine:1992wr,Ignatius:1993qn,Moore:1995si,Moore:1995ua,Moore:2000wx} and out of scope of the present analysis. Instead, we consider here $v_w$ as an input parameter.}. However, there is a note of caution that runway bubble walls are generally undesirable in view of the successful yield of a sizeable amount of EWBG \footnote{Recently, an improved analysis on bubble wall dynamics has reported that EWBG may be possible even for supersonic $v_w$ \cite{Cline:2020jre,Dorsch:2021nje,Dorsch:2021ubz} which is in contrast with our traditional notion. }.

The contribution to the total GW density from sound waves can be parameterized as \cite{Hindmarsh:2013xza,Hindmarsh:2016lnk,Hindmarsh:2017gnf,Guo:2020grp},
\beq\label{eq:GWswpart}
\Omega_{\rm sw} h^2 = 2.65 \times 10^{-6}\; \Upsilon(\tau_{\rm sw}) \left(  \frac{\beta}{H_{\ast}} \right)^{-1} v_w \left( \frac{\kappa_{\rm sw} \alpha}{1 + \alpha} \right)^2 \left( \frac{g^{\ast}}{100} \right)^{1/3} \left( \frac{f}{f_{\rm sw}} \right)^3 \left[ \frac{7}{4 + 3 \left( f/f_{\rm sw} \right)^2} \right]^{7/2},
\eeq
where $\kappa_{\rm sw}$ is the efficiency factor for the sound wave contribution representing the fraction of the energy (latent heat) that gets converted into the bulk motion of the plasma and subsequently emits gravitational waves as given by (in the limit $v_w\to 1$)
\beq\label{eq:kappasw}
\kappa_{\rm sw} \simeq \left[ \frac{\alpha}{0.73 + 0.083 \sqrt{\alpha} + \alpha} \right].
\eeq
The quantity $f_{\rm sw}$ corresponds to the present peak frequency for the sound wave contribution to the total GW energy density, expressed as
\beq\label{eq:PF2}
f_{\rm sw} = 1.9 \times 10^{-5} \left( \frac{1}{v_w} \right) \left( \frac{\beta}{H_{\ast}} \right) \left( \frac{T_n}{100 {\rm ~GeV}} \right) \left( \frac{g^{\ast}}{100} \right)^{1/6} {\rm Hz}.
\eeq
The parameter $\Upsilon (\tau_{\rm sw})$ appears due to the finite lifetime of the sound waves which suppresses their contributions to the GW energy density as written as
\beq\label{eq:swtimepart}
\Upsilon(\tau_{\rm sw}) = 1 - \frac{1}{\sqrt{1+2 \tau_{\rm sw} H_{\ast}}},
\eeq
with $\tau_{\rm sw}$ being the lifetime of the sound waves. The onset of the turbulence takes place at this timescale and disrupts the sound wave source. Following Ref. \cite{Hindmarsh:2017gnf}, we write $\tau_{\rm sw} \approx R_{\ast}/{\overline{U}_f}$, where $R_{\ast} = \left( 8 \pi \right)^{1/3} v_w/\beta$ and $\overline{U}_f = \sqrt{3 \kappa_{\rm sw} \alpha/4}$ are the mean bubble separation and the root-mean-squared fluid velocity which can be obtained from a hydrodynamic analysis, respectively.
\begin{center}
\begin{figure*}[h!]
\centering
\includegraphics[height=6cm,width=8cm]{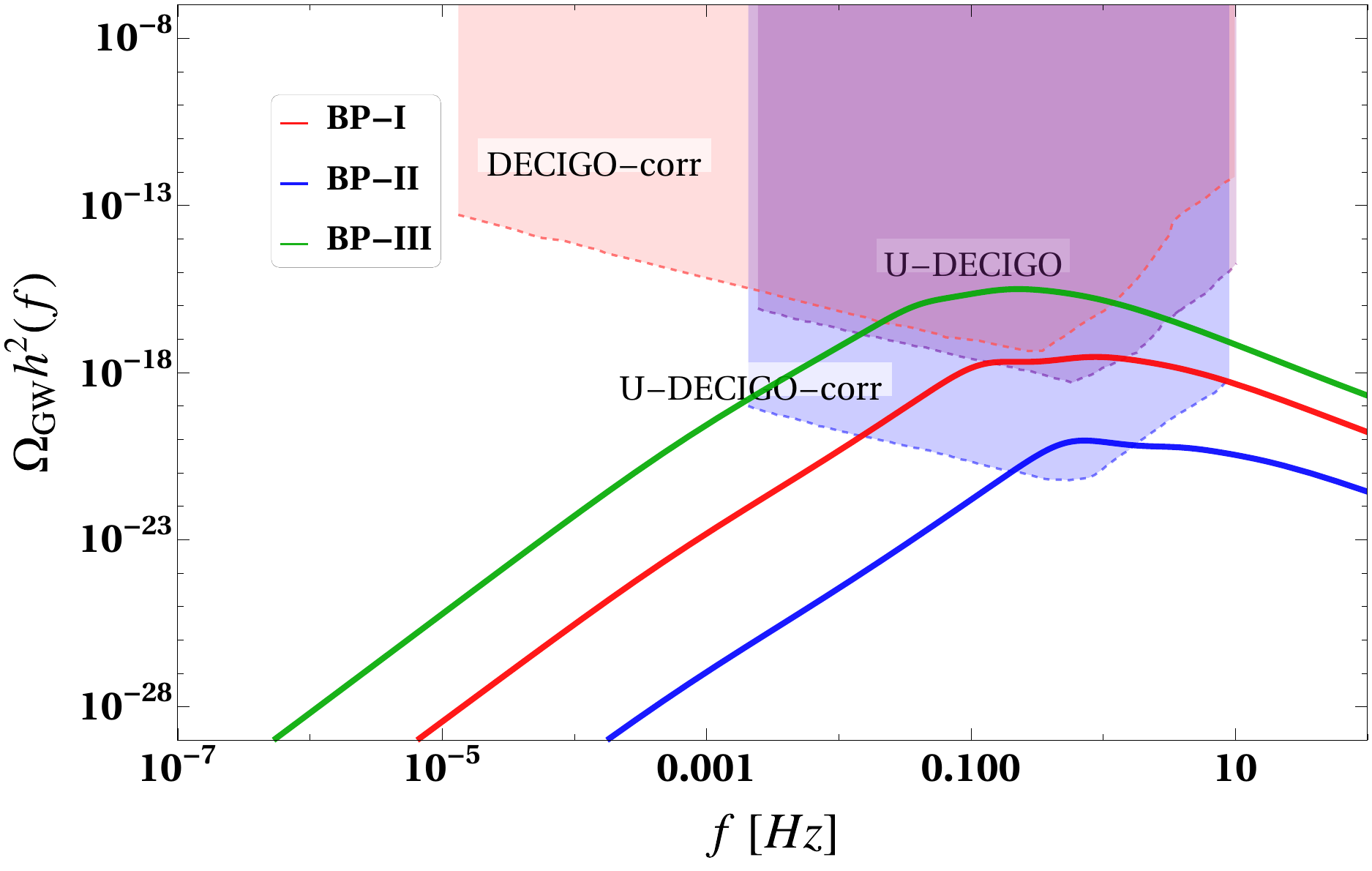}
\caption{\label{fig:resultsGW1} Prediction of GW energy density as a function of the frequency for the first three BPs as shown in Table\,\ref{tab:BPs}. We have also highlighted the regions that indicate the proposed sensitivities of the GW experiments namely U-DECIGO and U-DECIGO corr \cite{Kudoh:2005as,Yagi:2011wg}. The sensitivity curves for DECIGO and {U}-DECIGO with correlation analyses are taken from Ref. \cite{Nakayama:2009ce}.}.
\end{figure*}
\end{center}

At the time of PT, the plasma is fully ionized  and due to the resulting MHD turbulence, it leads to another source of GWs. The MHD turbulence contribution to the total GW energy density is modelled as \cite{Caprini:2009yp}
\beq\label{eq:GWturpart}
\Omega_{\rm tur} h^2 = 3.35 \times 10^{-4} \left( \frac{\beta}{H_{\ast}} \right)^{-1} v_w \left( \frac{\kappa_{\rm tur} \alpha}{1 + \alpha} \right)^{3/2} \left( \frac{100}{g^{\ast}}\right)^{1/3} \left[ \frac{\left( f/f_{\rm tur} \right)^3}{\left[ 1 + \left( f/f_{\rm tur} \right) \right]^{11/3} \left( 1 + \frac{8 \pi f}{h_{\ast}} \right)} \right],
\eeq
where $h_{\ast} = 16.5 \times \left( \frac{T_n}{100 {\rm ~GeV}} \right) \left( \frac{g^{\ast}}{100} \right)^{1/6} {\rm Hz}$, the inverse Hubble time during GW production, red-shifted to today. The peak frequency $f_{\rm tur}$ is given by,
\beq\label{eq:PF3}
f_{\rm tur} = 2.7 \times 10^{-5} \frac{1}{v_w} \left( \frac{\beta}{H_{\ast}} \right) \left( \frac{T_n}{100 {\rm~GeV}} \right) \left( \frac{g^{\ast}}{100} \right)^{1/6} {\rm Hz.}
\eeq
We set $\kappa_{\rm tur}=\epsilon \kappa_{\rm sw}$ where $\epsilon$ stands for the fraction of the bulk motion which is turbulent. Simulations suggest $\kappa_{\rm tur}=0.1 \kappa_{\rm sw}$ which we have considered in our numerical calculations.

\begin{figure*}[!ht]
\centering
\includegraphics[height=6cm,width=8cm]{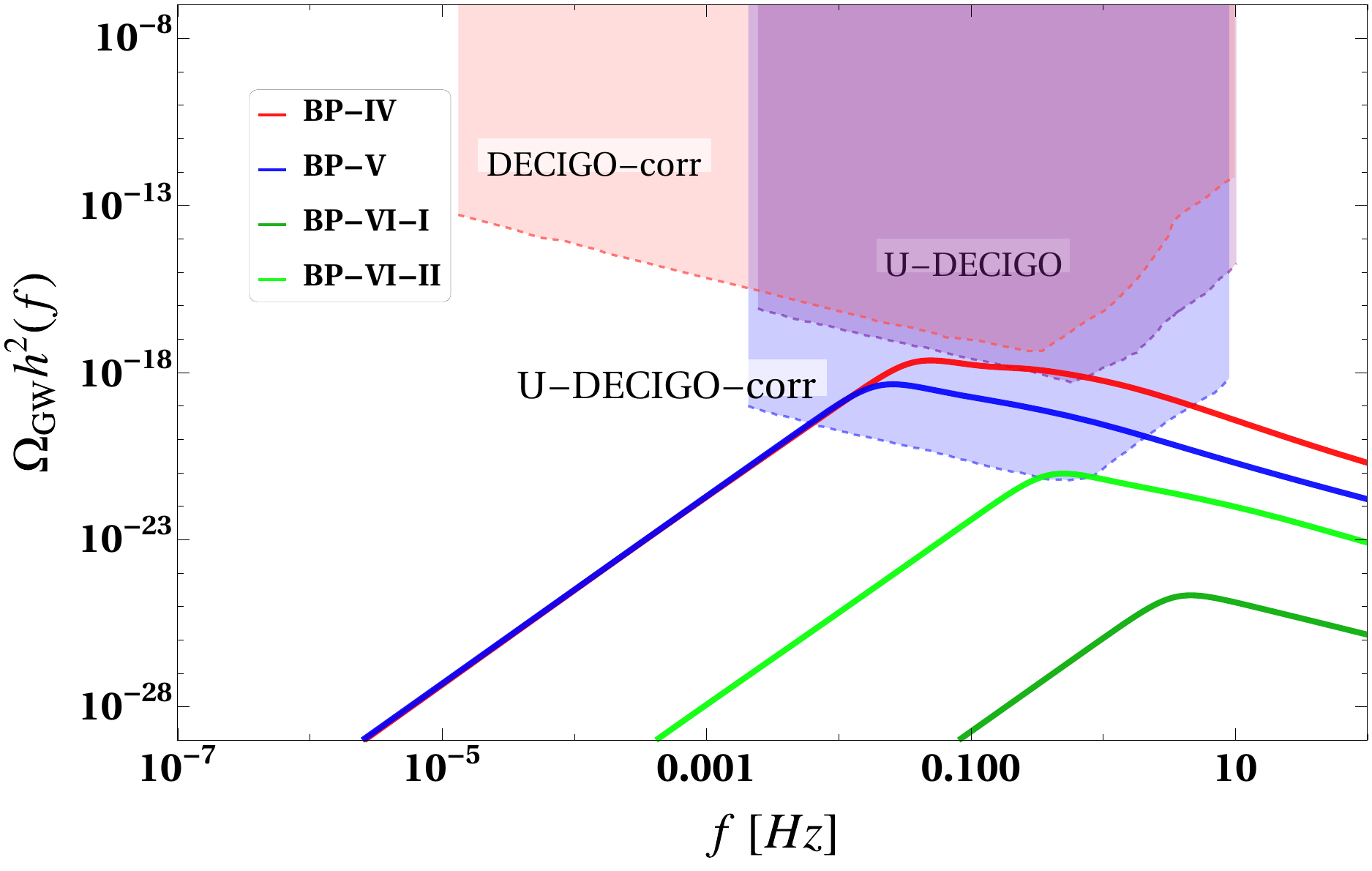}
\caption{\label{fig:resultsGW2}Prediction of GW energy density as a function of the frequency for the last three BPs from Table\,\ref{tab:BPs}. We have also highlighted the regions that indicate the proposed sensitivities of the GW experiments namely DECIGO-corr, U-DECIGO and U-DECIGO corr \cite{Kudoh:2005as,Yagi:2011wg}.}
\end{figure*}

With these details, in Figure\,\ref{fig:resultsGW1} we present the estimates of GW energy density spectrum as a function of frequency for the first three BPs as shown in Table\,\ref{tab:BPs}. The predictions of $\Omega_{\rm GW}h^2$ for the last three BPs of Table\,\ref{tab:BPs} are shown in Figure\,\ref{fig:resultsGW2}. We notice from Eq. (\ref{eq:GWcoldetails}), Eq. (\ref{eq:GWswpart}) and Eq. (\ref{eq:GWturpart}), that each individual contribution to the total GW energy density, $\Omega_{\rm GW} h^2$ (as defined in Eq.(\ref{eq:GWTotal})) is an increasing function of $\alpha$ \footnote{For $\alpha\gg 1$, $\Omega_{\rm col} h^2$, $\Omega_{\rm sw}h^2$ and $\Omega_{\rm tur}h^2$ are expected to turn insensitive to the change of $\alpha$.}. This feature in turn makes $\Omega_{\rm GW} h^2$ rise as well for a relatively larger $\alpha$.
In contrast, a larger $\frac{\beta}{H_*}$ reduces the amount of $\Omega_{\rm GW} h^2$. Earlier, in Table\,\ref{tab:GW_obs}, we observed that BP-III yields the largest value of $\alpha$ among the six BPs of Table\,\ref{tab:BPs} with relatively smaller $\frac{\beta}{H_*}$ ratio. Consequently, we find the corresponding peak amplitude of $\Omega_{\rm GW}h^2$ to be $\sim \mathcal{O}(10^{-17})$ for BP-III, which turns out to be the largest as well. 
This feature is depicted in Figure \ref{fig:resultsGW1}. The lowest peak amplitude of $\Omega_{\rm GW}h^2$ that we obtain is for the first-step PT of BP-VI which is $\sim \mathcal{O}(10^{-25})$ as shown in Figure \ref{fig:resultsGW2}. The massive suppression to $\Omega_{\rm GW}h^2$ for BP-VI-I is caused by the simultaneous presence of a large  $\frac{\beta}{H_*}$ value together with a small $\alpha$ value as shown in Table \ref{tab:GW_obs}. The second-step PT of BP-VI produces a peak having amplitude $\sim \mathcal{O}(10^{-22})$ which is relatively less suppressed due to a smaller value of $\frac{\beta}{H_*}$ compared to BP-VI-I as shown in Table \ref{tab:GW_obs}.

In view of such estimates, the proposed future GW interferometers namely U-DECIGO and U-DECIGO correlation have the required sensitivities to probe all the BPs, except BP-VI-I, considered in our analysis including BP-I which is preferred in order to address EWBG. We also find it pertinent to mention that the peak frequency of each contribution to GW energy density is linearly proportional to the ratio $\frac{\beta}{H_*}$ as evident from Eqs.\,(\ref{eq:PF1}), (\ref{eq:PF2}) and (\ref{eq:PF3}). It is numerically found that the frequency $f_{\rm max}$ where $\Omega_{\rm GW} h^2$ (see Eq.(\ref{eq:GWTotal})) attains maximum, also emerges to be an increasing function of $\frac{\beta}{H_*}$ ratio. As already noted in Table\,\ref{tab:GW_obs}, that BP-VI-I produces the largest $\frac{\beta}{H_*}$ ratio among all the BPs. This makes the peak frequency $f_{\rm max}$ of the corresponding GW spectrum for BP-VI-I the largest among all BPs.

\begin{figure*}[!h]
\centering
\subfigure{\includegraphics[height=6cm,width=7.4cm]{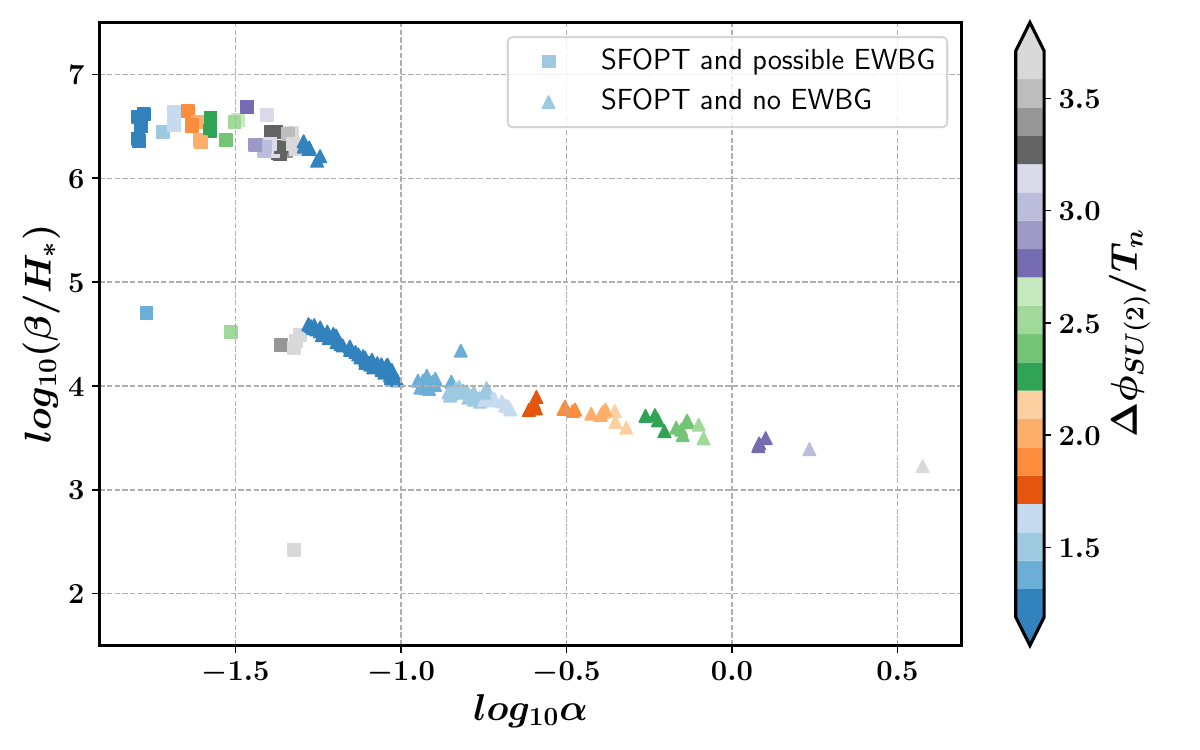}}
\subfigure{\includegraphics[height=6cm,width=7.4cm]{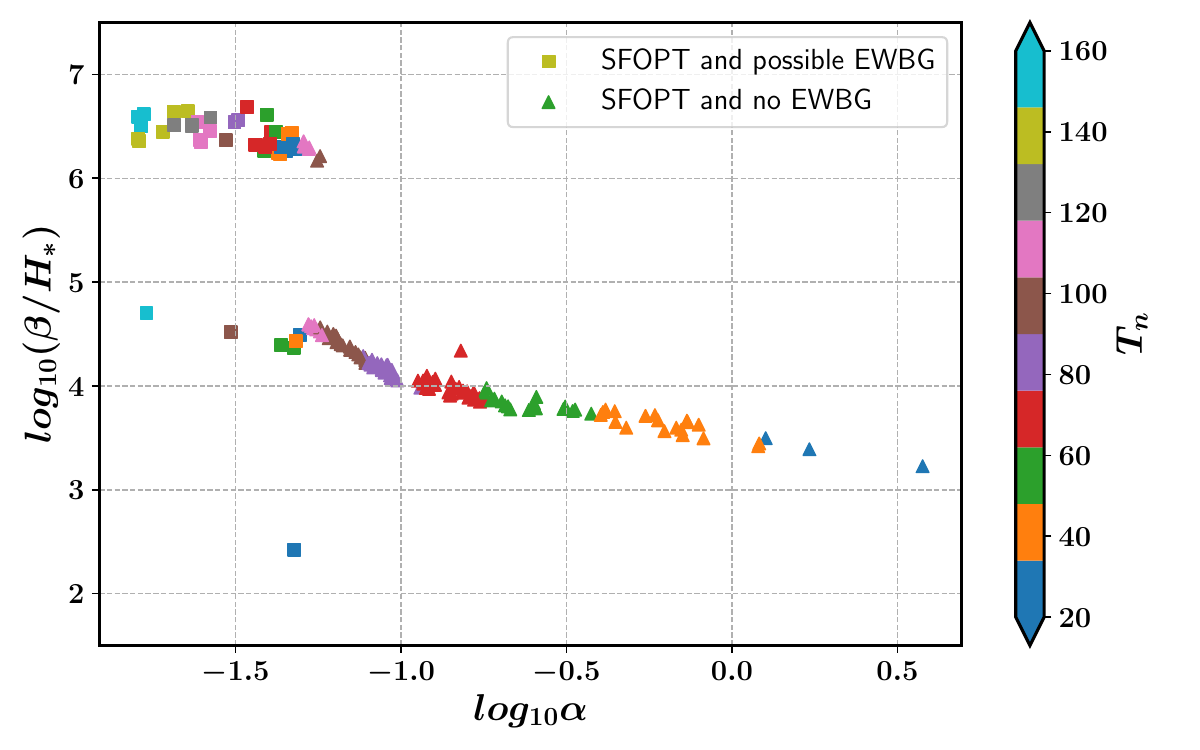}}
\caption{Values of $\alpha$ and $\frac{\b}{H_{*}}$ as a function of $\Delta \phi_{SU(2)}/T_n$ (left) and nucleation temperature {$T_n$} (right) for the points in Figure\,\ref{fig:resultsPT-scan2} that satisfy the criteria of SFOPT with possible EWBG (depicted by coloured `$\blacksquare$') and SFOPT without EWBG (depicted by coloured `$\blacktriangle$').}
\label{fig:alpha-betaH-S3}
\end{figure*}


Earlier in Figure\,\ref{fig:resultsPT-scan2} we have identified points in the $v_N-\lambda_N$ plane that exhibits strong PT along the $SU(2)_L$ doublet direction, i.e., $\Delta \phi_{SU(2)}/T_n >1$, with and without favouring EWBG as highlighted by coloured `$\blacksquare$' and `$\blacktriangle$' symbols, respectively. Recollect that, in order to prepare Figure\,\ref{fig:resultsPT-scan2}, we have utilised the fixed values of the other relevant independent parameters as in BP-I, except $v_N$ and $\lambda_N$. In Figure\,\ref{fig:alpha-betaH-S3}, we show the estimates of $\alpha$ and $\b/H_{*}$, corresponding to the same parameter corner, that is relevant to estimate $\Omega_{\rm GW}h^2$ as a function of $\Delta \phi_{SU(2)}/T_n$ (left) and the nucleation temperature {$T_n$} (right), respectively. Note that we are giving particular emphasis on analysing Figure\,\ref{fig:resultsPT-scan2} further to compute the GW energy density since it offers the scope of realising EWBG while exhibiting $\Delta \phi_{SU(2)}/T_n>1$ (traceable at GW interferometers) at the same time. The Figure\,\ref{fig:alpha-betaH-S3} illustrates the fact that the points, favoured for EWBG require relatively higher $\b/H_{*}$ and lower $\alpha$ values compared to the points that do not favour EWBG. This essentially suppresses the peak amplitude of $\Omega_{\rm GW} h^2$ for the points favouring EWBG and simultaneously increase the peak frequency $f_{\rm max}$. The right panel of Figure\,\ref{fig:alpha-betaH-S3} indicates that a lower $T_n$ tends to increase $\alpha$ which in turn enhance the $\Delta \phi_{SU(2)}/T_n$ leading to larger $\Omega_{\rm GW}^{\rm peak} h^2$. Such features are imprinted in Figure\,\ref{fig:GW-profile-EWBG} where we have shown the estimates of $\Omega_{\rm GW} h^2$ as a function of $f$ for both the coloured `$\blacksquare$' and `$\blacktriangle$' shaped points, present in Figure\,\ref{fig:resultsPT-scan2}. We clearly observe that the points which are not favoured for possible EWBG, produce a larger amount of $\Omega_{\rm GW}h^2$ at a particular $f$ and may even fall within the sensitivity curves of LISA \cite{Harry:2010zz} and BBO \cite{Corbin:2005ny}. However, the discovery scopes of those points purely depend on the signal-to-noise ratio of the corresponding experiments \cite{Kosowsky:1991ua}.

\begin{figure*}[!h]
\centering
\subfigure{\includegraphics[height=5.5cm,width=7.4cm]{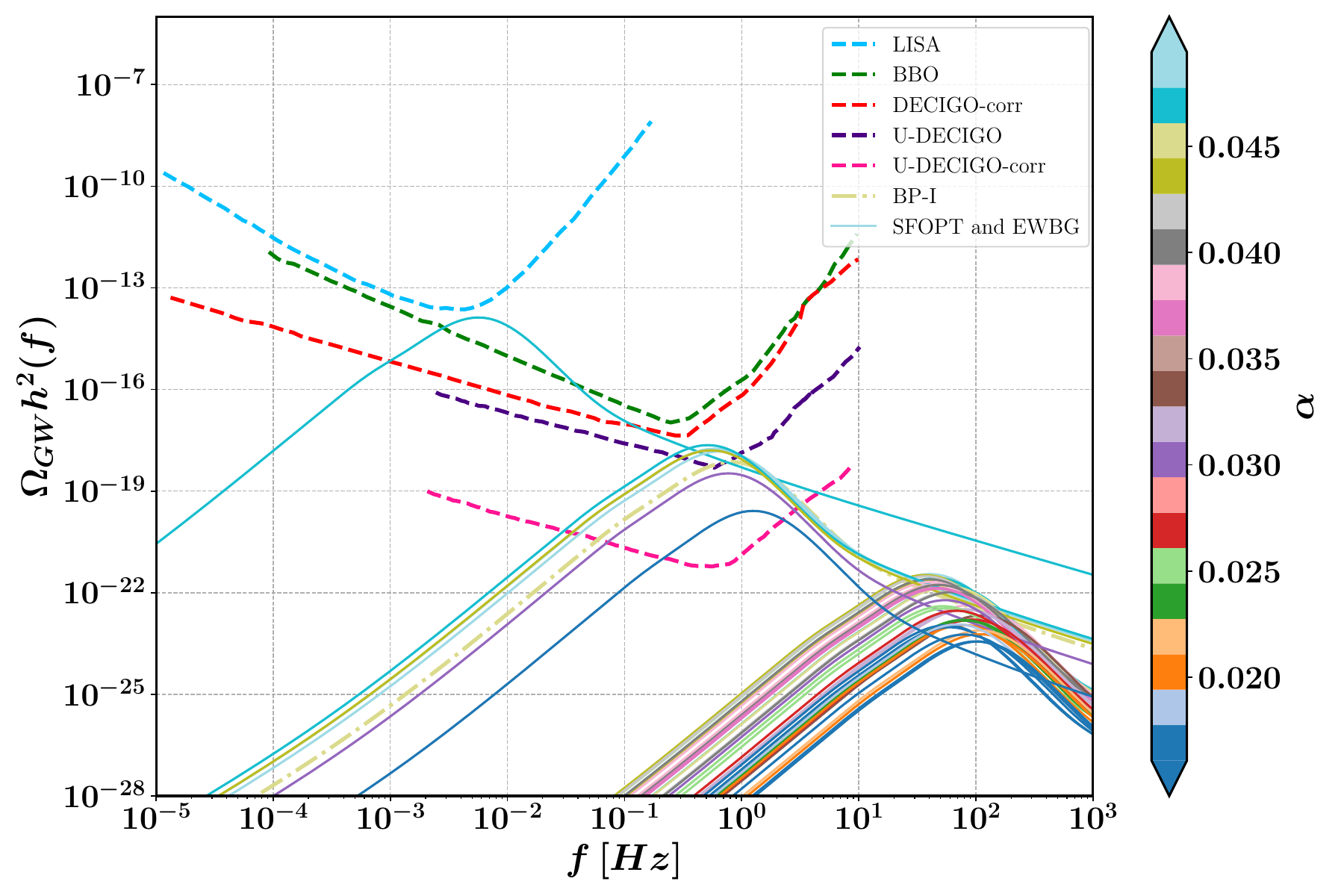}}
\subfigure{\includegraphics[height=5.5cm,width=7.4cm]{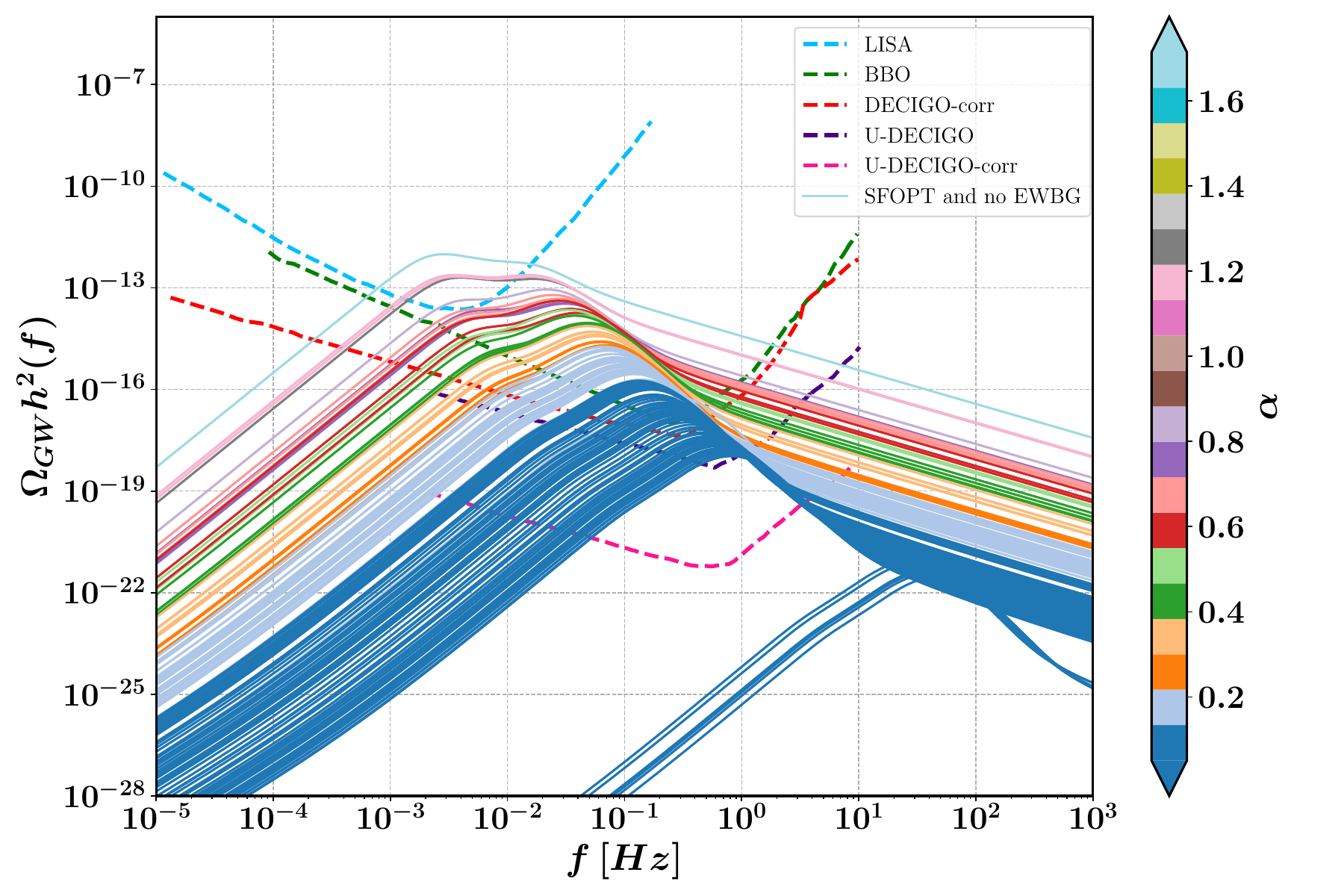}}
\caption{GW spectra for the points that show SFOPT in the $SU(2)_L$ doublet field directions with (left) and without (right) possible EWBG. Note that these points are marked by ` $\blacksquare$' and ` $\blacktriangle$' in Figure\,\ref{fig:resultsPT-scan2}. For both figures, we keep $\alpha$ as a {variable.}}
\label{fig:GW-profile-EWBG}
\end{figure*}

\section{Summary and Conclusion}\label{sec:conclu}
In the present work, we have addressed the properties of EWPT in the RHN superfield extended setup of $\mathbb{Z}_3$ invariant NMSSM. The RHN extended $\mathbb{Z}_3$ invariant NMSSM is captivating due to its ability to provide solutions to the $\mu-$problem of the MSSM and non-vanishing neutrino masses and mixing simultaneously. In particular, we consider the case where both the LH- and RH-sneutrino receive non-zero VEVs, leading to a spontaneous R-parity-violating scenario. We have worked in an effective field theory set-up by integrating out the heavier squarks, gluinos, as well as sleptons. Additionally, a simple parametrization of the TeV scale seesaw dictates the LH-sneutrino fields to weakly couple to the other relevant fields and thus, is expected to contribute negligibly to the PT dynamics. These facts effectively lead to a four-dimensional field space spanned by the four CP-even Higgses which is of interest in order to explore the PT characteristics in the present framework. 
	
Without going into the numerical details, one can naively anticipate that in the current setup having a four-dimensional field space, the PT dynamics is likely to be more involved than in the NMSSM where the relevant field space is three-dimensional. The EWPT properties and estimate of GW spectrum in the NMSSM have been extensively studied in literature where the roles of NMSSM parameters on the PT strength are also detailed. In this work, we scrutinize the role served by the new parameters that appear in theory due to the presence of the RHN superfield on the PT dynamics. In particular, we find that three new parameters $\lambda_N, A_{\lambda_N}$ and $v_N$ leave a non-trivial impact on determining the PT strength.
	
In the beginning, we describe the model details and successively develop the tools required to study the behaviour of the scalar potential as a function of temperature. We then demonstrate the possible experimental constraints that are of utmost importance to obtain a viable parameter space. Specifically, we undertake constraints arising from the validation of SM Higgs boson properties, BSM Higgs and SUSY searches at colliders, various flavour-violating processes, neutrino experiments and the muon anomalous magnetic moment. Since extensive scanning of full parameter space considering a four-dimensional field space, relevant for PT is numerically challenging, we first adopt a benchmark-based analysis. We provide six BPs that pass through all the experimental constraints and exhibit distinct kinds of FOPT patterns along the different field directions. We have discussed the PT dynamics corresponding to each BP in detail. 
	
An SFOPT is a prerequisite for EWBG with distinct high-temperature behaviour of the total scalar potential along the $SU(2)_L$ field directions. We have shown that BP-I is the preferred BP that exhibits the essential features required for a possible EWBG. On the other hand, BP-II - BP-V showing SFOPT along the different $SU(2)_L$ doublet and singlet field directions in single-step, however, are not suitable for successful EWBG. We find multi-step FOPT for BP-VI. All the BPs listed have one particular feature in common which is the preference for a lighter RH-sneutrino-dominated state below 125 GeV for the occurrence of a FOPT. Next, we utilize a few of the BPs to inquire about the role of new parameters on PT strength. Two of the new parameters $v_N$ and $\lambda_N$ show similar impacts on the PT strength along either of the $SU(2)_L$ doublet or singlet field directions. It turns out that the PT strength increases with the decrease of either $v_N$ or $\lambda_N$. The remaining parameter $A_{\lambda_N}$ has a minor role in the PT along $SU(2)_L$ doublet field directions whereas the PT strengths in the $SU(2)_L$ singlet field directions get enhanced with the increase of $|A_{\lambda_N}|$. The possible reasons for such unique properties are associated with the impact of the new parameters on the barrier height in the constituent field directions and also the lightness of the RH-sneutrino state, supported by our semi-analytic calculations as well.
	
Finally, we examine the testability of the BPs by computing the GW energy density corresponding to each BP. We have considered all possible sources that trigger GW emission in a FOPT namely, bubble wall collisions, sound waves and magneto-hydrodynamic turbulence. The highest peak amplitude of the GW energy density that we obtain is for BP-III which lies within the proposed sensitivity of DECIGO correlation data. The peak amplitude of $\Omega_{\rm GW} h^2$ for other BPs is relatively weaker, however, within the reach of U-DECIGO and U-DECIGO-corr sensitivities. It is to be noted that a TeV scale canonical seesaw model with RHN weakly coupled to SM particles is extremely difficult to probe at collider experiments. Our analysis infers an alternative albeit promising pathway to validate a TeV scale seesaw model at future GW interferometers beyond colliders.
	
In the present work, we have not performed an exact prediction of the baryon asymmetry of the Universe. Instead, we find the corner of the parameter space that shows SFOPT along the $SU(2)_L$ doublet field directions and facilitates EWBG. Improvement of our analysis is possible by precise computation of  bubble wall profile, bubble wall velocity, and CP-violation that decide the final amount of baryon asymmetry of the Universe, which is also  correlated with NMSSM + RHN model parameters. In an R-parity violating theory like the present one, gravitino can be a potential decaying dark matter candidate. Future works may also include investigating the correspondence between gravitino dark matter phenomenology and NMSSM + RHN parameter space, favouring an SFOPT.

\section*{Acknowledgements}
P. B. acknowledges the financial support received from the Indian Institute of Technology, Delhi (IITD) as a Senior Research Fellow. P. G. acknowledges the IITD SEED grant support 
{\bf{IITD/Plg/budget/2018-2019/21924}}, continued as {\bf{IITD/Plg/budget/2019-2020/173965}}, IITD Equipment Matching Grant {\bf{IITD/IRD/MI02120/208794}}, and Start-up Research Grant (SRG) support {\bf{SRG/2019/000064}} from the Science and Engineering Research Board (SERB), Department of Science and Technology, Government of India. A.K.S. is supported by NPDF grant {\bf{PDF/2020/000797}} from the SERB, Government of India. P.B. and A.K.S. also acknowledge Mikael Chala, Bo-Qiang Lu, Jiang Zhu and Kaius Loos for useful communications regarding {\tt cosmoTransitions} code. 

	
\appendix
\section{Field dependent mass matrices}\label{app:FM}
Our numerical studies are based on the field-dependent masses (see subsection \ref{susec:effective pot}). The corresponding scalar squared mass terms are evaluated at $T=0$ using the tree-level uncoloured scalar potential $V_{\rm scalar}$ (see below), including only the dominant higher-order contributions $\Delta V$ (see Eq. (\ref{eq:delV})). Mathematically, for the uncoloured scalar squared mass matrices
\beq\label{eq:scalarsqmassgen}
\mathcal{M}^2_{X,ij} = \mathcal{M}^2_{\phi_\alpha \phi_\beta}(H_{\rm SM}, H_{\rm NSM}, H_S, N_R) \equiv \frac{\partial^2 V_{\rm scalar}}{\partial \phi_\alpha \partial \phi_\beta} {\biggr\rvert}_{\phi_\alpha \neq 0},
\eeq
where $X=S$ (for the CP-even neutral scalar) or $A$ (for the CP-odd neutral scalar) and $i,\,j=1,.....,7$. Further, $\phi_{\alpha (\beta)}=H_{\rm SM},\, H_{\rm NSM},\, H_S,\, N_R$, $\Re(\widetilde{\nu}_{1,2,3})$ for the CP-even neutral scalar and $\phi_{\alpha (\beta)}=A_{\rm NSM},\, A_{\rm S},\, G^0,\, N_I$, $\Im(\widetilde{\nu}_{1,2,3})$ for the CP-even neutral scalar, respectively. For the uncoloured electrically charged scalar, $X=C$ with $i,\,j=1,.....,8$ and 
$\phi_{\alpha (\beta)} \equiv \mathcal{C}^+=$ $H^+,\, G^+$, $\widetilde{e}^+_L,\, \widetilde{\mu}^+_L,\, \widetilde{\tau}^+_L$, $\widetilde{e}^+_R,\, \widetilde{\mu}^+_R,\, \widetilde{\tau}^+_R$. Here, 
we have used 
\beq\label{eq:LHsneutRI}
\widetilde{\nu}_i = \frac{\Re \widetilde{\nu}_i + i \Im \widetilde{\nu}_i}{\sqrt{2}} \equiv \frac{\nu_{Ri} + i \widetilde{\nu}_{Ii}}{\sqrt{2}} \quad {\rm with}\quad i=1,2,3\equiv e,\mu,\tau.
\eeq
The full uncoloured scalar potential is given by\\
{\small{
\bea\label{eq:fullscalar_pot}
V_{\rm scalar} &=& \left| \sum^3_{i=1} Y^i_N \widetilde{\nu}_i \widetilde{N}
- \l S H^0_d\right |^2 + \left| \sum^3_{i,j=1} Y^{ij}_e \widetilde{l}_i \widetilde{e}^c_j -\l S H^0_u\right|^2 + \left|Y^i_N H^0_u \widetilde{N}-\sum^3_{j=1} Y^{ij}_e H^-_d \widetilde{e}^c_j\right|^2\nn\\
&+&\left| \l H_u \cdot H_d +\kappa S^2 + \frac{\l_N}{2} \widetilde{N}^2 \right|^2 
+ \left| \sum^3_{i=1} Y^i_N \widetilde{L}_i\cdot H_u + \l_N S\widetilde{N}\right|^2
+ \left|\sum^3_{i=1} Y^{ij}_e H_d \cdot \widetilde{L}_i\right|^2\nn\\
&+& \left|\l S H^+_u - \sum^3_{i,j=1} Y^{ij}_e\widetilde{\nu}_i \widetilde{e}^c_j\right|^2
+ \left|\l S H^-_d - \sum^3_{i=1} Y^i_N \widetilde{l}_i \widetilde{N}\right|^2
+ \left|\sum^3_{j=1} Y^{ij}_e H^0_d \widetilde{e}^c_j - Y^i_N H^+_u \widetilde{N}\right|^2\nn \\
&+&\frac{g^2_1 }{8}(|H_d|^2 - |H_u|^2 + |\widetilde{L}_i|^2-2 |\widetilde{e}^c_i|^2)^2 + \frac{g^2_2}{2}\sum^3_{a=1}\big(H_d^{\dagger} \frac{\tau^a}{2} H_d+H_u^{\dagger} \frac{\tau^a}{2} H_u+\widetilde{L}^{\dagger}_i \frac{\tau^a}{2} \widetilde{L}_i\big)^2\nn
\eea
}}
{\small{\bea
&+& m^2_{H_d}|H_d|^2 + m^2_{H_u}|H_u|^2 + m^2_{S}|S|^2 + M^2_N|\widetilde{N}|^2 + \sum^3_{i,j=1} m^2_{\widetilde{L}_{ij}} \widetilde{L}^{m^*}_i \widetilde{L}^m_j 
+ \sum^3_{i,j=1} m^2_{\widetilde{e}^c_{ij}} \widetilde{e}^{c^{m^*}}_i \widetilde{e}^{c^m}_j\nn\\
&+& \sum^3_{i=1} (A_e Y_e)^{ij} H_d \cdot \widetilde{L}_i \widetilde{e}^c_j
+ \l A_\l S H_u\cdot H_d + (A_N Y_N)^i \widetilde{L}_i \cdot H_u \widetilde{N} + \frac{\kappa A_\kappa}{3} S^3 + \frac{\l_N A_{\l_N}}{2} S \widetilde{N}^2\nn\\
&+& h.c. 
\eea}}
\hspace{-0.35cm}
Here $Y^{ij}_e$ belongs to $W'_{\rm MSSM}$ (see Eq. (\ref{eq:sup-nmssmRhn}))
and $m^2_{H_d},\,m^2_{H_u}$, $m^2_{\widetilde{L}_{ij}},\,m^2_{\widetilde{e}^c_{ij}}$,
$(A_e Y_e)^{ij}$ are encapsulated within $-\mathcal{L}'_{\rm soft}$ (see Eq. (\ref{eq:softSUSY})). Further, $i,j$ are generation indices, $\tau^a$s are Pauli spin matrices and $m=1,2$, as per the standard notation (see Refs. \cite{Haber:1984rc,Simonsen:1995cf,Drees:2004jm, Martin:1997ns, Nilles:1983ge, Sohnius:1985qm} for details).

In a similar way, one can derive field-dependent mass matrices for the uncoloured electrically neutral and electrically charged fermions, i.e., neutralinos and charginos, directly from the superpotential $W$ (see Eq. (\ref{eq:sup-nmssmRhn})). 
Mathematically, the generic mass term for the neutralino sector and the chargino sector are given by
\beq\label{eq:chi0chipmmassgen}
-\frac{1}{2} \left( \psi^{0^T}_i \mathcal{M}_{0_{ij}} \psi^{0}_j + {h.c.}\right),\,\,\,\,
-\frac{1}{2} (\psi^+,\, \psi^-)^T \mathcal{M}_{\chi_{\pm}} (\psi^+,\, \psi^-)+ {h.c.},
\eeq
respectively. Here basis for the neutralino sector is given by $\psi^{0^T}=\{\widetilde{B}^0, \widetilde{W}^0_3$, $\widetilde{H}^0_d, \widetilde{H}^0_u, \widetilde{S}, N$, $\nu_1, \nu_2, \nu_3\}$ involving neutral $U(1)_Y,\, SU(2)_L$ gauginos $(\widetilde{B}^0, \widetilde{W}^0_3)$, neutral higgsinos $(\widetilde{H}^0_d, \widetilde{H}^0_u)$, singlino $(\widetilde{S})$, RH-neutrino $(N)$ and LH-neutrinos $(\nu_{1,2,3})$. For charginos, including charged $SU(2)_L$
gauginos $(\widetilde{W}^\pm)$, charged higgsinos $(\widetilde{H}^+_u,\, \widetilde{H}^-_d)$
and charged leptons ($e^\pm_{L,\,R}$, $\mu^\pm_{L,\,R}$, $\tau^\pm_{L,\,R}$), one gets
$\psi^{+^T} =\;\{ \widetilde{W}^+, \widetilde{H}^+_u, e^+_R,\, \mu^+_R,\,\tau^+_R \}$ and 
$\psi^{-^T} =\;\{ \widetilde{W}^-, \widetilde{H}^-_d, e^-_L,\, \mu^-_L,\,\tau^-_L \}$, respectively. We will start with the scalar mass squared matrices and will discuss the fermionic sector subsequently.\footnote{While writing field-dependent masses, we
ignore terms that are quadratic in $v_i,\, Y^i_N$ and terms like $\sum \limits^3_{i=1} v_i Y^i_N$, keeping in mind their smallness. Besides, as already stated, these terms do not play any crucial role in the EWPT. Nevertheless, we have kept all these terms in our numerical analysis.}

\subsection{CP-even neutral scalars squared mass matrix}\label{ap:CPeSFdM}
In the basis $H_{\rm SM},\, H_{\rm NSM},\, H_S,\, N_R$, $\Re(\widetilde{\nu}_{1,2,3})$, non-zero entries of the symmetric $\mathcal{M}^2_{S,ij}$ are
\bea\label{M2S11}
\mathcal{M}^2_{S,11} & \simeq & \frac{1}{16 v_u v_d}\;\Big\{8 \lambda  v_S v^2  \left(A_{\lambda }+\kappa  v_S\right)+2 \lambda ^2 v_u v_d \left(-4 \left(v^2 +2 v_S^2\right)+H_{\text{NSM}}^2+4 H_S^2+3 H_{\text{SM}}^2\right)\nn\\
&&+v_u v_d\left(3 \Delta \lambda _2+ \mathcal{G}^2\right)  \left(H_{\text{NSM}}^2+3 H_{\text{SM}}^2\right)\nn\\
&&-4 \cos 2 \beta\; \left( v^2\cos2\beta\left(2 \lambda  v_S \left(A_{\lambda }+\kappa  v_S\right)+v_u v_d \left(\mathcal{G} -2 \lambda ^2\right) v_u\right)+3 \Delta \lambda _2 v_u v_d H_{\text{SM}}^2 \right)\nn\\
&&+v_u v_d \Big\{(4 \sin 2 \beta\; \left(3 \Delta \lambda _2 H_{\text{NSM}} H_{\text{SM}}-2 \lambda  H_S \left(\sqrt{2} A_{\lambda }+\kappa  H_S\right)\right)\nn\\
&&-3 \left(\Delta \lambda _2 + \mathcal{G} -2 \lambda ^2\right) \left(2 \sin 4 \beta\; H_{\text{NSM}} H_{\text{SM}}+\cos 4 \beta\; \left(H_{\text{NSM}}^2-H_{\text{SM}}^2\right)\right)\Big\}\Big\}\nn\\
&& {-\frac{1}{2 v}\;\Big\{ \frac{\l_N v}{2} \sin 2\beta \left(N_R^2-2 v_N^2\right)}\Big\},\\
\mathcal{M}^2_{S,12} & \simeq & \frac{1}{16 v_u v_d} \Big\{\frac{ 2 v^2 \sin 4\beta \left(2 \lambda  v_S \left(A_{\lambda }+\kappa  v_S\right)+v_u v_d \left(\mathcal{G}-2 \lambda ^2\right)\right)}{v_u v_d}\nn\\
&&-8 \lambda  \cos 2 \beta\; H_S \left(\sqrt{2} A_{\lambda }+\kappa  H_S\right)+3 \sin 4 \beta\; \left(\Delta \lambda _2+ \mathcal{G}-2 \lambda ^2\right) \left(H_{\text{NSM}}^2-H_{\text{SM}}^2\right)\nn\\
&&-6 \cos 4 \beta\; H_{\text{NSM}} H_{\text{SM}} \left(\Delta \lambda _2+ \mathcal{G} -2 \lambda ^2\right)+2 H_{\text{NSM}} H_{\text{SM}} \left(3 \Delta \lambda _2+ \mathcal{G} +2 \lambda ^2\right)\nn\\
&&+6 \Delta \lambda _2 \sin 2 \beta\; \left(H_{\text{NSM}}^2+H_{\text{SM}}^2\right)\Big\} -\frac{1}{4}\;\Big\{\lambda \cos 2 \beta\; \lambda _N \left(N_R^2-2 v_N^2\right)\Big\},\\
\mathcal{M}^2_{S, 13} & \simeq & \lambda ^2 H_S H_{\text{SM}}-\frac{1}{2} \lambda  \left(\sqrt{2} A_{\lambda }+2 \kappa  H_S\right) \left(\cos 2 \beta\; H_{\text{NSM}}+\sin 2 \beta\; H_{\text{SM}}\right),\\
\mathcal{M}^2_{S, 14} & \simeq & {-\frac{1}{2} \lambda  \lambda _N N_R \left(\cos 2 \beta\; H_{\text{NSM}}+\sin 2 \beta\; H_{\text{SM}}\right)},\\
\mathcal{M}^2_{S, 1\,(4+i)} & \simeq & {\frac{1}{2} N_R Y^i_N \left(\sqrt{2} A_N \sin \beta\;+H_S \left(\lambda  \cos \beta\ +\l_N\sin \beta \right)\right)},\\
\mathcal{M}^2_{S, 22} & \simeq & \frac{1}{16 v_u v_d}\;\Big\{8 \lambda  v_S v^2 \left(A_{\lambda }+\kappa  v_S\right)+v_u v_d \left(3 \Delta \lambda _2+ \mathcal{G}\right) \left(3 H_{\text{NSM}}^2+H_{\text{SM}}^2\right)\nn\\
&&+2 \lambda ^2 v_u v_d \left(-4 \left( v^2 +2 v_S^2\right)+3 H_{\text{NSM}}^2+4 H_S^2+H_{\text{SM}}^2\right)\nn \\
&&+4 \cos 2 \beta\; \left( v^2 \cos2\beta \left(2 \lambda  v_S \left(A_{\lambda }+\kappa  v_S\right)+v_d v_u \left(\mathcal{G}-2 \lambda ^2\right) \right)+3 \Delta \lambda_2 v_d H_{\text{NSM}}^2 v_u\right)\nn \\
&&+v_d v_u \Big\{4 \sin 2 \beta\; \left(2 \lambda  H_S \left(\sqrt{2} A_{\lambda }+\kappa  H_S\right)+3 \Delta \lambda _2 H_{\text{NSM}} H_{\text{SM}}\right)\nn\\
&&+3 \left(\Delta \lambda _2+\mathcal{G}-2 \lambda ^2\right) \left(2 \sin 4 \beta\; H_{\text{NSM}} H_{\text{SM}}+\cos 4 \beta \left(H_{\text{NSM}}^2-H_{\text{SM}}^2\right)\right)\Big\}\Big\}\nn\\
&&+ \frac{1}{8 v}\;\Big\{\cos\beta \cot\beta \lambda _N \Big(\lambda  v (\cos 4\beta+3) \sec ^3\beta v_N^2+4 \lambda  v \sin\beta \tan\beta\; N_R^2\Big)\Big\},
\eea
\bea
		\mathcal{M}^2_{S, 23} & \simeq & \frac{1}{2} \lambda  \left(\sqrt{2} A_{\lambda }+2 \kappa  H_S\right) \left(\sin 2 \beta\; H_{\text{NSM}}-\cos 2 \beta\; H_{\text{SM}}\right)+\lambda ^2 H_{\text{NSM}} H_S,\\
		\mathcal{M}^2_{S, 24} & \simeq & {\frac{1}{2} \lambda  \lambda _N N_R \left(\sin 2 \beta\; H_{\text{NSM}}-\cos 2 \beta\; H_{\text{SM}}\right)},  \\
		\mathcal{M}^2_{S, 2\,(4+i)} & \simeq & {\frac{1}{2} N_R Y^i_N \left(\sqrt{2} A_N \cos \beta\;+H_S \left(\cos \beta\; \lambda _N-\lambda  \sin \beta\;\right)\right)},
\eea
\bea
\mathcal{M}^2_{S, 33} & \simeq & \frac{\lambda  v_u v_d \left(A_{\lambda }+2 \kappa  v_S\right)}{v_S}+\kappa  \left(A_{\kappa } \left(\sqrt{2} H_S-v_S\right)+\kappa  \left(3 H_S^2-2 v_S^2\right)\right) - \l^2 v^2\nn \\
&&+\frac{\lambda ^2}{2} \left(H_{\text{NSM}}^2+H_{\text{SM}}^2\right)-\l\kappa   \cos 2 \beta\; H_{\text{NSM}} H_{\text{SM}}+\frac{\l\kappa}{2} \sin2\beta  \left(H_{\text{NSM}}^2-H_{\text{SM}}^2\right)\nn\\
&& +{\frac{1}{2 v_S}\;\Big\{\lambda _N \left(\left(\kappa +\lambda _N\right) v_S \left(N_R^2-2 v_N^2\right)-v_N^2 A_{\lambda _N}\right)} \Big\}, \label{M2Sij2}\\
\mathcal{M}^2_{S, 34} & \simeq & {\frac{1}{2} \lambda _N N_R \left(\sqrt{2} A_{\lambda _N}+2  \left(\kappa +\lambda _N\right) H_S\right)},\\
\mathcal{M}^2_{S, 3\,(4+i)}& \simeq & {\frac{1}{2} Y^i_N  N_R \left(H_{\text{NSM}} \left(\l_N\cos \beta\-\lambda  \sin \beta\;\right)+H_{\text{SM}} \left(\lambda  \cos \beta+ \l_N\sin \beta\ \right)\right)},\\
\mathcal{M}^2_{S, 44} & \simeq & {\frac{1}{4 v_N} \Big\{  -2 \lambda\l_N  \cos 2 \beta\; H_{\text{NSM}} H_{\text{SM}} v_N}{+\lambda  \sin 2 \beta\; v_N \lambda _N \left(H_{\text{NSM}}^2-H_{\text{SM}}^2+2 v^2\right)}\nn\\
&&{+\l_N v_N \left(2 A_{\lambda _N} \left(\sqrt{2} H_S-2 v_S\right)+2  \left(\kappa +\lambda _N\right)H_S^2\right)}\nn\\
&&{+\lambda_N v_N \left(\lambda _N \left(3 N_R^2-2 v_N^2\right)-4  \left(\kappa +\lambda _N\right)v_S^2\right)\Big\}},\\
\mathcal{M}^2_{S, 4\,(4+i)} & \simeq & {\frac{1}{2} Y^i_N \Big\{\sqrt{2} A_N \left(\cos \beta\; H_{\text{NSM}}+\sin \beta\; H_{\text{SM}}\right)+H_S \Big(H_{\text{NSM}} \left(\lambda _N\cos \beta-\lambda  \sin \beta\right)}\nn\\
&&{+H_{\text{SM}} \left(\lambda  \cos \beta+\lambda _N\sin \beta\right)\Big)\Big\}},\\
\mathcal{M}^2_{S, (4+i)\,(4+j)} & \simeq & {\frac{\delta_{ij}}{8}\;\Big\{-2 \mathcal{G}  \sin 2 \beta\; H_{\text{NSM}} H_{\text{SM}}-\mathcal{G} \cos 2 \beta\; \left(H_{\text{NSM}}^2-H_{\text{SM}}^2\right)}\nn\\
&&{-\frac{8 v_N Y^i_N \left(v_u \left(A_N+\lambda _N v_S\right)+\lambda  v_d v_S\right)}{v_i}- 2\mathcal{G} v^2 \cos2\beta\Big\}}\nn\\
&&{-\frac{1}{4} g_2^2 \left(\sin \beta\; H_{\text{NSM}}-\cos \beta\; H_{\text{SM}}\right){}^2},
\eea
where we have used $\mathcal{G}=g^2_1+g^2_2$, $v^2_u+v^2_d=v^2$ and $i=1,\,2,\,3$ are generational indices.
\subsection{CP-odd neutral scalars squared mass matrix}\label{ap:CPoSFdM}
In the basis $A_{\rm NSM},\, A_{\rm S},\, G^0,\, N_I$, $\Im(\widetilde{\nu}_{1,2,3})$, non-zero entries of the symmetric $\mathcal{M}^2_{A,ij}$ are
\bea\label{M2A11}
\mathcal{M}^2_{A, 11} & \simeq & \frac{1}{16 v_d v_u}\Big\{8 \lambda  v_S v^2 \left(A_{\lambda }+\kappa  v_S\right)+\mathcal{G} v_d v_u \left(H_{\text{NSM}}^2-H_{\text{SM}}^2\right)\nn\\
&&+2 \lambda ^2 v_d v_u \left(-4 \left(v^2+2 v_S^2\right)+H_{\text{NSM}}^2+4 H_S^2+3 H_{\text{SM}}^2\right)+\Delta \lambda _2 v_d v_u \left(3 H_{\text{NSM}}^2+H_{\text{SM}}^2\right)\nn\\
&&+4 \cos 2 \beta \left(v^2 \cos2\beta \left(2 \lambda  v_S \left(A_{\lambda }+\kappa  v_S\right)+v_u v_d \left(\mathcal{G}-2 \lambda ^2\right)\right)+\Delta \lambda _2 v_u v_d H_{\text{NSM}}^2\right)\nn\\
&&+v_u v_d \Big\{4 \sin 2 \beta \left(2 \lambda  H_S \left(\sqrt{2} A_{\lambda }+\kappa  H_S\right)+\Delta \lambda _2 H_{\text{NSM}} H_{\text{SM}}\right)\nn\\
&&+\left(\Delta \lambda _2+ \mathcal{G} -2 \lambda ^2\right) \left(2 \sin 4 \beta\; H_{\text{NSM}} H_{\text{SM}}+\cos 4 \beta\; \left(H_{\text{NSM}}^2-H_{\text{SM}}^2\right)\right)\Big\}\Big\}\nn\\
&& + \frac{1}{8 v}\Big\{ \cos \beta\; \cot \beta\; \lambda _N \Big[\lambda  v (\cos 4 \beta\;+3) \sec ^3\beta\; v_N^2 +4 \lambda  v \sin \beta\; \tan \beta\; N_R^2\Big] \Big\},
\eea
\bea\label{M2Aij1}
 \mathcal{M}^2_{A, 12} & \simeq & \frac{1}{2} \lambda  H_{\text{SM}} \left(\sqrt{2} A_{\lambda }-2 \kappa  H_S\right),\\
\mathcal{M}^2_{A, 13} & \simeq & \frac{1}{16}\;\Big\{ \frac{ 2 v^2 \sin 4\beta \left(2 \lambda  v_S \left(A_{\lambda }+\kappa  v_S\right)+ \left(\mathcal{G}-2 \lambda ^2\right)v_u v_d \right)}{v_u v_d}\nn\\
&&-8 \lambda  \cos 2 \beta\; H_S \left(\sqrt{2} A_{\lambda}+\kappa  H_S\right)+2 \Delta \lambda _2 \sin 2 \beta\; \left(H^2_{\rm NSM}+(1-2 \sin^2\b) H^2_{\rm SM}\right)\nn\\
&&+2 H_{\text{NSM}} H_{\text{SM}} \left(\Delta \lambda _2+ \mathcal{G} -2 \lambda ^2\right)-2 \cos 4 \beta\; H_{\text{NSM}} H_{\text{SM}} \left(\Delta \lambda _2+\mathcal{G}-2 \lambda ^2\right)\nn\\
&&+\sin 4 \beta\; \left(\Delta \lambda _2+\mathcal{G}-2 \lambda ^2\right) \left(H_{\text{NSM}}-H_{\text{SM}}\right) \left(H_{\text{NSM}}+H_{\text{SM}}\right)\Big\}\nn\\
&&-\frac{1}{4 v}\;\Big\{\lambda \l_N v \cos 2 \beta \left(N_R^2-2 v_N^2\right)\Big\},\\
\mathcal{M}^2_{A, 14} & \simeq & {-\frac{1}{2} \lambda \l_N H_{\text{SM}} N_R},\\
\mathcal{M}^2_{A, 1\,(4+i)} & \simeq &  {-\frac{1}{2} Y^i_N N_R\left(\sqrt{2} A_N \cos \beta\;+H_S \left(\cos \beta\; \lambda _N-\lambda  \sin \beta\;\right)\right)},
\eea
\bea\label{M2Aij2}
\mathcal{M}^2_{A, 22} & \simeq & \frac{\lambda  v_u v_d \left(A_{\lambda }+2 \kappa  v_S\right)}{v_S}-\kappa  A_{\kappa } \left(\sqrt{2} H_S+v_S\right)-\frac{1}{2} \lambda ^2 \left(2 v^2 +H_{\text{NSM}}^2+H_{\text{SM}}^2\right)\nn\\
&&+\l\kappa  \cos 2 \beta\; H_{\text{NSM}} H_{\text{SM}}+\l\kappa  \sin \beta\; \cos \beta\; \left(H_{\text{SM}}^2-H_{\text{NSM}}^2\right)+\kappa ^2 \left(H_S^2-2 v_S^2\right)\nn\\
&& {-\frac{1}{2 v_S}\;\Big\{\lambda _N \Big[(v_N^2 A_{\lambda _N}+v_S \left(2 v_N^2 \left(\kappa +\lambda _N\right)+N_R^2 \left(\kappa -\lambda _N\right)\right)\Big]\Big\}},\\
\mathcal{M}^2_{A, 23} & \simeq & -\frac{1}{2} \lambda  H_{\text{NSM}} \left(\sqrt{2} A_{\lambda }-2 \kappa  H_S\right),\\
\mathcal{M}^2_{A, 24} &=& {-\frac{1}{2} \lambda _N N_R \left(\sqrt{2} A_{\lambda _N}-2 \kappa  H_S\right)},\\
\mathcal{M}^2_{A, 2\,(4+i)} & \simeq &  {\frac{Y^i_N}{2} N_R  \left(H_{\text{NSM}} \left(\l_N \cos \beta -\lambda  \sin \beta\right)+H_{\text{SM}} \left(\lambda  \cos \beta +\l_N \sin \beta \right)\right)},\\
\mathcal{M}^2_{A, 33} & \simeq & \frac{1}{16 v_u v_d}\Big\{8 \lambda  v_S v^2 \left(A_{\lambda }+\kappa  v_S\right)- \mathcal{G} v_u v_d \left(H_{\text{NSM}}^2-H_{\text{SM}}^2\right)\nn\\
&&+2 \lambda ^2 v_u v_d \left(-4 \left(v^2+2 v_S^2\right)+3 H_{\text{NSM}}^2+4 H_S^2+H_{\text{SM}}^2\right)+\Delta \lambda _2 v_u v_d \left(H_{\text{NSM}}^2+3 H_{\text{SM}}^2\right)\nn\\
&&-4 \cos 2 \beta \left(v^2 \cos 2\beta \left(2 \lambda  v_S \left(A_{\lambda }+\kappa  v_S\right)+v_u v_d \left(\mathcal{G}-2 \lambda ^2\right) \right)+\Delta \lambda _2 v_d H_{\text{SM}}^2 v_u\right)\nn\\
&&+v_u v_d \Big\{4 \sin 2 \beta\; \left(\Delta \lambda _2 H_{\text{NSM}} H_{\text{SM}}-2 \lambda  H_S \left(\sqrt{2} A_{\lambda }+\kappa  H_S\right)\right)\nn\\
&&-\left(\Delta \lambda _2+\mathcal{G} -2 \lambda ^2\right) \left(2 \sin 4 \beta\; H_{\text{NSM}} H_{\text{SM}}+\cos 4 \beta\; \left(H_{\text{NSM}}^2-H_{\text{SM}}^2\right)\right)\Big\}\Big\}\nn\\
&& {-\frac{1}{2 v}\;\Big\{\lambda\l_N  v \sin \beta\; \cos \beta \left(N_R^2-2 v_N^2\right)} \Big\},\\
\mathcal{M}^2_{A, 34} & \simeq & {\frac{1}{2} \lambda \l_N H_{\text{NSM}}  N_R},\\
\mathcal{M}^2_{A, 3\,(4+i)} & \simeq &{-\frac{Y^i_N}{2} N_R \left(\sqrt{2} A_N \sin \beta\;+H_S \left(\lambda  \cos \beta\;+\sin \beta\; \lambda _N\right)\right)},
\eea
\bea\label{M2Aij3}
\mathcal{M}^2_{A, 44} & \simeq & {\frac{1}{4 v_N}\Big\{2 \lambda\l_N  \cos 2 \beta H_{\text{NSM}} H_{\text{SM}} v_N }\nn\\
&&{+ \lambda\l_N  \sin 2 \beta v_N  \left(-H_{\text{NSM}}^2+H_{\text{SM}}^2+2 v^2\right)}\nn\\
&&{+\l_N v_N \Big[-2 A_{\lambda _N} \left(\sqrt{2} H_S+2 v_S\right)+2  \left(\lambda _N-\kappa \right)H_S^2\Big]}\nn\\
 &&{+\l_N v_N \Big[\lambda _N \left(N_R^2-2 v_N^2\right)-4 v_S^2 \left(\kappa +\lambda _N\right)\Big]\Big\}},\\ 
\mathcal{M}^2_{A, 4\,(4+i)} & \simeq & \frac{Y^i_{N}}{2}\;\Big\{	H_S \Big[H_{\text{NSM}} \left(\lambda  \sin \beta +\l_N\cos \beta\right)+H_{\text{SM}} \left(\lambda_N\sin \beta -\lambda  \cos \beta\right)\Big]\nn\\
&& -\sqrt{2} A_N \left(\cos \beta H_{\text{NSM}}+\sin \beta\; H_{\text{SM}}\right)\Big\},\\
\mathcal{M}^2_{A,\, (4+i)(4+j)} & \simeq & {-\frac{\delta_{ij}}{8 v_j}\;\Big\{\mathcal{G} \cos 2 \beta v_j \left(H_{\text{NSM}}^2-H_{\text{SM}}^2+2 v^2\right)+2 \mathcal{G} v_j \sin 2 \beta\; H_{\text{NSM}} H_{\text{SM}}}\nn\\
&&+ {8 v \sin \beta\; v_N Y^j_N \left(A_N+\lambda _N v_S\right)+8 \lambda  v \cos \beta\; v_N Y^j_N v_S\Big\}}\nn\\
&& {-\frac{1}{4} g_2^2 \left(\sin \beta\; H_{\text{NSM}}-\cos \beta\; H_{\text{SM}}\right)^2},\\
\mathcal{M}^2_{A,\, 56} & \simeq & \frac{Y^1_N Y^2_N}{2}  \left(\cos \beta H_{\text{NSM}}+\sin \beta H_{\text{SM}}\right)^2,\\
\mathcal{M}^2_{A,\, 57} & \simeq & \frac{Y^1_N Y^3_N}{2}  \left(\cos \beta H_{\text{NSM}}+\sin \beta H_{\text{SM}}\right)^2 \\
\mathcal{M}^2_{A, 67} & \simeq & \frac{Y^2_N Y^3_N}{2}  \left(\cos \beta\; H_{\text{NSM}}+\sin \beta\; H_{\text{SM}}\right)^2.
\eea
where we have used $\mathcal{G}=g^2_1+g^2_2$, $v^2_u+v^2_d=v^2$ and $i=1,\,2,\,3$ are generational indices. At the physical vacuum, i.e., $\Big\{\langle H_{\rm SM} \rangle, \langle H_{\rm NSM} \rangle, \langle H_{\rm S} \rangle, \langle N_{\rm R} \rangle  \Big\} = \left\{\sqrt{2} v, 0, \sqrt{2} v_S, \sqrt{2} v_N  \right\}$, neglecting terms like 
$v^2_i,\, Y^{i^2}_N, \sum\limits^3_{i=1} v_i Y^i_N$, the Goldstone mode appears massless and decouples from the other CP-odd states.

\subsection{Uncoloured charged scalars squared mass matrix}\label{ap:CharSFdM}
Non-zero entries for the uncoloured symmetric charged scalar mass squared matrix, i.e., 
$\mathcal{C}^+ {\mathcal{M}_{C}} \mathcal{C}^-$, in the basis $\mathcal{C}^+=$ $H^+,\, G^+$, $\widetilde{e}^+_L,\, \widetilde{\mu}^+_L,\, \widetilde{\tau}^+_L$, $\widetilde{e}^+_R,\, \widetilde{\mu}^+_R,\, \widetilde{\tau}^+_R$ are 
\bea\label{M2Cij1}
\mathcal{M}^2_{C,11} & \simeq & \frac{1}{16} \Big\{ 2 \cos2\beta g_1^2 \left(2 \sin2\beta H_{\rm NSM} H_{\rm SM} + \cos2\beta(2 v^2+H^2_{\rm NSM}-H^2_{\rm SM}) \right)\nn\\
&&+ g_2^2\Big[  (1+\cos4\beta)H^2_{\rm NSM}+2\sin4\beta H_{\rm NSM} H_{\rm SM}-(-3+\cos4\beta)H^2_{\rm SM} +2v^2(1+\cos4\beta)\nn\\
&&+4\cos2\b \nn\Big] - 4v^2 \l^2(3+\cos4\b) +2 \l^2(4 H^2_{\rm S}+(-1+\cos4\b)H^2_{\rm SM})-16 \l^2 v^2_s\nn\\
&& +2 \Delta \l_2 \sin^2 2\b H^2_{\rm SM}\nn\\
&&+4 (\l^2 \sin^2 2\b +\Delta \l_2 2\cos^4\b)H^2_{\rm SM}+4 (\l^2\sin4\b -4 \Delta\l_2 \cos^3\b \sin\b) H_{\rm NSM}H_{\rm SM}\nn\\
&&+4 \l v_S A_\l (3+\cos4\b) \csc\b \sec\b + 4 \l \sin\b \left(2 H_{\rm S}(\sqrt{2} A_\l+\k H_{\rm S}+\l_N N^2_{\rm R})\right)\nn\\
&&+4 \l (3+\cos4\b)\csc2\b (2\k v^2_S+v^2_N\l_N)\Big\},
\eea
\bea
\mathcal{M}^2_{C,12} & \simeq & \frac{1}{16} \Big\{ \left(2 \cos4\b(2\l^2-\mathcal{G}-\Delta \l_2)
+ 2  (2 \l^2+g_1^2-g_2^2+\Delta \l_2) \right) H_{\rm NSM} H_{\rm SM}\nn\\
&& +2 \sin2\b \Delta \l_2 (H^2_{\rm NSM}+(1-2 \sin^2\b) H^2_{\rm SM})  \nn\\
&& + \sin4\b \left( (\mathcal{G}-2\l^2)(2 v^2+H^2_{\rm NSM}-H^2_{\rm SM})+(H^2_{\rm NSM}-H^2_{SM})\Delta \l_2 \right)\nn\\
&&+8 \l \cos2\b \left(  \k H^2_S + A_\l (\sqrt{2} H_S -2v_S)-2\k v^2_S+\frac{\l_N}{2}(N^2_R-2 v^2_N) \right)\Big\},\,\,\\
\mathcal{M}^2_{C1,\,(2+i)} & \simeq & \frac{\delta_{ij}}{4} \Big\{ v_j \Big( \left(g_2^2 \cos 2\beta+2 (Y^{ij}_e)^2 \sin ^2\beta\right) H_{\text{NSM}}+\sin 2\beta \left(g_2^2-(Y^{ij}_e)^2\right) H_{\text{SM}}\Big)\nn\\
&&-2 Y^j_{N} N_R \Big(\sqrt{2} A_N \cos \beta+\cos \beta \Big(\lambda _N H_S -Y^{ij}_e (\sin \beta H_{\text{NSM}}\nn\\
&&- \cos \beta H_{\text{SM}})\Big)+\lambda  \sin \beta H_S\Big)\Big\},\\
\mathcal{M}^2_{C, 1\,(5+i)} & \simeq & -\frac{1}{\sqrt{2}} A_e Y^{ij}_e v_j \sin\beta-\frac{1}{2} Y^{ij}_e Y^j_{N} \sin\b (N_R) \left(\cos \beta H_{\text{NSM}}+\sin \beta H_{\text{SM}}\right),\\
\mathcal{M}^2_{C, 22} & \simeq & \frac{1}{16} \Bigg\{ 2 \mathcal{G} \cos^2\b (H^2_{SM}-2 v^2)+4 \l^2 \sin^2 2\b(H^2_{\rm SM} -2 v^2) +8 \l^2(H^2_{\rm S} -2v^2_S)\nn\\
&&+\Big( -2 g_1^2 \cos^2 2\b + g_2^2(\cos4\b-3)+(\cos4\b-1)(2\l^2-\Delta \l_2) \Big)H^2_{\rm NSM}\nn\\
&&+2 \left(  \sin4\b (2\l^2-\mathcal{G}) 8 \cos\b \sin^3\b \Delta \l_2 \right) H_{\rm SM} H_{\rm NSM}+4 \l \sin2\b \Big( -2 \k H^2_S \nn\\
&& +4 \k v^2_S+A_\l \left(  -2 \sqrt{2} H_{\rm S} +4 v_S -\l_N (N^2_R -2 v^2_N) \right) \Big) \Bigg\},\\
\mathcal{M}^2_{C2,\,(2+i)} & \simeq & \frac{\delta_{ij}}{4} \Big\{ v_j \Big( \left(-g_2^2 \cos 2\beta-2 (Y^{ij}_e)^2 \sin ^2\beta\right) H_{\text{SM}}+\sin 2\beta \left(g_2^2-(Y^{ij}_e)^2\right) H_{\text{NSM}}\Big)\nn\\
&&-2 Y^j_{N} N_R \Big(\sqrt{2} A_N \sin \beta+\sin \beta \Big(\lambda _N H_S -Y^{ij}_e (\sin \beta H_{\text{NSM}}\nn\\
&&- \cos \beta H_{\text{SM}})\Big)-\lambda  \cos \beta H_S\Big)\Big\},
\eea
\bea\label{M2Cij2}
\mathcal{M}^2_{C, 2\,(5+i)} &\simeq& -\frac{(A_e Y_e)^{ij}}{\sqrt{2}}  v_j \cos\beta-\frac{1}{2} Y^{ij}_e Y^j_{N} N_R \left(\cos^2 \beta H_{\text{NSM}}+ \frac{\sin2\b}{2} H_{\rm SM}\right),\\
\mathcal{M}^2_{C, (2+i)(2+j)} &\simeq& m^2_{\widetilde{L}_{ij}} +  \frac{\delta_{ij}}{8} \Bigg\{ (g_1^2-g_2^2) (\cos2\b (H^2_{\rm SM}-H^2_{\rm NSM})-2\sin2\b H_{\rm SM}H_{\rm NSM}) 
\nn\\
&&+4 (Y^{ij}_e)^2 (\cos\b H_{\rm SM} - \sin\b H_{\rm NSM})^2  \Bigg\}, \\
\mathcal{M}^2_{C, (2+i)(5+j)} &\simeq& \frac{\delta_{ij}(A_e Y_e)^{ij}}{\sqrt{2}}  (\cos\b H_{\rm SM}-\sin\b H_{\rm NSM}) \nn\\
&& -\frac{\delta_{ij}\l Y^{ij}_e}{2} (\cos\b H_{\rm NSM} +\sin\b H_{\rm SM}) H_S,\\
\mathcal{M}^2_{C, (5+i)(5+j)} &\simeq&   m^2_{\widetilde{e}^c_{ij}} -\frac{\delta_{ij}}{4} \Bigg\{g_1^2 (\cos2\b (H^2_{\rm SM}-H^2_{\rm NSM})-2\sin2\b H_{\rm SM}H_{\rm NSM}) \nn\\
 &&- 2 (Y^{ij}_e)^2 (\cos\b H_{\rm SM} - \sin\b H_{\rm NSM})^2 \Bigg\},
\eea
where we have used $\mathcal{G}=g^2_1+g^2_2$, $v^2_u+v^2_d=v^2$ and $i=1,\,2,\,3$ are generational indices. At the physical vacuum, i.e., $\Big\{\langle H_{\rm SM} \rangle, \langle H_{\rm NSM} \rangle, \langle H_{\rm S} \rangle, \langle N_{\rm R} \rangle  \Big\} = \left\{\sqrt{2} v, 0, \sqrt{2} v_S, \sqrt{2} v_N  \right\}$, neglecting terms like 
$v^2_i,\, Y^{i^2}_N, \sum\limits^3_{i=1} v_i Y^i_N$, the Goldstone mode appears massless and decouples from the other charged states.
\subsection{Neutralino mass matrix}\label{ap:neutFdM}
In the basis of $\psi^{0^T}=\{\widetilde{B}^0, \widetilde{W}^0_3$, $\widetilde{H}^0_d, \widetilde{H}^0_u, \widetilde{S}, N$, $\nu_1, \nu_2, \nu_3\}$, the matrix $\mathcal{M}_0$ (see Eq. (\ref{eq:chi0chipmmassgen})) is given as
\begin{equation}\label{eq:chi0block}
		\mathcal{M}_0 =
		\begin{pmatrix}
			\mathcal{M}_{6 \times 6}  & m_{6 \times 3}\\
			& & \\
			m^T_{3 \times 6} & 0_{3 \times 3}\\
		\end{pmatrix},
\end{equation}
where we have used $\langle \widetilde{\nu}_i\rangle=v_i$ (see Eq. (\ref{eq:VEVs})) as the LH-sneutrinos are not dynamical in nature (see subsection \ref{susec:effective pot}). Further, matrices $m^T_{3 \times 6}$ and $\mathcal{M}_{6 \times 6}$,
using Eq. (\ref{eq:Higgsbasis}), are given as
\begin{equation}\label{eq:chi0m3by6}
m^T_{3 \times 6}=
\begin{pmatrix}
\frac{- g_1 v_e}{\sqrt{2}} & \frac{g_2 v_e}{\sqrt{2}} & 0 & \frac{Y^1_N N_R}{\sqrt{2}} & 0 & \frac{Y^1_N}{\sqrt{2}}\mathcal{Y} \\ \\
\frac{- g_1 v_\mu}{\sqrt{2}} & \frac{g_2 v_\mu}{\sqrt{2}} & 0 & \frac{Y^2_N N_R}{\sqrt{2}} & 0 & \frac{Y^2_N}{\sqrt{2}}\mathcal{Y} \\ \\
\frac{- g_1 v_\tau}{\sqrt{2}} & \frac{g_2 v_\tau}{\sqrt{2}} & 0 & \frac{Y^3_N N_R}{\sqrt{2}} & 0 & \frac{Y^3_N}{\sqrt{2}}\mathcal{Y}
\end{pmatrix},
\end{equation}
with $\mathcal{Y}=\big[s_{\beta} H_{SM} + c_{\beta} H_{NSM}\big]$ and the symmetric matrix ${\cal M}_{6 \times 6}$ is given as,
\beq\label{eq:chi0m6by6}
\begin{pmatrix}
M_1 & 0 & -\frac{g_1}{{2}}\mathcal{X}  & \frac{g_1}{{2}}\mathcal{Y} & 0 & 0 \\
& & & & & &\\
& M_2 & \frac{g_2}{{2}}\mathcal{X} & -\frac{g_2}{{2}}\mathcal{Y} & 0 & 0\\
& & & & & &\\
& & 0 & -\frac{\l}{\sqrt{2}}H_S & -\frac{\l}{\sqrt{2}} \mathcal{Y} & 0 \\
& & & & & &\\
& & & 0 & -\frac{\l}{\sqrt{2}} \mathcal{X} & 0 \\
& & & & & &\\
& & & & \sqrt{2} \kappa H_S & \frac{\l_N}{2\sqrt{2}} N_R \\
& & & & & &\\
& & & & & \frac{\l_N}{\sqrt{2}} H_S
\end{pmatrix},
\eeq
where we have omitted symmetric entries, i.e., $\mathcal{M}_{0_{ij}}=\mathcal{M}_{0_{ji}}$
for $\neq j$ and $\mathcal{X}=\big[c_{\beta} H_{SM} - s_{\beta} H_{NSM}\big]$.
\subsection{Chargino mass matrix}\label{ap:charFdM}
Using a similar approach, in the basis $\psi^{+^T} =\{ \widetilde{W}^+, \widetilde{H}^+_u, e^+_R,\, \mu^+_R,\,\tau^+_R \}$ and $\psi^{-^T}$ $ =\{ \widetilde{W}^-$, $\widetilde{H}^-_d, e^-_L,\, \mu^-_L,\,\tau^-_L \}$, the matrix $\mathcal{M}_{\pm}$ is given as
\beq\label{eq:chipmblock}
\mathcal{M}_{\pm} = 
		\begin{pmatrix}
			0 & X^T\\
			X & 0
		\end{pmatrix},
\eeq
where the $5\times 5$ matrix $X$ is given by
{\footnotesize
\beq\label{eq:chipm5by5}
\begin{pmatrix}
M_2 & \frac{g_2}{\sqrt{2}}\mathcal{Y} & 0 & 0 & 0 \\ \\
\frac{g_2}{\sqrt{2}}\mathcal{X} & \frac{\l}{\sqrt{2}} H_S & -Y^{11}_e v_e & -Y^{22}_e v_e & -Y^{33}_e v_\tau\\ \\
g_2 v_e & -\frac{Y^1_N N_R}{\sqrt{2}} & \frac{Y^{11}_e}{\sqrt{2}} \mathcal{X} & 0 & 0 \\ \\
g_2 v_\mu & -\frac{Y^2_N N_R}{\sqrt{2}} & 0 & \frac{Y^{22}_e}{\sqrt{2}}\mathcal{X}  & 0 \\ \\
g_2 v_\tau & -\frac{Y^3_N N_R}{\sqrt{2}} & 0 & 0 & \frac{Y^{33}_e}{\sqrt{2}}\mathcal{X}  \\
\end{pmatrix}.
\eeq
}
Here we have used $Y^{ij}_e=Y^{ii}_e \delta_{ij}$.
\section{Neutral scalar mass matrices after the EWSB} \label{ap:MassmMatricesEWSB}
Weak couplings among the LH-handed sneutrino states and the remaining states, as already discussed in section \ref{sec:Model}, suggest that one can safely decouple the LH-sneutrino-dominated states from the CP-even and CP-odd scalar squared mass matrices without any loss of generality. After the aforesaid detachment, both CP-even and CP-odd scalar squared mass matrices appear to be $4\times 4$ in size. The full $7\times7$ squared mass matrices are given in subsections \ref{ap:CPeSFdM} \& \ref{ap:CPoSFdM}, including LH-sneutrino states. In this section, squared mass matrices of the CP-even and the CP-odd Higgses are given after the EW symmetry breaking, i.e., using relations given in subsections \ref{ap:CPeSFdM} \& \ref{ap:CPoSFdM} and considering $\langle H_{\rm SM}\rangle=\sqrt{2}v$, $\langle H_{\rm NSM} \rangle=0$, $\langle H_{S}\rangle=\sqrt{2}v_S$, $\langle N_{R}\rangle=\sqrt{2}v_N$, $\langle A_{\rm NSM} \rangle=0$, $\langle A_{S}\rangle=0$, $\langle N_{I}\rangle=0$ (see subsection \ref{susec:effective pot}). For the CP-even states, we consider the \{$H_{\rm{SM}},H_{\rm{NSM}},H_{\rm S},N_{\rm R}$\} basis while for the CP-odd ones we use \{$A_{\rm{NSM}},A_{\rm S},G^0,A_{\rm N}$\} basis.

\subsection{CP-even mass squared elements} \label{ap:CPeMassmMatricesEWSB}
\bea\label{eq:MassCPeven}
\mathcal{M}^2_{S, 11} &=& \l^2v^2 \sin^2 2\beta + \frac{(g^2_1+g^2_2)v^2}{2}\cos^22\beta,\,\,\,
\mathcal{M}^2_{S, 12} = \frac{\l^2v^2}{2}\sin4\beta - \frac{(g^2_1+g^2_2)v^2}{4}\sin4\beta,\nn\\
\mathcal{M}^2_{S, 13} &=& 2\l^2 v v_S - \l v (A_\l + 2 \kappa v_S) \sin 2\beta,\,\,\,\, 
\mathcal{M}^2_{S, 14} =  -\l \l_N v_N \sin 2\beta,\nn\\
\mathcal{M}^2_{S, 22} &=& 2\l v_S (A_\l + \kappa v_S) \csc 2\beta + \l \l_N v^2_N \csc2\beta
-\l^2v^2 \sin^2 2\beta + \frac{(g^2_1+g^2_2)v^2}{2}\sin^2 2\beta,\nn\\
\mathcal{M}^2_{S, 23} &=&  -\l v (A_\l + 2\kappa v_S) \cos 2\beta,\,\,\,\,
\mathcal{M}^2_{S, 24} =  -\l\l_N v v_N \cos2\beta,\nn\\ 
\mathcal{M}^2_{S, 33} &=& \kappa v_S (A_\kappa + 4\kappa v_S) + \frac{\l v^2 A_\l}{2 v_S} \sin2\beta -\frac{\l_N v^2_N A_{\l_N}}{2 v_S},\nn\\
\mathcal{M}^2_{S, 34} &=&  \l_N v_N A_{\l_N} + 2\l_N \kappa v_S v_N 
+ 2\l^2_N v_S v_N, \,\,\,\,
\mathcal{M}^2_{S, 44} = \lambda _N^2 v^2_N,
\eea
where we have used the symmetric nature of these entries, i.e., $\mathcal{M}^2_{S, ij} = \mathcal{M}^2_{S, ji}$ for $i \neq j$.
\subsection{CP-odd mass squared elements} \label{ap:CPoMassmMatricesEWSB}
\bea\label{eq:MassCPodd}
\mathcal{M}^2_{A, 11} &=& \l\l_N v^2_N \csc2\beta + 2\l v_S (A_\l + \kappa v_S) \csc2\beta,
\,\,\,\, \mathcal{M}^2_{A, 12}= \l v A_\l - 2 \l \kappa v v_S,\nn\\
\mathcal{M}^2_{A, 13} &=& 0,\,\,\,\,\mathcal{M}^2_{A, 14} = - \lambda \l_N v v_N, \nn\\
\mathcal{M}^2_{A, 22} &=& \kappa(2\l v^2\sin 2\beta- 3 v_S A_\kappa)
+ \frac{\l v^2 A_\l}{2 v_S}\sin 2\beta - \frac{\l_N v^2_N}{2 v_S} (A_{\l_N} + 4 \kappa v_S),\nn\\
\mathcal{M}^2_{A, 23} &=& 0, \,\,\,\, \mathcal{M}^2_{A, 24} = 2 \l_N \kappa v_S v_N - \l_N v_N A_{\l_N},\,\,\,\, \mathcal{M}^2_{A, 33}=0,\,\,\,\, \mathcal{M}^2_{A, 34} = 0,\nn\\
\mathcal{M}^2_{A, 44} &=& \l \l_N v^2 \sin 2\beta - 2\l_N v_S (A_{\l_N}+\kappa v_S),
\eea
where we have used the symmetric nature of these entries, i.e., $\mathcal{M}^2_{A, ij} = \mathcal{M}^2_{A, ji}$ for $i \neq j$.


\section{Counter terms}\label{ap:Counter}
{As already addressed in subsection \ref{susec:higher-order-pot}, after including Coleman-Weinberg contributions (see Eq. (\ref{eq:T0Veff})), counter terms are necessary to restore the original physical minima and masses. These terms are encapsulated within $V_{ct}$ which is written as
\begin{eqnarray}
V_{ct} &&= \delta_{m_{H_d}^2}\; |H_d|^2 + \delta_{m_{H_u}^2}\; |H_u|^2 + \delta_{m_{S}^2}\; |S|^2 + {\delta_{M_N^2}\; |\widetilde{N}|^2} + \delta_{\l A_{\l}}\; (S H_u\cdot H_d + h.c.)\nn\\
&& \quad+ {\delta_{\l_N A_{\l_N}}\; (S \widetilde{N} \widetilde{N}+ h.c.)} + \frac{\delta \l_2}{2}\;|H_u|^4,
\end{eqnarray}
where $\delta_{m_{H_d}^2},\,\delta_{m_{H_u}^2},\,\delta_{m_{S}^2},\,\delta_{M_N^2},\,\delta_{\l A_{\l}},\,\delta_{\l_N A_{\l_N}},\,\delta \l_2$ are counter terms that have to be included in Eq.(\ref{eq:T0Veff}). Entries corresponding to $\delta_{m_{H_d}^2},\,\delta_{m_{H_u}^2}$ are encapsulated within $-\mathcal{L}'_{\rm soft}$ of Eq. (\ref{eq:softSUSY}). In order to maintain the location of the physical minima solutions for the counter-terms must satisfy the following relations:}

\begin{eqnarray}\label{eq:CTall}
\delta_{m_{H_d}^2} &&=\frac{1}{\sqrt{2} v}\;  \left(\tan\beta  \frac{\partial V_{\rm eff}}{\partial H_{NSM}}-\frac{\partial V_{\rm eff}}{\partial H_{SM}}\right)+
\frac{\mu  \sec ^2\beta}{2\l v}  \frac{\partial^2 V_{\rm eff}}{\partial H_S \partial H_{SM}}, \nn\\
\delta_{m_{H_u}^2} &&= \frac{\csc ^2\beta}{4v\l} \frac{\partial }{\partial H_{SM}} \left(\sqrt{2} \lambda  (\cos 2\beta -2)\;V_{\rm eff}+2 \mu  \frac{\partial V_{\rm eff}}{\partial H_S}+2 \lambda  v \frac{\partial V_{\rm eff}}{\partial H_{SM}}\right)\nn\\
&& - \frac{1}{\sqrt{2}v}  \cot\beta \frac{\partial V_{\rm eff}}{\partial H_{NSM}},\nn\\
\delta_{m_{S}^2} &&= \frac{\l}{2 \m}\;  \frac{\partial }{\partial H_S} \left(v \frac{\partial V_{\rm eff}}{\partial H_{SM}}+v_N \frac{\partial V_{\rm eff}}{\partial N_R}-\sqrt{2}\; V_{\rm eff}\right),\nn\\
\delta_{M_N^2} &&= -\frac{1}{2 v_N}\;\frac{\partial }{\partial N_R} \left(\sqrt{2}\;V_{\rm eff}-\frac{2 \mu  }{\lambda }\frac{\partial V_{\rm eff}}{\partial H_S}\right),\nn\\
\delta_{\l A_{\l}} &&= \frac{\csc 2\beta}{v}\; \frac{\partial^2 V_{\rm eff}}{\partial H_S \partial H_{SM}},\nn\\
\delta_{\l_N A_{\l_N}} &&= -\frac{1}{2 v_N}\;\frac{\partial^2 V_{\rm eff}}{\partial H_S \partial N_R},\nn\\
\delta \l_2 &&= \frac{\csc ^4\beta}{4 v^3}\;\frac{\partial }{\partial H_{SM}} \left(\sqrt{2}\; V_{\rm eff}-2 v \frac{\partial V_{\rm eff}}{\partial H_{SM}}\right). 
\end{eqnarray}

Identifying $\delta \l_2$ as a counter term for $\Delta \l_2$, a quartic coupling among $H_u$ as given in Eq. (\ref{eq:delV}), seems inconsistent. However, in reality, $\Delta\l_2$ is connected to the soft SUSY-breaking terms as the estimation of $\Delta \l_2$ includes soft SUSY-breaking terms of the stop sector (see Eq. (\ref{eq:dellambda2})).

\section{Daisy coefficients}\label{ap:Daisy}
The Daisy coefficients \cite{Dolan:1973qd,Kirzhnits:1976ts,Parwani:1991gq,Espinosa:1992gq,Arnold:1992rz}, $c_i$, using Eq. (\ref{Eq:Daisy_coeff_ci}) is given by
\begin{equation}\label{eq:daisyformula1}
c_i = \frac{m_i^2({\phi_\alpha},\,T)-m_i^2({\phi_\alpha})}{T},
\end{equation}
and can be estimated using the high-temperature limit, i.e., $T^2\gg m^2$  
($m$ depicts a generic mass term involved in the calculation) \cite{Dolan:1973qd}, of the thermal corrections from $V^{T\neq 0}_{\mathrm{1-loop}}$ (see Eq. (\ref{eq:VatTnotZero})) as
\begin{equation}\label{eq:daisyformula2}
\frac{1}{T^2} \frac{\partial^2 V^{\rm 1-loop}_{T\neq 0}}{\partial \phi_i \partial \phi_j}.
\end{equation}
Daisy coefficients are calculated at the $T^2\gg m^2$ limit which helps to efface gauge dependence for these coefficients although $V^{\rm 1-loop}_{T\neq 0}$, as already discussed in subsection \ref{susec:finiteT}, has explicit gauge dependence. The form of Eq. (\ref{eq:daisyformula2}), except the $1/T^2$ factor, looks similar to relations that are conventionally used for the computation of $i,\,j$-th entry of the different scalar mass matrices from the concerned potential. For the calculation of Daisy coefficients we use $V^{\rm 1-loop}_{T\neq 0}$ as a function of $m_i^2({\phi_\alpha})$ and not as a function of $m_i^2({\phi_\alpha},\,T)$.  However, while computing $V^{\rm 1-loop}_{T\neq 0}$ and $V'^{\rm 1-loop}_{\rm CW}$ (see Eq. (\ref{eq:totPotR})) we use thermal masses $m_i^2({\phi_\alpha},\,T)$. Expanding thermal function $J_{B/F}$ (see Eq. (\ref{eq:thermalF})), in the limit $T^2 \gg m^2$, one gets in the leading order \cite{Arnold:1992rz,Curtin:2016urg} 
\begin{equation}\label{eq:daisyformula3}
V^{T\neq 0}_{\mathrm{1-loop}} \quad \sim \quad \frac{T^2}{48}\; \left(2\;\sum_{i=B}\;n_im^2_i + \sum_{i=F}\;n_im^2_i\right),
\end{equation}
where $B (F)$ represents boson (fermion) and $n_i$ depicts the associated degrees of freedom, as already detailed in subsection \ref{susec:higher-order-pot}. It is also apparent from Eq. (\ref{eq:daisyformula3}) that contributions from the bosonic sources are the leading ones. Also, as detailed in Ref.\cite{Arnold:1992rz}, cubic contributions in the $V^{T\neq 0}_{\mathrm{1-loop}}$ appears only via bosons. Further, quartic contributions from fermions are suppressed compared to the same from bosons and do not affect the shift in the VEVs \cite{Arnold:1992rz}. Thus, we neglect contributions from the relevant fermionic sources (see Ref. \cite{Athron:2019teq} for a similar discussion in the context of the NMSSM.). In light of Eq. (\ref{eq:daisyformula2}) and Eq. (\ref{eq:daisyformula3}), non-zero Daisy coefficients  are given below where field-dependent masses are considered as a function of all bosonic degrees of freedom.
\bea\label{Eq:Daisy_coeff}
c_{H_{\rm SM} H_{\rm SM}} &=& c_{G^0 G^0} =\frac{\lambda^2}{4} + \frac{(3 m^2_Z+4 m^2_W)}{8 v^2} + \frac{m^2_Z}{4 v^2} {\sin^2 \theta_w \cos^2\beta} + \frac{m^2_t}{4 v^2}+\frac{\Delta \lambda_2}{4 v^2}, \nn\\
c_{H_{\rm SM} H_{\rm NSM}} &=& c_{H_{\rm NSM} G^0} =\frac{m^2_t}{4 v^2} \frac{1}{\tan^2\beta} + \frac{\Delta \lambda_2 \sin 2\beta}{8} - \frac{m^2_Z}{8 v^2} {\sin^2\theta_w \sin 2\beta}, \nn \\
c_{H_{\rm NSM} H_{\rm NSM}} &=& c_{A_{\rm NSM} A_{\rm NSM}} = \frac{\lambda^2}{4} + \frac{(m^2_Z + 4 m^2_W)}{8 v^2} + \frac{m^2_t}{4 v^2 \tan^2\beta} + \frac{m^2_Z}{4 v^2} {\sin^2\theta_w \sin^2\beta}\nn \\
&+& \frac{\Delta \lambda_2}{4} {\cos^2\beta}, \nn\\
c_{H_{\rm S} H_{\rm S}} &=& \frac{\lambda^2 + \kappa^2}{2} + \frac{\lambda^2_N}{8}, \,\,\,
c_{A_{\rm S} A_{\rm S}} = \frac{\lambda^2 + \kappa^2}{3} + \frac{\lambda^2_N}{12}, \,\,\,
c_{N_{\rm R} N_{\rm R}} = \frac{\lambda^2_N}{4},\,\,\,
c_{N_{\rm I} N_{\rm I}} = \frac{\lambda^2_N}{6}, \nn\\
c_{H^{\rm +} H^{\rm -}}	&=& \frac{\lambda^2}{6} + \frac{(m^2_Z + 8 m^2_W)}{24 v^2} - \frac{m^2_Z}{4 v^2} {\sin^2\theta_w \sin^2\beta} + \frac{m^2_t}{4 v^2 \tan^2\beta} \frac{1}{\rm \tan^2\beta} + \frac{\Delta \lambda_2}{4} { \cos^2\beta}, \nn\\
c_{H^{\rm +} G^{\rm -}}	&=& \frac{m^2_t}{4 v^2 \tan^2\beta} \frac{1}{\rm tan^2\beta} + \frac{\Delta \lambda_2 \sin 2\beta}{8} - \frac{m^2_Z}{8 v^2} {\sin^2\theta_w} \sin 2\beta, \nn\\
c_{G^{\rm +} G^{\rm -}}	&=& \frac{\lambda^2}{6} + \frac{(7 m^2_Z + 8 m^2_W)}{24 v^2} - \frac{m^2_Z}{4 v^2} {\sin^2\theta_w \sin^2\beta} + \frac{m^2_t}{4 v^2} + \frac{\Delta \lambda_2}{4} {\sin^2\beta},
\eea
where $m_W,\,m_Z$ represent masses for the $W^\pm,\,Z^0$ bosons, respectively and $\theta_w$ is Weinberg angle \cite{ParticleDataGroup:2022pth}.

Longitudinal modes of the massive gauge bosons also yield non-zero Daisy coefficients
\cite{Carrington:1991hz,Comelli:1996vm}
\beq\label{eq:Daisy-Gaugebosons}
c_{W^{+}_L W^{-}_L} = c_{W^{3}_L W^{3}_L} = \frac{5}{2} g^2_2, \quad  c_{B_L B_L} = \frac{13}{6} g^2_1,
\eeq
where $W^\pm_L,\,W^3_L,\,B_L$ correspond to longitudinal modes of the SM $SU(2)_L,\,U(1)_Y$
gauge bosons. These results are the same as the $\mathbb{Z}_3$-invariant NMSSM as gauge
sector of the chosen NMSSM + one RH-neutrino framework remains exactly the same as the $\mathbb{Z}_3$-invariant NMSSM. Finally, at $T\neq0$ the photon ($\gamma$) also gets a temperature-dependent mass, i.e., a non-vanishing longitudinal component, which should also be included in the field-dependent mass matrix used to evaluate eigenvalues of the electrically neutral EW gauge bosons, $\gamma, Z^0$ at $T\neq0$. 
\beq\label{eq:Daisy-photons}
m^2_{Z_L\gamma_L} (H_{\rm SM},\,H_{\rm NSM}, H_{S},\,N_{R},\,T) =  
\begin{pmatrix}
g^2_2 \frac{H^2_{\rm SM} + H^2_{\rm NSM}}{4} + \frac{5}{2} g^2_2 T^2\; & -g_1 g_2 \frac{H^2_{\rm SM} + H^2_{\rm NSM}}{4} \\[3ex]
-g_1 g_2 \frac{H^2_{\rm SM} + H^2_{\rm NSM}}{4} & g^2_1 \frac{H^2_{\rm SM} + H^2_{\rm NSM}}{4} + \frac{13}{6} g^2_1 T^2
\end{pmatrix}.
\eeq

\section{Minimization conditions}\label{ap:minmcond}
As already stated in section \ref{sec:param_space} that one can trade different soft-masses,
i.e., $m_{H_u}^2,\,m_{H_d}^2$, $m^2_{\widetilde{L}_{ij}},\, m^2_S,\, M^2_N$ (see Eq. (\ref{eq:softSUSY})) with the corresponding VEVs (see Eq. (\ref{eq:VEVs})) using minimization conditions of the $V_{\rm tree}$ (see Eq. (\ref{eq:TotScalar})). One can also use the neutral part of $V_{\rm scalar}$ as depicted in Eq. (\ref{eq:fullscalar_pot}). Mathematically, the minimization condition gives a set of equations like
\beq\label{eq:minimcondgen}
\biggl< \frac{\partial V_{\rm tree}}{\partial X_i} \biggr>\biggr|_{X=\langle X \rangle} =0,
\eeq
where $X_i=H^0_u,\, H^0_d,\, \widetilde{\nu_i},\, S,\, \widetilde{N}$, and $\langle X \rangle$
represents all the concerned VEVs as given in Eq. (\ref{eq:VEVs}). In detail, assuming 
all superpotential couplings (see Eq. (\ref{eq:sup-nmssmRhn})) to be real, one gets
\bea\label{eq:minmcond1}
\biggl< \frac{\partial V_{\rm tree}}{\partial H^0_u} \biggr>\biggr|_{\rm VEVs} &=& 
\lambda v_d \left (\l v_u v_d - \kappa v^2_S - \frac{\l_N}{2} v^2_N \right) + Y^{i^2}_N v^2_N v_u
+ \l^2 v^2_S v_u + m^2_{H_u} v_u \nn\\
&+& \sum^3_{j=1} Y^j_N v_j \left (\sum^3_{i=1} Y^i_N v_i v_u + \l_N v_S v_N  \right)
+ \frac{g^2_1+g^2_2}{4} \left( v^2_d + \sum^3_{i=1} v^2_i - v^2_u \right) v_u \nn\\
&+& \l A_\l v_S v_d + \sum^3_{i=1} (A_N Y_N)^i v_i v_N, 
\eea
\bea\label{eq:minmcond2}
\biggl< \frac{\partial V_{\rm tree}}{\partial H^0_d} \biggr>\biggr|_{\rm VEVs} &=& 
\lambda v_u \left (\l v_u v_d - \kappa v^2_S- \frac{\l_N}{2} v^2_N \right)
+ \l v_S \left( \l v_S v_d - \sum^3_{i=1} Y^i_N v_i v_N  \right)\nn\\
&+& \frac{g^2_1+g^2_2}{4} \left( v^2_d + \sum^3_{i=1} v^2_i - v^2_u \right) v_d
+ m^2_{H_d} v_d + \l A_\l v_S v_u, 
\eea
\bea\label{eq:minmcond3}
\biggl< \frac{\partial V_{\rm tree}}{\partial \widetilde{\nu_i}} \biggr>\biggr|_{\rm VEVs} &=& 
Y^i_N v_u \left(\sum^3_{j=1} Y^j_N v_j v_u + \l_N v_S v_N \right)
+ Y^i_N v_N \left(\sum^3_{j=1} Y^j_N v_j v_N - \l v_d v_S\right)\nn\\
&+& \frac{g^2_1+g^2_2}{4} \left( v^2_d + \sum^3_{i=1} v^2_i - v^2_u \right) v_i
+ (A_N Y_N)^i v_u v_N + \sum^3_{j=1} m^2_{\widetilde{L}_{ij}} v_j, 
\eea
\bea\label{eq:minmcond4}
\biggl< \frac{\partial V_{\rm tree}}{\partial S} \biggr>\biggr|_{\rm VEVs} &=& 
2\kappa v_S \left (-\l v_u v_d + \kappa v^2_S + \frac{\l_N}{2} v^2_N \right)
+ \l v_d \left(\l v_S v_d - \sum^3_i Y^i_N v_i v_N \right) \nn\\
&+& \l_N v_N \left( \sum^3_{i=1} Y^i_N v_i v_u + \l_N v_S v_N \right) + \l^2 v^2_u v_S
+ m^2_S v_S + \l A_\l v_u v_d \nn\\
&+& \kappa A_\kappa v^2_S + \frac{\l_N A_{\l_N}}{2} v^2_N, 
\eea
\bea\label{eq:minmcond5}
\biggl< \frac{\partial V_{\rm tree}}{\partial \widetilde{N}} \biggr>\biggr|_{\rm VEVs} &=& 
\l_N v_N \left( -\l v_u v_d + \kappa v^2_S + \frac{\l_N}{2} v^2_N\right)
+ \l_N v_S \left( \sum^3_{i=1} Y^i_N v_i v_u + \l_N v_S v_N\right)\nn\\
&+& \sum^3_{j=1} Y^j_N v_j \left( \sum^3_{i=1} Y^i_N v_i v_N -\l v_S v_d \right)
+ Y^{i^2}_N v^2_u v_N + M^2_N v_N  \nn\\
&+&\sum^3_{i=1} (A_N Y_N)^i v_i v_u +\l_N A_{\l_N} v_S v_N.
\eea

\bibliographystyle{JHEP}
\bibliography{NMSSM1RHNv3}  
	
\end{document}